\documentclass[fleqn,usenatbib]{rasti}

\usepackage{newtxtext,newtxmath}
\usepackage[T1]{fontenc}

\DeclareRobustCommand{\VAN}[3]{#2}
\let\VANthebibliography\thebibliography
\def\thebibliography{\DeclareRobustCommand{\VAN}[3]{##3}\VANthebibliography}

\usepackage{graphicx}	% Including figure files
\usepackage{amsmath}	% Advanced maths commands
\usepackage[version=4]{mhchem}
\usepackage{comment}
\usepackage{subcaption}
\usepackage{lineno}
\usepackage{ulem}

\title[Detection of hydrocarbons in Titan with HRCCS]{Detection of hydrocarbons in Titan using high-resolution cross-correlation spectroscopy}

\author[Maria Coelho et al.]{
Maria Coelho,$^{1}$
Rafael Rianço-Silva,$^{2,3,4}$\thanks{E-mail: rafael.silva@ucl.ac.uk}
Diogo Gonçalves,$^{5}$
Pedro Machado$^{2,3}$ and 
Zita Martins$^{5}$
\\
$^{1}$Physics Department, Instituto Superior Técnico, University of Lisbon, Portugal\\
$^{2}$Instituto de Astrofísica e Ciências do Espaço, Universidade de Lisboa, OAL, Edifício Leste, Tapada da Ajuda, PT1349--018 Lisbon,
Portugal\\
$^{3}$Departamento de Física, Faculdade de Ciências, Universidade de Lisboa, Edifício C8, Campo Grande, PT1749--016 Lisbon, Portugal\\
$^{4}$Department of Physics and Astronomy, University College London, Gower Street, WC1E 6BT London, United Kingdom\\
$^{5}$Centro de Química Estrutural, Institute of Molecular Sciences, and Department of Chemical Engineering, Instituto Superior Técnico, University of Lisbon, Portugal
}

\date{Accepted XXX. Received YYY; in original form ZZZ}

\pubyear{\the\year{}}

\begin{document}
%\pagewiselinenumbers
\label{firstpage}
\pagerange{\pageref{firstpage}--\pageref{lastpage}}
\maketitle

% Abstract of the paper
\begin{abstract}
High-resolution cross-correlation spectroscopy (HRCCS) is a powerful technique for detecting molecules whose individual spectral lines are too weak to be identified directly, but its sensitivity is limited by the availability and quality of high-resolution opacity data. Many molecules of atmospheric and astrobiological interest lack complete line lists, restricting traditional template-based searches. In this work, we use Titan as a controlled testbed to develop and validate a new cross-section-based methodology for HRCCS template construction. We analyse K-band CRIRES+ observations of Titan (1.99--2.48 \textmu m), and compute cross-correlation functions using both line-by-line and cross-section-based templates. Our analysis recovers known hydrocarbons such as methane (\ce{CH4}) and acetylene (\ce{C2H2}), and yields the first HRCCS detection of ethane (\ce{C2H6}) with a significance of SNR$_{\text{peak}}=5.17\pm0.07$. The ethane detection was made possible exclusively through cross-section-based templates, as no high-resolution line list currently exists for this molecule. These results demonstrate that cross-section-based template construction is a practical and powerful strategy for extending HRCCS to molecules that currently lack reliable line lists, and establish Titan as a benchmark for calibrating molecular detection techniques that can be applied to both solar system and exoplanet atmospheres. Future applications of this approach to other ground-based high-resolution spectrographs, as well as to JWST's highest-resolution modes and next-generation facilities such as the ELT, could significantly expand the inventory of molecules detectable in planetary atmospheres.
%This is a simple template for authors to write new RASTI papers.
%The abstract should briefly describe the aims, methods, and main results of the paper.
%It should be a single paragraph not more than 250 words.
%No references should appear in the abstract.
\end{abstract}

% Include between one and six keywords.

%%%%%%%%%%%%%%%%%%%%%%%%%%%%%%%%%%%%%%%%%%%%%%%%%%

%%%%%%%%%%%%%%%%% BODY OF PAPER %%%%%%%%%%%%%%%%%%

\section{Introduction}

\subsection{The high-resolution cross-correlation technique for (exo)planetary atmospheres}

High‑resolution spectroscopy (HRS) has become one of the most powerful techniques for probing the chemical diversity of planetary atmospheres \citep{Snellen_2025}. Nearly 6,000 exoplanets have been discovered in the past three decades, and spectroscopy of their atmospheres is thought to be the main tool to address fundamental questions about planetary composition, formation, evolution and, ultimately, habitability \citep{Yurchenko_2025}. Among the various observational techniques, high‑resolution cross‑correlation spectroscopy (HRCCS) has emerged as a particularly effective method for detecting molecules whose individual spectral lines are too weak to be identified directly \citep{Snellen_2025, Yurchenko_2025}. HRCCS has been used to confirm the presence of individual atomic and molecular species \citep{Snellen_2010, Birkby_2013, Nugroho_2017, Guilluy_2019, Hoeijmakers_2019, Yan_2019, Giacobbe_2021, Pelletier_2023, Flagg_2023}, providing a qualitative inventory of atmospheric constituents, and is increasingly being applied to extract quantitative information on temperature structure, metallicity, and elemental ratios such as the carbon-to-oxygen ratio (C/O), which are considered to be important tracers of planet formation and early evolution \citep{Oberg_2011, Mordasini_2016, Lothringer_2021}.

The HRCCS method is based on the cross-correlation of the observed high-resolution spectrum with a pre-calculated model spectrum of the planetary atmosphere \citep{Snellen_2010, Brogi_2012, Birkby_2013}. The method uses an accurate radiative transfer model to amplify the faint signature of the exoplanet, leveraging the Doppler shift caused by the planet’s orbital motion to distinguish exoplanetary spectroscopic features from the telluric spectra \citep{Birkby_2018}. Typical HRCCS studies use resolving powers greater than \(R\approx25{,}000\), often reaching \(R\) = 50,000--100,000 with ground-based telescopes in the optical and near-infrared \citep{Yurchenko_2025}.

Despite its strengths, HRCCS often operates ‘in the dark’, with little or no prior information about the molecular composition of the target atmosphere, requiring the consideration of a much broader set of possible species \citep{Yurchenko_2025}. As a result, HRCCS depends entirely on the quality of the model spectrum even for the initial detection step, since an incorrect spectral model would produce no (or negligible) correlation \citep{Yurchenko_2025, Birkby_2018}. This strong reliance on accurate models introduces tangible challenges for HRCCS studies. In this context, solar system targets provide a valuable opportunity to overcome some of these limitations, since atmospheric composition is often better constrained and the signal‑to‑noise ratio is significantly higher \citep{Branco_2024, Niraula_2025}. These conditions allow HRCCS techniques to be tested, validated, and refined before being applied to exoplanets.

Because HRCCS ultimately relies on accurate radiative‑transfer models, detection sensitivity is governed by how well the spectral template captures the molecule’s absorption pattern within the observed spectrum. Optimising these templates is therefore essential for improving detection sensitivity. Yet, many molecules of astrobiological interest lack high‑resolution line‑by‑line opacity data \citep{Tennyson_2016, Rothman_2010, Gordon_2026, Tennyson_2022}, preventing traditional template‑based searches. In contrast, laboratory absorption cross‑sections are far easier to obtain and are available for a much wider range of molecules \citep{Gordon_2026}, even though they depend on specific temperature and pressure conditions and are therefore not as universally applicable as line lists. Cross‑sections contain the combined opacity of all transitions at a given state, and despite including both broad continuum structure and narrow absorption features, careful baseline removal and peak selection could allow the extraction of templates suitable for HRCCS. This makes cross‑section‑based template construction a practical and powerful alternative for expanding molecular searches beyond the limited set of species with complete line lists.

Once a suitable template is computed, whether from line lists or cross‑sections, the detection significance \(\sigma\) is evaluated using statistical measures that quantify the match between the observed signal and the expected molecular pattern, typically expressed as the signal‑to‑noise ratio (SNR) of the cross-correlation function (CCF) peak \citep{Snellen_2010, Brogi_2012, Yurchenko_2025}. This metric has been central to establishing the growing list of molecular detections obtained with HRCCS. At the time of writing, many atomic and molecular species have been detected in exoplanet atmospheres using HRCCS, including CO \citep{Snellen_2010}, \ce{H2O} \citep{Birkby_2013}, TiO \citep{Nugroho_2017}, HCN \citep{Guilluy_2019, Cabot_2019, Giacobbe_2021}, \ce{CH4} \citep{Guilluy_2019}, \ce{NH3} \citep{Guilluy_2019}, \ce{C2H2} \citep{Giacobbe_2021}, OH \citep{Nugroho_2021}, VO \citep{Pelletier_2023} and CrH \citep{Flagg_2023}, as well as numerous atoms and atomic ions, such as Fe, Ti, Mg, and Ca \citep{Hoeijmakers_2019, Yan_2019}. Among these, hydrocarbons are of particular interest: they trace a planet’s C/O ratio and metallicity and may also be produced photochemically in the upper atmosphere \citep{Madhusudhan_2012}. Despite their importance, detections of hydrocarbons other than methane remain scarce. For instance, only one claim of \ce{C2H2} has been reported so far: a 6.1\(\sigma\) detection in HD 209458 b using GIANO \citep{Giacobbe_2021} with the ExoMol molecular line list of \citet{Chubb_2020}. The scarcity of detections also reflects the strong spectral degeneracies within the hydrocarbon family. Many hydrocarbons exhibit highly similar absorption patterns, particularly in the 3.0--3.5 \textmu m region dominated by the C--H stretching mode, making it difficult to distinguish between species even when a signal is present \citep{Sousa_Silva_2019, Gasman_2022, Niraula_2025}. These degeneracies reduce the ability to uniquely identify individual hydrocarbons, especially in low-resolution observations \citep{Niraula_2025}. By contrast, HRCCS isolates individual spectral lines and is therefore intrinsically more sensitive to subtle differences between hydrocarbon line patterns, provided that accurate, high‑resolution templates are available.

Methane provides another illustrative example of the challenges associated with hydrocarbon detections. Before the launch of the James Webb Space Telescope (JWST), bona fide detections of molecular species in exoplanet atmospheres from space were mostly limited to \ce{H2O} \citep{Sing_2016}, since the restricted wavelength coverage and coarse resolving power (\(R \sim\) 30--200) of space-borne instruments prevented the identification of other molecules \citep{Esparza_Borges_2025, Brogi_2019}. JWST has transformed this landscape by providing, for the first time, a low‑to‑moderate resolution (\(R \sim\) 100--3500) and a broad and continuous spectral coverage (0.6--12 \textmu m) sensitive to a wide range of molecular species \citep{Greene_2016}. Within only three years, the catalogue of atmospheric detections has expanded significantly, enabling several detections of \ce{CH4}, including in WASP‑80 b \citep{Bell_2023}, K2‑18 b \citep{Madhusudhan_2023} and TOI‑270 d \citep{Holmberg_2024}. JWST now enables detections of multiple species, typically through computationally intensive retrieval models that estimate abundances, temperatures and elemental ratios \citep{Esparza_Borges_2025}. Cross‑correlation functions (CCFs) offer a complementary approach \citep{Snellen_2010, Brogi_2012}: while less efficient than retrievals at constraining atmospheric parameters, they are more time‑efficient at detecting molecular features and less sensitive to systematics in transmission spectra \citep{Esparza_Borges_2025}. CCFs are routinely used in very high‑resolution ground‑based observations, and have now also been demonstrated on JWST’s highest‑resolution modes, particularly in NIRSpec/G395H transmission and direct‑imaging observations \citep{Esparza-Borges_2023, Gandhi_2023}.

In parallel with low-resolution spectroscopy, HRCCS searches for methane have been ongoing since the early 2000s \citep{Wiedemann_2001} and continued systematically but largely unsuccessfully during the 2010s, partly due to the hotter temperatures targeted \citep{Yurchenko_2025}. The first evidence (4.1\(\sigma\)) of methane came from HD 102195 b using GIANO (\(R\) = 50,000, 0.95--2.45 \textmu m) \citep{Guilluy_2019} and the HITRAN 2016 line list \citep{Gordon_2016}. Yet, the detection was not confirmed when reanalysed \citep{Gandhi_2020} with the more extensive HITEMP 2020 line list \citep{Hargreaves_2020}. HITEMP 2020 has since been used to detect methane in HD 209458 b (5.6\(\sigma\)) \citep{Giacobbe_2021}, WASP‑69 b (4.9\(\sigma\)) \citep{Guilluy_2022} and WASP‑80 b (4.2\(\sigma\)) \citep{Carleo_2022}. Recent efforts have also focused on extending methane line lists into the visible to improve HRCCS sensitivity in optical wavelengths \citep{RiancoSilva_2024, RiancoSilva_2026b}, underscoring the need for complete and accurate opacity data for this molecule.

The next decade will see a major expansion of HRCCS capabilities with the arrival of the Extremely Large Telescope (ELT) and its high‑resolution spectrographs, such as ANDES (0.4--1.8~\textmu m) \citep{Marconi_2021} and METIS (3--13~\textmu m) \citep{Brandl_2021}. With its 39‑m primary mirror, advanced adaptive‑optics system, and unprecedented collecting power, the ELT will enable studies of smaller and cooler exoplanets, including temperate sub‑Neptunes and potentially rocky worlds \citep{Padovani_2023}. In the solar system, it will enable repeated high‑resolution imaging and spectroscopy of planets and moons with evolving atmospheres and surfaces, probe volcanic activity and water plumes on the Jovian and Saturnian moons, and characterise faint outer‑system bodies such as Uranus, Neptune, and Kuiper Belt objects \citep{Padovani_2023}. The enhanced sensitivity of the ELT will also amplify the importance of accurate opacity data: HRCCS searches will be able to target smaller, cooler, and more chemically diverse planets, including species for which no high‑resolution line lists currently exist. In this context, the new cross‑section‑based template construction method we present here provides a practical pathway for fully exploiting the chemical reach of next‑generation instruments.

Taken together, these research avenues point to the need for template‑construction methods that improve detection sensitivity and extend HRCCS to a wider range of hydrocarbons. In this work, we address three main challenges: (i) the lack of complete high‑resolution line lists for many molecules of atmospheric and astrobiological interest, (ii) the resulting difficulty in constructing accurate HRCCS templates, and (iii) the spectral degeneracies that hinder the identification of hydrocarbons, particularly in low-resolution observations. To overcome these limitations, we apply HRCCS to Titan as a controlled testbed to calibrate and improve molecular detection techniques, combining traditional line-by-line templates with a new cross‑section‑based methodology. Using CRIRES+ K-band observations of Titan (1.99--2.48 \textmu m), we demonstrate that this approach not only recovers known hydrocarbons such as methane (\ce{CH4}) and acetylene (\ce{C2H2}), but also enables the first HRCCS detection of ethane (\ce{C2H6}). This result establishes cross‑section‑based template construction as a powerful and broadly applicable tool for high‑resolution atmospheric studies, opening the door to systematic searches for molecules that currently lack reliable line lists.

\subsection{Limitations of line lists and the need for alternative opacity data}

Line lists provide the frequencies, intensities, and quantum assignments of individual transitions and are the foundation of line‑by‑line radiative‑transfer modelling. For increasingly complex molecules (i.e. hydrocarbons), the computational extraction of line lists through \textit{ab initio} calculations becomes progressively challenging, with millions to billions of transitions contributing to the opacities \citep{Yurchenko_2025}. Thus, line lists for heavier molecules are often incomplete or entirely unavailable, and constructing fully assigned, accurate, and comprehensive line lists experimentally is hindered by the sheer volume of data, the difficulty of obtaining absolute line strengths, and the need for broad wavelength coverage \citep{Tennyson_2012}. In practice, line lists for molecules with more than about six atoms are currently unattainable \citep{Yurchenko_2025}. As an example, the molecule with the most atoms with line lists available at ExoMol is \ce{C2H4} with six atoms \citep{Mant_2018}.

Due to these constraints, for molecules larger than six atoms, our knowledge of their spectral structure is solely based on laboratory measurements -- what we call here absorption cross-sections \citep{Yurchenko_2025}. Cross‑sections have been part of HITRAN since 1986 \citep{Rothman_1987}, and their importance has grown substantially with each new release. In HITRAN2024, absorption cross‑sections are available for 644 molecules, nearly doubling the coverage of HITRAN2020 \citep{Gordon_2022}, reflecting the difficulty of generating reliable line‑by‑line parameters for heavy polyatomic species. Many of these molecules are of high scientific interest: they include complex hydrocarbons, nitriles, alcohols, ethers, and oxygenated organics relevant to planetary atmospheres, climate studies, and even biosignature searches \citep{Gordon_2026}. For most of these species, cross‑sections are the only available opacity data. Other spectroscopic databases, such as MOLLIST \citep{Bernath_2019}, provide high‑resolution cross‑sections and selected line lists for a subset of hydrocarbons and nitriles species.

Despite their advantages, cross‑sections have limitations. Because they are measured at specific temperature-pressure conditions, they lack the flexibility of line lists, which can be used to simulate atmospheric spectra across diverse physical conditions. Nevertheless, for many complex molecules, especially hydrocarbons, cross‑sections remain the only available opacity format. In this context, cross‑section‑based template construction offers a powerful complement to traditional line-by-line models. By extracting narrow absorption features from high‑resolution cross‑sections, it becomes possible to build templates suitable for high-resolution cross-correlation spectroscopy (HRCCS) for large molecules that would otherwise be inaccessible. This motivates the methodology developed in this work, which leverages cross‑sections to expand the chemical reach of HRCCS.

\subsection{Titan as a testbed for exoplanets}

Titan, Saturn’s largest moon, provides an ideal environment for developing and testing observational techniques dedicated to studying exoplanet atmospheres \citep{Niraula_2025, Chubb_2024}, including HRCCS. Its atmospheric composition is well characterised thanks to decades of observations from Cassini-Huygens and ground‑based facilities \citep{Hanel_1981, Kunde_1981, Samuelson_1981, Samuelson_1983, Lutz_1983, Bezard_1993, Coustenis_1998, Coustenis_2007}, offering a reliable reference against which new methods can be validated \citep{Niraula_2025}. Titan’s atmosphere hosts an exceptionally rich photochemistry, producing a wide variety of hydrocarbons, nitriles, and other organic molecules relevant to prebiotic chemistry \citep{Nixon_2024, Horst_2017}. High‑resolution spectra of Titan are available in both the optical and infrared, and the signal‑to‑noise ratio achievable for this bright solar system target is far superior to that of exoplanets. As a result, any detection obtained through HRCCS is significantly easier to interpret. Titan also remains a promising environment for the discovery of new molecular species, and for improving our knowledge of those already known but so far poorly constrained or only detected at low spectral resolution \citep{Nixon_2024}, making it an excellent laboratory for refining HRCCS methodologies.

Beyond its observational advantages, Titan hosts one of the most chemically complex and Earth‑analogous atmospheres in the solar system, composed predominantly of nitrogen (\ce{N2}) and methane (\ce{CH4}). Titan's atmosphere supports a remarkable combination of physical and chemical processes, including surface liquids \citep{Mastrogiuseppe_2019}, cloud formation \citep{Hanson_2025, Trenquelleon_2026}, organic chemistry \citep{MacKenzie_2021}, and dynamic climate patterns \citep{Vuitton_2024}. Titan is the only moon known to possess a substantial atmosphere and the only solar system body besides Earth with stable liquid reservoirs, albeit of methane-ethane mixtures rather than water \citep{Mastrogiuseppe_2019, Hayes_2018}.

Besides molecular nitrogen and methane, 22 additional molecular species have been identified, including ten hydrocarbons (\ce{C2H2}, \ce{C2H4}, \ce{C2H6}, \ce{c-C3H2}, \ce{CH2CCH2}, \ce{CH3CCH}, \ce{C3H6}, \ce{C3H8}, \ce{C4H2}, and \ce{c-C6H6}), eight cyanides (HCN, HNC, \ce{HC3N}, \ce{C2N2}, \ce{CH3CN}, \ce{C2H3CN}, \ce{C2H5CN}, and \ce{CH3C3N}), three oxygen‑bearing species (CO, \ce{CO2}, and \ce{H2O}), and molecular hydrogen (\ce{H2}) \citep{Nixon_2024}. These molecules were initially detected through a combination of remote‑sensing and astronomical techniques from both ground‑based and space‑based platforms \citep{Hanel_1981, Kunde_1981, Samuelson_1981, Samuelson_1983, Lutz_1983, Bezard_1993, Coustenis_1998, Coustenis_2007}. Most recently, \ce{C3}, a photochemical species mostly produced on Titan’s upper atmosphere by the reaction of \ce{C2H2} with atomic carbon, has been detected with ground-based high-resolution spectroscopy in the optical wavelength range \citep{RiancoSilva_2024, RiancoSilva_2026}.

Although all major hydrocarbon families have been detected, several nitrogen‑bearing chemical families predicted by photochemical models remain unobserved (including amines, imines, azines, and N-heterocyclic rings) \citep{Nixon_2024}. Moreover, no oxygen‑bearing organic compounds have yet been detected in Titan, limiting the scope of astrobiological molecules (i.e., those with the elements CHON in a variety of functional groups) \citep{Nixon_2024}.

The chemical reaction pathways among Titan’s atmospheric constituents have been incorporated into increasingly complex computational models that have largely succeeded in reproducing observed gas abundances \citep{Nixon_2024}. Early models, developed prior to Cassini-Huygens, focused mainly on replicating neutral gas concentrations observed by Voyager and the Infrared Space Observatory (ISO), with some also addressing ionospheric processes \citep{Lara_1994, Lara_1996, English_1996, Wilson_2004}. The extensive dataset returned by Cassini-Huygens, particularly its ionospheric sampling, enabled the development of numerous new and refined atmospheric models \citep{Lavvas_2008I, Lavvas_2008II, Robertson_2009, Krasnopolsky_2009, Krasnopolsky_2012, Dobrijevic_2014, Dobrijevic_2016,  Willacy_2016, Vuitton_2019}.

By providing a new method to identify chemical species, our work also offers a pathway to refine Titan’s atmospheric models. Many of Titan’s hydrocarbons are detected only at low spectral resolution, while some remain poorly constrained \citep{Nixon_2024}. HRCCS detections could help to validate or revise model predictions and test the completeness of current photochemical networks, ultimately informing \textit{in-situ} measurements by NASA’s upcoming Dragonfly \citep{Barnes_2021} and by later missions to Titan's hydrocarbon lakes or evaporitic deposits \citep{Goncalves_2026, Martins_2024, Mitri_2021, Lorenz_2018, Stofan_2013}.

Overall, Titan’s well‑characterised atmospheric structure, extensive molecular inventory, and uniquely favourable observing conditions make it an exceptional analogue for exoplanet atmospheres \citep{Niraula_2025}. Thus, Saturn's largest moon serves not only as a natural laboratory for atmospheric chemistry but also as a critical testing ground for advancing molecular detection strategies that will ultimately enhance our ability to characterise the atmospheres of distant worlds, within and beyond our solar system.

\section{Observations \& data reduction}

Titan was observed with the CRyogenic InfraRed Echelle Spectrograph Upgrade Project (CRIRES+) at the ESO Very Large Telescope (VLT) over three nights, each Titan exposure being paired with the G‑type calibration star HD 198802, observed at a similar airmass (Program ID: 110.23UK, PI: Bruno Bézard). CRIRES+ is a cross-dispersed, high-resolution infrared spectrograph that offers multiple wavelength settings across the near‑infrared \citep{Dorn_2023}. For this dataset, the K‑band K2166 configuration was used, which covers six spectral orders between $\sim1.99-2.48$ \textmu m, each divided into three detectors. This results in a total of 18 spectral bands per exposure, with small wavelength gaps between adjacent detectors. Table \ref{tab:observations} summarises the observations performed in 3 nights of October 2022 and the wavelength coverage of each spectral order.

Titan’s resolved disk subtends approximately 0.8" during the observations, while the CRIRES+ slit width is 0.2" and the slit length is 10". The slit was oriented at a position angle of 6.6\textdegree, corresponding to an almost North--South alignment in the sky. As a result, the spatial direction along the slit samples both hemispheres of Titan and includes contributions from the limbs. Titan’s equatorial rotation speed is \(\sim\)12 m~s\(^{-1}\), implying a maximum Doppler difference of \(\lesssim 25\) m~s\(^{-1}\) between opposite limbs, which is far below the CRIRES+ resolution element (\(R \approx\) 100,000, \(\Delta v \approx 3\) km~s\(^{-1}\)). Consequently, Titan’s rotational velocity gradient does not broaden or shift the atmospheric lines in any measurable way, and the slit orientation has no impact on the HRCCS detections.

Before describing the data reduction, it is important to highlight how Titan observations differ fundamentally from the exoplanet transmission and emission spectra for which most high-resolution cross-correlation spectroscopy (HRCCS) pipelines were originally developed. In exoplanet spectroscopy, the planetary signal is embedded within the much brighter stellar spectrum and is further contaminated by telluric absorption \citep{Snellen_2025}. Exoplanet data analysis, therefore, begins with the removal of stellar and telluric features through normalisation, high-pass filtering, and detrending algorithms such as \texttt{SysRem} \citep{Tamuz_2005}, which mitigate time‑variable throughput, airmass effects, and subtle wavelength‑solution drifts \citep{Snellen_2025}.

In contrast, Titan is a bright, spatially resolved Solar System target whose spectrum is dominated by Titan’s own atmospheric absorption. For this reason, applying a standard \texttt{SysRem} algorithm to remove telluric contribution would introduce unnecessary complexity and risk distorting Titan’s intrinsic spectral features. Instead, only a wavelength-dependent airmass correction is applied, as described later in this section. Although telluric lines remain in the data, they do not interfere with the HRCCS analysis (a detailed discussion of residual telluric contamination and its impact on the HRCCS analysis is provided in Section \ref{sec:telluric_contamination}). After shifting each spectrum into Titan’s rest frame, Titan’s intrinsic lines align across all nights, whereas telluric lines, fixed in the observatory frame, appear at different apparent wavelengths due to the changing Titan--Earth relative velocity \citep{RiancoSilva_2024}. In velocity space, this Doppler correction places any Titan-originating signal near \(v=0\)~km~s\(^{-1}\) in the CCF, while residual telluric features are displaced away from zero velocity and therefore do not add coherently, contributing only to the noise budget. This preserves Titan’s intrinsic line information without the distortions that would arise from dividing by a telluric standard or applying blind detrending algorithms. The reduction is therefore substantially simpler than in exoplanet HRCCS and avoids the risk of removing Titan’s atmospheric signal.

For completeness, we note that Titan’s near‑infrared spectrum contains solar Fraunhofer lines imprinted on the backscattered sunlight, but these atomic features do not correlate with the molecular templates used in this work. Furthermore, the solar Fraunhofer lines are Doppler‑shifted by Titan’s heliocentric velocity, which varies between nights, so they behave analogously to the telluric lines in the HRCCS analysis and contribute only noise to the CCF.

\begin{table}
    \centering
    \caption{Summary of CRIRES+ observations of Titan and calibration star (Program ID: 110.23UK, PI: Bruno Bézard). Each night consists of a Titan--star pair observed over six spectral orders, each split across three detectors.}
    \label{tab:observations}
    \begin{tabular}{lllll}
        \hline
        Night & Target & Time & Airmass & $v_{\mathrm{Titan-Earth}}$ (km/s) \\
        \hline
        2022-10-09 & Titan & 02:15:52.7484 & 1.067 & 22.40 \\
                    & Star  & 03:14:21.9955 & 1.348 &        \\
        2022-10-20 & Titan & 01:20:41.7385 & 1.052 & 31.89 \\
                    & Star  & 02:28:04.0027 & 1.334 &        \\
        2022-10-24 & Titan & 00:45:39.3763 & 1.031 & 27.88 \\
                    & Star  & 01:43:14.7724 & 1.221 &        \\
        \hline
    \end{tabular}
    \vspace{0.5cm}
    \begin{tabular}{lll}
        \hline
        Spectral order & Wavelength range (nm) & Detectors \\
        \hline
        1 & 1988--2035 & 3 \\
        2 & 2060--2110 & 3 \\
        3 & 2140--2190 & 3 \\
        4 & 2228--2275 & 3 \\
        5 & 2320--2370 & 3 \\
        6 & 2420--2475 & 3 \\
        \hline
    \end{tabular}
\end{table}

The raw CRIRES+ data are processed by the standard ESO CR2RES pipeline recipe, returning one-dimensional, wavelength-calibrated spectra. The spectra are first grouped by target (Titan or star) and by night, and then divided into wavelength bands corresponding to the six CRIRES+ spectral orders. Within each order, the spectrum is further split into three detector segments, yielding a set of bands indexed by \texttt{(order, detector)}. To avoid detector edge artefacts, a fixed number of pixels is removed from both ends of each detector segment (10 pixels on each side).

Before any cross-correlation analysis, the spectra are corrected for atmospheric extinction and normalised. Atmospheric extinction is treated using Bouguer’s law, which expresses the observed magnitude as a linear function of airmass, 
\begin{equation}
    m_{\mathrm{obs}}(\lambda) = m(\lambda) + k(\lambda)X, 
\end{equation}
where \(m_{\mathrm{obs}}(\lambda)\) is the magnitude as observed inside the atmosphere, \(m(\lambda)\) is the magnitude outside the atmosphere, \(k(\lambda)\) is the wavelength-dependent extinction coefficient and \(X\) is the airmass \citep{Chromey_2010}. The extinction coefficients are computed using the Paranal \texttt{AtmosphericExtinction} model \citep{Patat_2011} implemented in the \texttt{specreduce} package. Using \(m=-2.5 \log_{10} F\), this relation is expressed in flux units as
\begin{equation}
    F_{\mathrm{corr}}(\lambda) = F_{\mathrm{obs}}(\lambda)\,10^{0.4\,k(\lambda)\,X},
\end{equation}
where \(F_{\mathrm{obs}}(\lambda)\) is the observed flux and \(F_{\mathrm{corr}}(\lambda)\) is the extinction-corrected flux. The same correction is applied to the associated uncertainties. The normalisation combines a percentile-based continuum estimate, which traces the upper envelope of the spectrum, with Savitzky-Golay smoothing to produce a stable continuum \citep{Savitzky_1964}. The procedure is applied independently to each spectral order and detector segment to prevent discontinuities from propagating across the spectrum. The normalised flux is obtained by dividing the corrected spectrum by the continuum,
\begin{equation}
    f_{\mathrm{norm}}(\lambda) = \frac{F_{\mathrm{corr}}(\lambda)}{C(\lambda)},
\end{equation}
and the measurement errors are scaled accordingly. All subsequent analysis is performed in Titan's rest frame. Because Titan's radial velocity relative to Earth varies between observing nights, the Doppler correction is applied after the spectra have been fully corrected and normalised. For each exposure, the wavelength array $\lambda_{\text{obs}}$ is transformed according to
\begin{equation}
    \lambda_{\mathrm{rest}} = \frac{\lambda_{\mathrm{obs}}}{1 + \dfrac{v_{\mathrm{Titan-Earth}}}{c}}.
\end{equation}
where $v_{\mathrm{Titan-Earth}}$ is the Titan--Earth radial velocity listed in Table~\ref{tab:observations}.

\section{Methods}

\subsection{Cross-correlation functions}

To identify molecular absorption in the CRIRES+ spectra, we compute the cross-correlation function (CCF) between the observed spectrum and a high-resolution molecular template. Let the continuum-normalised observed flux be $f_i$ at wavelengths $\lambda_i$, and let $t(\lambda_i - \Delta\lambda_{i,j})$ be the template evaluated at the same wavelengths after being Doppler-shifted by a trial radial velocity $v_j$. The wavelength shift is
\begin{equation}
    \Delta\lambda_{i, j} = \lambda_i \frac{v_j}{c},
\end{equation}
where $c$ is the speed of light. The discrete cross-correlation at velocity $v_j$ is then
\begin{equation}
    CC(v_j) = \sum_{i=1}^{N} \alpha_i \, f_i \, t(\lambda_i - \Delta\lambda_{i,j}),
\end{equation}
where $N$ is the number of spectral points and $\alpha_i$ are optional weights (set to unity in this work). This formulation corresponds to the standard discrete cross-correlation implemented in the \texttt{PyAstronomy} routine \texttt{crosscorrRV} \citep{pya}, and is widely used in high-resolution exoplanet and planetary spectroscopy.

In practice, the CCF is computed independently for each CRIRES+ spectral band. For each band, we evaluate the CCF over a radial-velocity range of $-150$ to $+150$~km~s$^{-1}$, sampled in steps of 1.0~km~s$^{-1}$. At each trial velocity, the template is Doppler-shifted and linearly interpolated onto the wavelength grid of the observed spectrum before the CCF is evaluated.

Because each CRIRES+ detector covers a separate and discontinuous wavelength interval, the spectra from different detectors cannot be merged prior to cross-correlation \citep{Birkby_2017}. Each detector segment is therefore cross-correlated independently. However, since all CCFs share the same velocity grid, they can be summed directly \citep{Birkby_2017}. For each night, the CCFs from all available bands are added together, producing a nightly CCF. This summation coherently reinforces Titan’s molecular signal while averaging down uncorrelated noise. Noise statistics are estimated from the far wings of the CCF ($|v| > 50$~km~s$^{-1}$), allowing the CCF to be normalised into units of SNR. Let \(CC(v_j)\) be the raw cross‑correlation value at velocity \(v_j\), and let \(\langle CC \rangle_{\mathrm{noise}}\) and \(\sigma_{\mathrm{noise}}\) be the mean and standard deviation of the CCF evaluated in the noise window (\(|v| > 50\)~km~s\(^{-1}\)). The normalised CCF is then
\begin{equation}
\mathrm{CC}_{\mathrm{SNR}}(v_j)
    = \frac{CC(v_j) - \langle CC \rangle_{\mathrm{noise}}}
           {\sigma_{\mathrm{noise}}}.
\end{equation}
The same procedure is repeated across all nights. The nightly CCFs are then summed to produce a master CCF \citep{Birkby_2017}. This final CCF represents the combined molecular signal across all spectral orders, detectors, and nights.

A molecular detection manifests as a statistically significant peak at Titan’s rest-frame velocity. To quantify the statistical significance of each detection, we perform a Monte Carlo Markov chain (MCMC) resampling procedure following \citet{Esparza_Borges_2025}, allowing us to propagate the observed flux uncertainty to CCF space. Each point in the normalised spectrum is perturbed randomly within its flux uncertainty, generating 2000 realisations of the transmission spectrum. For each realisation, the CCF is computed using the same methodology described above. The flux at every wavelength point is perturbed according to a Gaussian distribution centred on the measured flux and with a standard deviation equal to its propagated uncertainty. This ensures that most realisations remain close to the observed spectrum while still sampling the full range of noise fluctuations permitted by the data. For each sampled CCF, we identify the SNR value of the central peak (SNR$_{\text{peak}}$) and the corresponding radial velocity (RV). To determine the statistical significance of each detection, we measured the median value and percentiles on each distribution. The final CCF of the ensemble is the median value computed at each RV point.

\subsection{Direct and indirect method for template construction}

To perform the cross-correlation analysis, we implemented two complementary template-generation strategies: a direct and an indirect method. Although Titan is not a transiting exoplanet, its atmosphere can still be modelled within the same radiative‑transfer framework since the depth of the lines and their relative depth ratios in the template spectra are governed largely by the composition of the model atmosphere and its temperature-pressure profile \citep{Birkby_2018}, which are well constrained. In contrast, parameters such as planetary radius and surface gravity primarily affect the absolute continuum level and the overall scaling of the transit radius spectrum, neither of which influences the relative line‑by‑line pattern used in high-resolution cross-correlation spectroscopy (HRCCS). After continuum normalisation, the template retains only the differential line structure, making the \texttt{petitRADTRANS} \citep{Molliere_2019} forward model fully applicable despite Titan’s different size and gravity. Moreover, although Titan’s spectrum is reflected sunlight modified by atmospheric scattering, these scattering processes affect only the smooth continuum level and slope. The narrow molecular absorption features used in HRCCS are set by Titan’s atmospheric composition and temperature-pressure profile and are unaffected by continuum-level scattering. HRCCS is not sensitive to the continuum or broad low-resolution features, but instead it is sensitive to the high-resolution line structure of the spectrum. After continuum normalisation, any modulation caused by scattering to the continuum is removed, leaving only the differential line-by-line structure relevant for cross-correlation.

The direct approach uses \texttt{petitRADTRANS} to compute high-resolution line-by-line spectra under Titan-like atmospheric conditions, activating only one molecule at a time, with a sampling comparable to the CRIRES+ resolving power. These model spectra contain the full radiative-transfer complexity of the molecule, including blended and low-contrast features. Because the observed CRIRES+ spectra are continuum-normalised, the model transit radius is likewise normalised using a Gaussian filter to isolate the relative line structure. The resulting template therefore preserves the physically accurate pattern of line contrasts predicted by the forward model, making it the most complete representation of the molecular fingerprint.

The indirect method is designed to isolate the information content that is most relevant for cross-correlation, while avoiding unnecessary spectral complexity. Its physical basis rests on two observations. First, in a normalised high-resolution spectrum, the detectability of a molecular line is determined primarily by its absolute line depth. Lines with larger absorption depths contribute more strongly to the cross-correlation signal, whereas very shallow lines contribute mostly noise. For this reason, we identify all absorption features in the model transit spectrum and rank them by decreasing absolute intensity. This ordering naturally prioritises the lines that arise from the highest-opacity regions of the spectrum: those that dominate the molecular fingerprint and are least sensitive to noise.

Second, once the spectrum is normalised, only the relative depths of the lines matter for the cross-correlation. The absolute continuum level is removed, and the CCF responds only to the pattern of line contrasts. To reflect this, each selected line is replaced by a Gaussian whose amplitude matches the relative depth of the corresponding feature in the model spectrum, and whose width is set by the instrumental resolving power. This ensures that the indirect template reproduces the expected line shapes in the CRIRES+ data while retaining only the physically meaningful contrast information. Because the template is constructed from relative intensities, it remains consistent with the normalised observational spectrum, in which the continuum has been divided out.

\begin{figure*}
    \centering
    \includegraphics[width=0.9\linewidth]{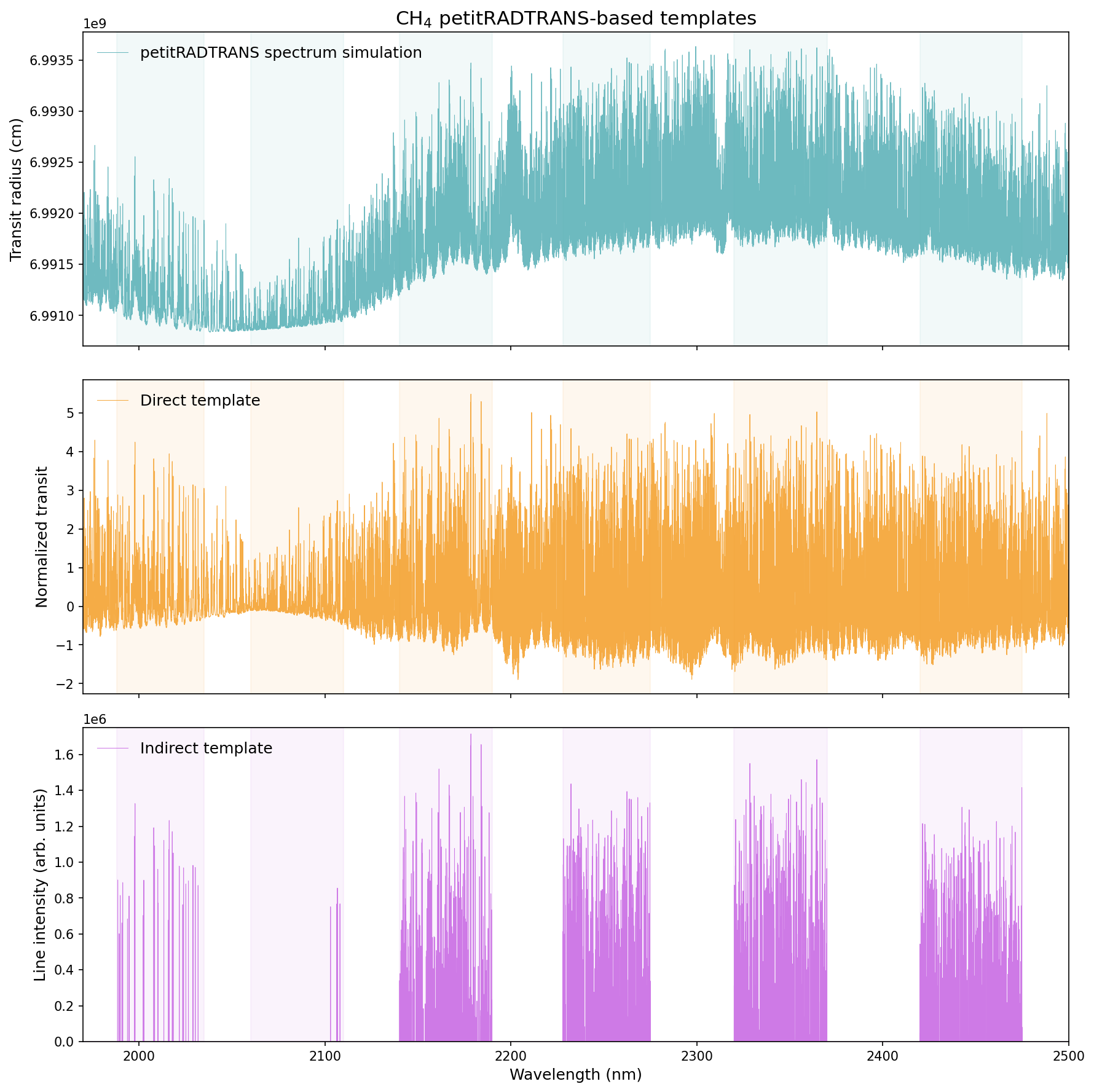}
    \caption{\ce{CH4} \texttt{petitRADTRANS} line-by-line transit spectrum and corresponding direct and indirect templates. The top panel shows the transit spectrum computed with the radiative‑transfer model. The middle panel displays the corresponding direct template, obtained by normalising the transit radius with a Gaussian filter. The bottom panel shows the indirect template constructed from the strongest absorption features (75\% of lines included here). Only lines that fall within the CRIRES+ spectral orders are included in this step, ensuring that the intensity‑based ranking reflects the subset of lines actually used in the cross‑correlation analysis. The line intensities in the indirect template are expressed in arbitrary units proportional to the transit-radius contrast; only their relative pattern is used in the cross-correlation. The shaded regions indicate the wavelength coverage of each CRIRES+ spectral order.}
    \label{fig:CH4_temp}
\end{figure*}

By adjusting the fraction of ranked lines that are retained, we can tune the complexity of the template to the spectral richness of each molecule and to the noise properties of the data. In the limit where all identified lines are retained, the indirect template converges toward the direct template, differing only by the Gaussian reconstruction of each feature. This tunability is a key advantage of the indirect method, allowing the template to be optimised for cross-correlation performance. Figure~\ref{fig:CH4_temp} shows the \ce{CH4} \texttt{petitRADTRANS} line-by-line model and the corresponding direct and indirect templates.

Once both templates are obtained, each normalised CRIRES+ band is cross-correlated against the direct and indirect templates over a range of radial velocities. The resulting CCFs quantify the similarity between the observed spectrum and the molecular fingerprint encoded in the template.

\subsection{Absorption cross-sections: extending CCFs to complex molecules}

For several molecules of astrobiological interest, no high‑resolution line lists are currently available in databases such as HITRAN \citep{Gordon_2026} or ExoMol \citep{Tennyson_2016}. Line lists provide the frequency and intrinsic strength of every individual transition, which is precisely the information required for line‑by‑line radiative‑transfer modelling. In contrast, absorption cross‑sections represent the cumulative opacity obtained when all those transitions are thermally and pressure‑broadened and blended into a ready‑to‑use spectrum at a specific temperature and pressure. Although cross‑sections do not list individual transitions, high‑resolution cross‑sections retain narrow absorption peaks at the correct wavelengths because they encode the cumulative opacity of all blended lines. Temperature and pressure modify the strengths and widths of these peaks, not their central wavelengths, while the number of resolved peaks depends on the degree of line blending, set by temperature, pressure, and the spectral resolution of the cross-section. Since the cross‑correlation is sensitive primarily to the pattern of line contrasts rather than to absolute flux levels, this preserved structure is sufficient to construct simplified templates. These templates capture the dominant molecular fingerprint even in the absence of individual transition data, enabling a physically meaningful extension of the cross‑correlation analysis to molecules for which no line‑by‑line information exists. Importantly, these cross-section spectra are not used as inputs to \texttt{petitRADTRANS}; instead, they are used to construct indirect templates that are compared to the observational data.

\begin{figure*}
    \centering
    \includegraphics[width=.7\linewidth]{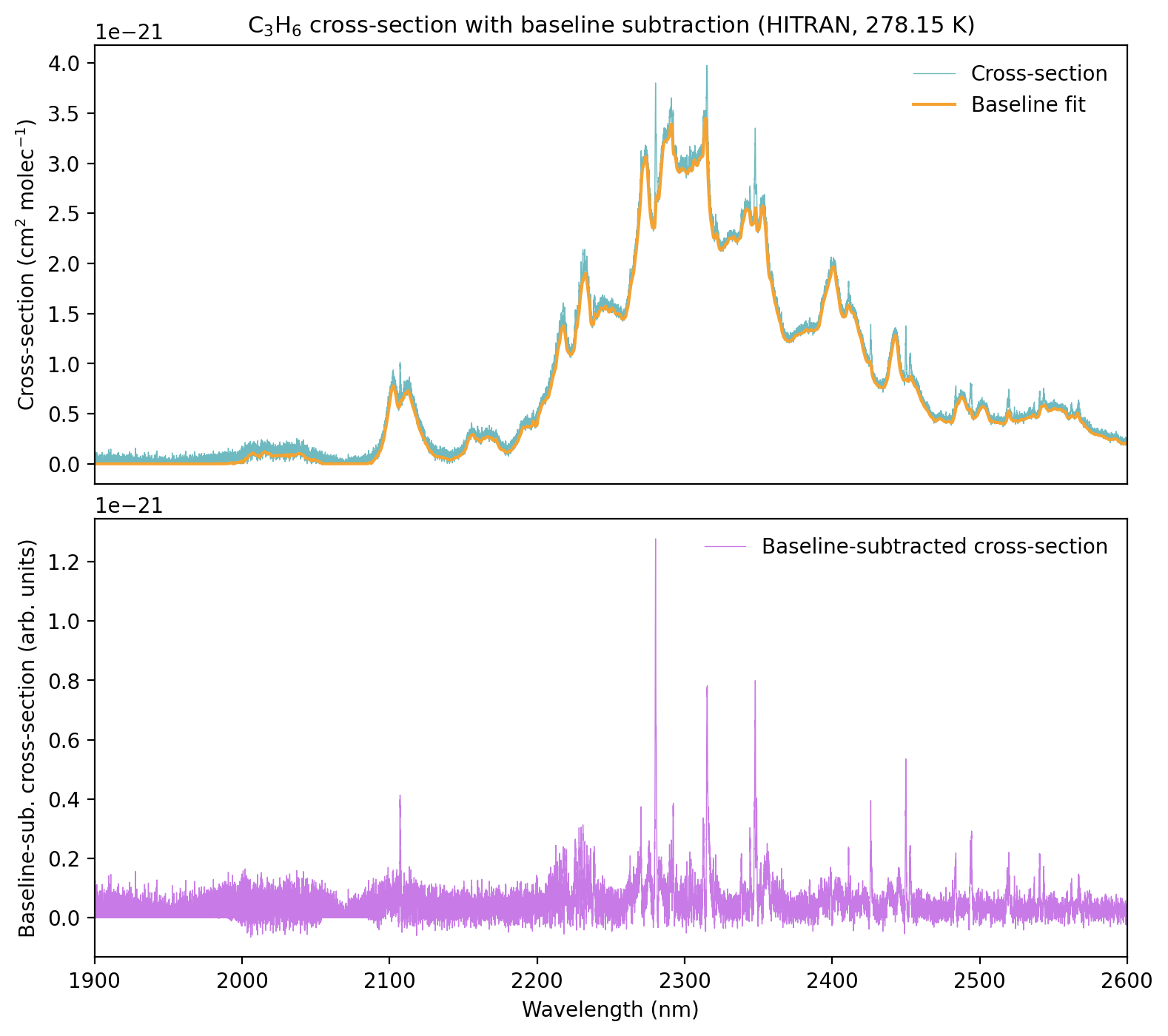}
    \caption{Baseline fitting and subtraction applied to the HITRAN absorption cross‑section of propene (\ce{C3H6}). The top panel shows the raw absorption cross‑section spectrum together with the fitted baseline, which captures the broad-scale structure underlying the molecular features. The bottom panel displays the baseline‑subtracted cross‑section, isolating the narrow absorption peaks that carry the spectroscopic information used in the cross‑correlation analysis.}
    \label{fig:baseline}
\end{figure*}

Because cross‑section spectra contain both broad-scale variations and narrow absorption peaks, an accurate baseline removal is essential. We therefore fitted a smooth baseline using a low‑percentile filter, which traces the lower envelope of the spectrum, followed by Savitzky-Golay smoothing. Subtracting this baseline isolates the relative absorption structure and enhances the contrast of narrow features, allowing them to be treated analogously to individual spectral lines. An example of a baseline-subtracted cross-section is shown in Figure~\ref{fig:baseline}. After baseline subtraction, we apply the same indirect‑template methodology used for the \texttt{petitRADTRANS} models. Absorption peaks are identified and ranked by decreasing absolute intensity, and each peak is replaced by a Gaussian whose width corresponds to the instrumental resolving power. The template is then constructed by retaining a chosen fraction of the ranked peaks. This fraction‑based selection provides a way to tune the template complexity: including only the strongest lines yields a sparse template dominated by the most prominent spectral features, while including a larger fraction incorporates progressively weaker lines. In the limit where all identified peaks are retained, the cross‑section‑based template approaches the full spectral structure encoded in the baseline-subtracted cross‑section, differing only in the Gaussian reconstruction of each feature. The accuracy of the baseline subtraction is therefore critical, as residual curvature can distort or suppress weak peaks and thereby reduce the number of usable features in the final template.

\subsection{Opacity data selection}

To assess the detectability of each molecular species in the CRIRES+ spectrum, we performed a Monte Carlo sampling procedure in which 2000 realisations of the normalised spectrum were generated by perturbing each pixel within its flux uncertainty. For every realisation, the CCF was recomputed using the appropriate molecular template, and the distribution of central-peak SNR values (SNR$_\mathrm{peak}$) was used to quantify both the detection significance and its associated uncertainty.

Four distinct sources of spectroscopic data were used to construct the templates, depending on the availability and quality of line lists or cross-sections for each molecule:

\begin{enumerate}
    \item \textbf{Line-by-line \texttt{petitRADTRANS} templates.}  
    For molecules with high-resolution line lists available (\ce{CH4} \citep{Hargreaves_2020}, $^{13}$\ce{CH4} \citep{Gordon_2022}, \ce{C2H2} \citep{Rothman_2012}, CO \citep{Rothman_2010} and HCN \citep{Harris_2006}), we generated line-by-line model spectra using \texttt{petitRADTRANS} \citep{Molliere_2019} under Titan-like atmospheric conditions. The models were computed assuming an isothermal temperature of 150~K, a vertically extended pressure grid spanning Titan’s surface pressure to thermospheric pressures, and a gas mixture dominated by N$_2$ with the remaining fraction assigned to the target species. N$_2$--N$_2$ collision-induced absorption was included. At Titan-like temperatures, the \ce{C2H2} HITRAN line list is considered complete, but at higher temperatures additional hot-band transitions become important; these are covered by high-temperature databases such as HITEMP \citep{Rothman_2010} and ExoMol \citep{Tennyson_2016}.

    \item \textbf{HITRAN laboratory cross-sections.}  
    The HITRAN \citep{Gordon_2026} cross-sections used in this study were retrieved for each molecule using N$_2$ broadening at the lowest available temperature (278.15~K) \citep{Sharpe_2004}, with a native resolution of 0.112~cm\(^{-1}\). Although this temperature is warmer than Titan’s stratosphere, these data provide the only available pressure-broadened representation of the absorption structure. The cross-section-based method was applied to search for propadiene (\ce{C3H4}), propene (\ce{C3H6}), propane (\ce{C3H8}), 1,3-butadiene (\ce{C4H6}), 1-butyne (\ce{C4H6}), 1-butene (\ce{C4H8}), butane (\ce{C4H10}), benzene (\ce{C6H6}), acrylonitrile (\ce{C2H3CN}) and ethyl cyanide (\ce{C2H5CN}). Because these molecules lack detailed line lists, the cross-section method provides the only feasible way to construct templates for high-resolution analysis. No detections were obtained for these species, as shown in the CCFs from Figures \ref{fig:non_detections_hydrocarbons} and \ref{fig:non_detections_nitriles}.

    \item \textbf{ExoMol computed cross-sections.} The ExoMol \citep{Tennyson_2016} cross-sections used here are opacity tables computed from the underlying ExoMol line lists, distinct from laboratory-measured cross-sections such as those in HITRAN. To validate the cross-section-based template method, we used CH$_4$ and C$_2$H$_2$ cross-sections computed under the same temperature and resolution as the HITRAN data (280~K, 0.112~cm$^{-1}$). The cross-sections were retrieved from \url{https://www.exomol.com/data/data-types/xsec/}, with the underlying line lists described in \citet{Yurchenko_2024} and \citet{Chubb_2020}. Contrary to the HITRAN laboratory cross-sections, the ExoMol cross-sections used here are computed at zero pressure, representing the intrinsic molecular line structure without collisional broadening. Cross‑sections at non‑zero pressure can be computed using ExoCross (\url{https://github.com/ExoMol/ExoCross}), and additional pressure‑broadened cross‑sections (including \ce{H2}/He broadening) are available at \url{https://www.exomol.com/data/data-types/opacity/} \citep{Chubb_2021}. 
    
    The method successfully recovered both \ce{CH4} and \ce{C2H2} [Figures~\ref{fig:CH4_results} and \ref{fig:C2H2_results}], confirming that the cross-section method can successfully extract molecular information from high-resolution spectra, provided that the underlying absorption structure is sufficiently rich within the observed wavelength range.

    We then tested whether template fidelity improves the results by using ExoMol cross-sections computed under Titan-like conditions (150~K) and at a higher resolution (0.05~cm$^{-1}$), comparable to the CRIRES+ resolving power in wavenumber space. For both CH$_4$ and C$_2$H$_2$, the resulting SNR$_\mathrm{peak}$ values increased relative to the ones obtained from the ExoMol cross-sections computed at 280~K [Figures~\ref{fig:CH4_results} and \ref{fig:C2H2_results}].

    \item \textbf{MOLLIST laboratory cross-sections for ethane.}  
    For \ce{C2H6}, we used the high-resolution (0.01~cm$^{-1}$), low-temperature (205~K), N$_2$-broadened cross-section of \citet{Hewett_2020}, available in the MOLLIST database \citep{Bernath_2019}. Because the intrinsic resolution of the cross-section exceeds that of CRIRES+, we first deconvolved it to match the instrumental resolving power before constructing the template. The \citet{Hewett_2020} dataset provides the most physically appropriate representation of ethane available for HRCCS.
\end{enumerate}

\section{Results} 

Below, we present the results for each molecule, together with a detailed description of how the corresponding template was constructed. For each molecule, we explored several template configurations using both the direct and indirect methods. In the sections below, we present the configurations that provided the most consistent and physically reliable results [Figures~\ref{fig:CH4_results}, \ref{fig:C2H2_results}, \ref{fig:13CH4_results} and \ref{fig:C2H6_results}]. Table~\ref{tab:snr_summary} summarises the detection significances obtained. Figure~\ref{fig:non_detections_petitRADTRANS} presents the results of the non-optimal \texttt{petitRADTRANS} template configurations for \ce{CH4}, \(^{13}\)\ce{CH4} and \ce{C2H2}, as well as the results for CO and HCN using the indirect template.

\subsection{\ce{CH4}}

Both the direct and indirect methods were tested for \ce{CH4}. The optimal configuration was obtained using the indirect method applied to the \texttt{petitRADTRANS} line‑by‑line model under Titan‑like conditions. In this approach, absorption peaks were identified in the modelled transit spectrum, ranked by decreasing depth, and only a chosen fraction of the strongest features was retained. The indirect approach is motivated by the fact that the full line list contains a very large number of weak lines that contribute little to the CCF signal but add noise when included in the template. By excluding these weak features and retaining only the strongest lines, the template yields a more significant detection. The upper panel from Figure~\ref{fig:CH4_results} shows the resulting median CCF and the corresponding SNR distribution, which yields a high-significance detection of methane with SNR$_\mathrm{peak} = 33.72^{+0.29}_{-0.28}$.

To assess the reliability of the cross-section-based method, we performed a validation test using the ExoMol cross-section (i.e., opacity table generated from the ExoMol line lists) at 280 K and 0.112 cm\(^{-1}\), matching the HITRAN temperature and resolution. This provides a controlled test case for evaluating the performance of the indirect method when applied to cross-sections. The middle panel from Figure~\ref{fig:CH4_results} shows the median CCF and the SNR distribution, yielding SNR$_\mathrm{peak} = 17.11^{+0.09}_{-0.08}$.

Finally, to evaluate the impact of template fidelity, we repeated the analysis using the ExoMol cross-section computed at a resolution of 0.05~cm$^{-1}$ and a temperature of 150~K, representative of Titan’s stratosphere. The lower panel from Figure~\ref{fig:CH4_results} shows the resulting CCF, with SNR$_\mathrm{peak} = 22.71\pm0.12$. As expected, the cross-section computed under conditions closer to Titan’s atmosphere outperforms the 280~K ExoMol cross-section template.

The agreement between the \ce{CH4} CCFs obtained with the \texttt{petitRADTRANS} line-by-line template and the cross-section-based templates shows that the cross-section method retains the molecular fingerprint required for HRCCS.

\subsection{\ce{C2H2}}

For \ce{C2H2} both the direct and indirect methods were tested. As with \ce{CH4}, the indirect template constructed from the \texttt{petitRADTRANS} line‑by‑line model provided the optimal result. The upper panel of Figure~\ref{fig:C2H2_results} shows the median CCF and the corresponding Monte Carlo SNR\(_\mathrm{peak}\) distribution, which confirms a detection of acetylene at SNR$_\mathrm{peak} = 5.30\pm0.05$. 

To validate the cross‑section-based method, we repeated the analysis using the ExoMol cross‑section computed at 280 K, 0.112 cm\(^{-1}\). The middle panel of Figure~\ref{fig:C2H2_results} shows the resulting CCF, with a weaker but still identifiable peak at the expected velocity (SNR$_\mathrm{peak} = 3.23\pm0.04$) and a CCF shape closely resembling the result obtained using the \texttt{petitRADTRANS} line-by-line template. 

Finally, to assess the impact of template fidelity, we used the ExoMol cross‑section computed at 150 K, with a resolution of 0.05 cm\(^{-1}\). The lower panel of Figure~\ref{fig:C2H2_results} shows the median CCF, with $\mathrm{SNR}_\mathrm{peak}=4.30\pm0.05$. Once more, the 150~K cross‑section template yields a higher SNR\(_\mathrm{peak}\) value than the 280~K case. 

The consistency between the CCFs obtained with the \texttt{petitRADTRANS} line-by-line template and the cross-section-based templates confirms that the cross‑section method accurately captures the spectral fingerprint of \ce{C2H2} required for HRCCS.

\subsection{$\mathbf{^{13}}$\ce{CH4}}

\begin{figure*}
    \centering
    \includegraphics[width=\linewidth]{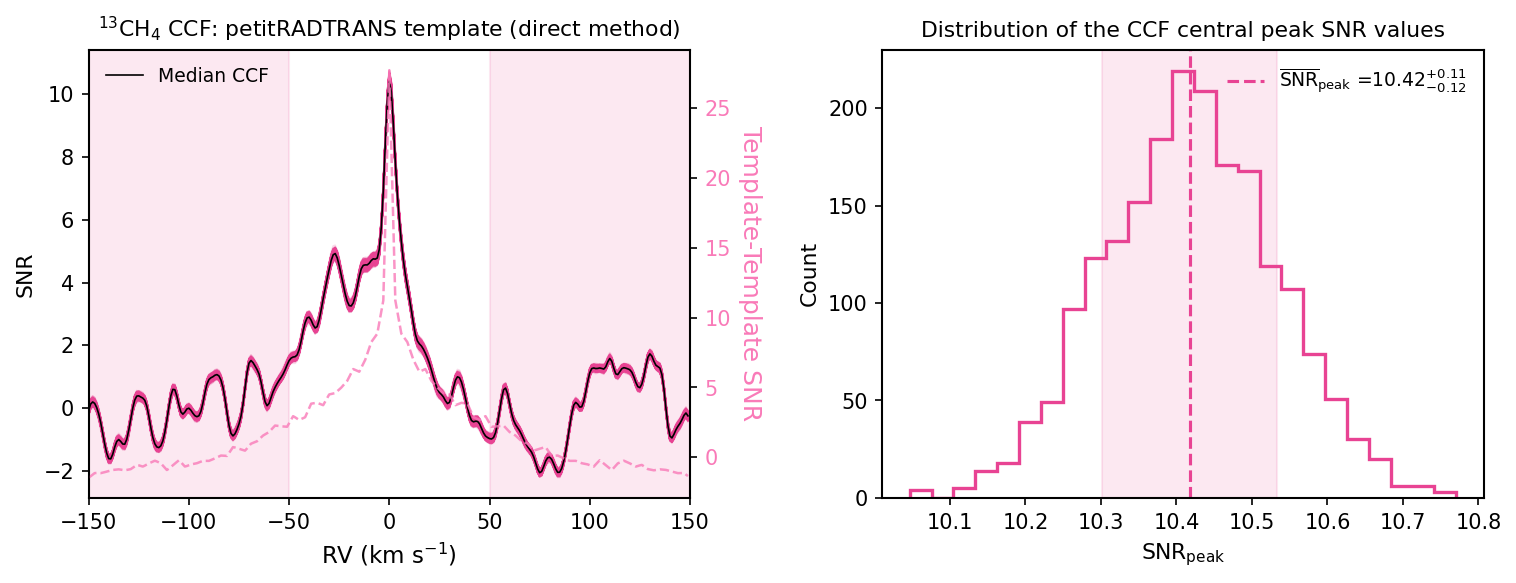}
    \caption{Cross-correlation results for $^{13}$\ce{CH4} obtained from 2000 Monte Carlo realisations using the direct line-by-line \texttt{petitRADTRANS} template. The black solid line shows the median CCF, while the dashed line shows the template autocorrelation. The shaded region marks the velocity interval used to estimate noise statistics. Right panel: distribution of the central-peak SNR values (SNR$_{\text{peak}}$). Since no high-resolution cross-sections exist for $^{13}$\ce{CH4}, only the result obtained with the \texttt{petitRADTRANS} line-by-line template is shown. The blue-shifted bump (\(v \approx -30\)~km~s\(^{-1}\)) adjacent to the main peak could be a result of telluric contamination of the CCF (Earth's radial velocity relative to Titan is of the order of this value, as shown in the second row of Figure~\ref{fig:ccf_geo}), which nonetheless is separate from the main absorption signal at Titan's rest frame.}
    \label{fig:13CH4_results}
\end{figure*}

Both the direct and indirect methods were evaluated for $^{13}$\ce{CH4}. Although the indirect method performs well when all spectral orders are included (as shown in Figure~\ref{fig:CCF_13CH4_all_orders}), the direct \texttt{petitRADTRANS} template provides the most reliable result when the analysis is restricted to the strongest absorption band (order 5), as it preserves the full radiative-transfer structure of the isotopologue. In the case of $^{13}$\ce{CH4}, the line-by-line model in order~5 forms a dense and homogeneous forest in which individual lines have similar intensity. The indirect method, which selects only a subset of peaks, therefore removes lines that contribute equally to the band profile and does not represent the spectral structure as accurately as the direct model. As a result, the direct template outperforms the indirect one in spectral order~5. Figure \ref{fig:13CH4_results} shows the resulting CCF, and the Monte Carlo distribution confirms a statistically significant detection of $^{13}$\ce{CH4} in Titan’s atmosphere, with SNR$_\mathrm{peak} = 10.42^{+0.11}_{-0.12}$. 

When all CRIRES+ spectral orders are combined, the indirect template nevertheless yields a significant detection (SNR \(\approx\) 6), as shown in Figure~\ref{fig:CCF_13CH4_all_orders}. In contrast, the direct template performs poorly in this configuration (SNR \(\approx\) 2). The inclusion of many weak lines in the direct template introduces additional noise into the CCF. The indirect method suppresses these features and retains only the strongest lines, improving the overall correlation signal.

\subsection{\ce{C2H6}}

\begin{figure*}
    \centering
    \includegraphics[width=\linewidth]{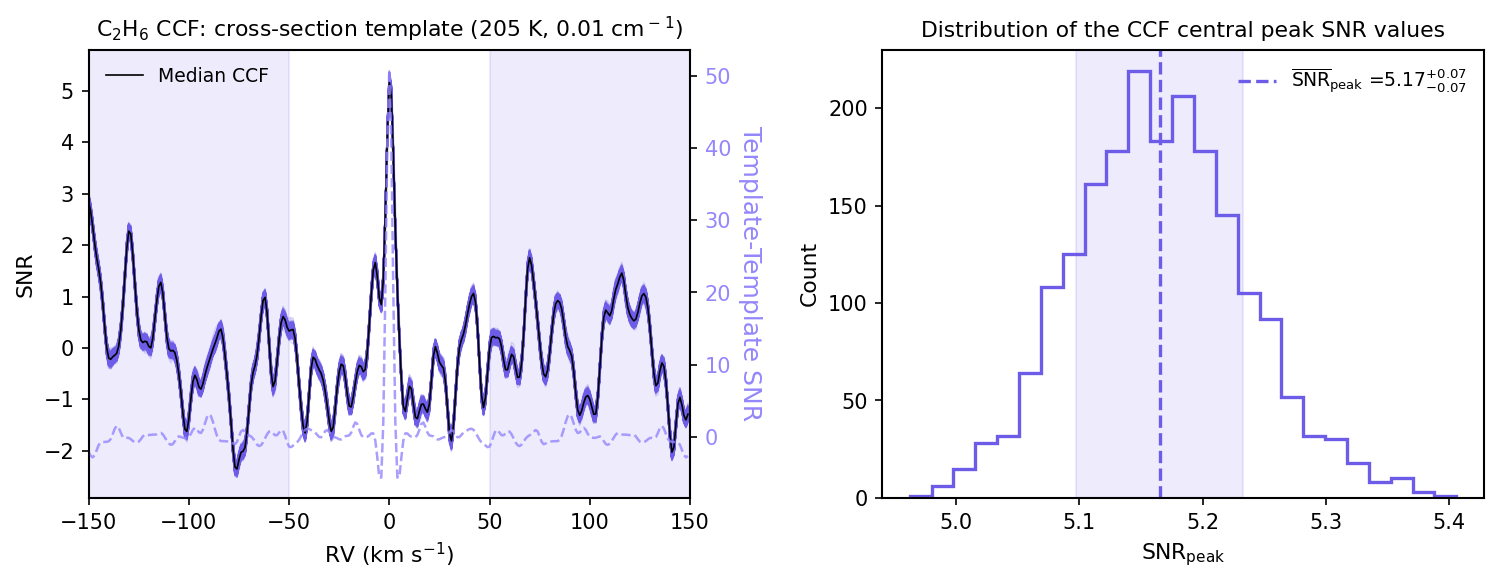}
    \caption{Cross‑correlation results for \ce{C2H6} obtained using the template constructed from the MOLLIST laboratory cross‑section of \citet{Hewett_2020}. Left panel: CCFs obtained from 2000 Monte Carlo realisations. The black solid line shows the median CCF, while the dashed line represents the autocorrelation of the template used to compute the cross‑correlation. The shaded region marks the velocity interval used to estimate the noise statistics. Right panel: distribution of the central‑peak SNR values (SNR$_{\text{peak}}$).}
    \label{fig:C2H6_results}
\end{figure*}

For \ce{C2H6}, no high‑resolution line list is available for constructing a direct \texttt{petitRADTRANS} template. We therefore used the indirect method applied to the high‑resolution MOLLIST laboratory cross‑section of \citet{Hewett_2020}, computed at a high spectral resolution of 0.01~cm$^{-1}$, a temperature of 205~K, and a pressure of 30~Torr, including N$_2$ broadening. Because the intrinsic resolution of this cross-section exceeds that of CRIRES+, we first deconvolved it to match the instrumental resolving power. The indirect template method was then applied to the deconvolved spectrum. Figure~\ref{fig:C2H6_results} shows the resulting CCF and the Monte Carlo distribution, which yields a SNR$_\mathrm{peak} = 5.17\pm0.07$. This constitutes the first detection of ethane using high-resolution cross-correlation spectroscopy (HRCCS), demonstrating the capability of cross-section-based templates for cross-correlation studies.

\renewcommand{\arraystretch}{1.3}
\begin{table}
    \centering
    \caption{Molecular detections and corresponding SNR$_\mathrm{peak}$ values. Multiple entries for the same molecule correspond to different opacity datasets. All templates were obtained using the indirect method except for $^{13}$\ce{CH4}, for which the direct line‑by‑line template yields the highest significance.}
    \label{tab:snr_summary}
    \begin{tabular}{l l l l l}
    \hline
    Molecule & Opacity data & T (K) & SNR$_\mathrm{peak}$ \\
    \hline
    CH$_4$ & \texttt{petitRADTRANS} line-by-line model & 150 & \ensuremath{33.72^{+0.29}_{-0.28}} \\
    & ExoMol cross-section & 280 & \ensuremath{17.11^{+0.09}_{-0.08}} \\
    & ExoMol cross-section & 150 & \ensuremath{22.71^{+0.12}_{-0.12}} \\
    C$_2$H$_2$ & \texttt{petitRADTRANS} line-by-line model & 150 & \ensuremath{5.30^{+0.05}_{-0.05}} \\
    & ExoMol cross-section & 280 & \ensuremath{3.23^{+0.04}_{-0.04}} \\
    & ExoMol cross-section & 150 & \ensuremath{4.30^{+0.05}_{-0.05}} \\
    $^{13}$CH$_4$ & \texttt{petitRADTRANS} line-by-line model & 150 & \ensuremath{10.42^{+0.11}_{-0.12}} \\
    C$_2$H$_6$ & MOLLIST laboratory cross-section & 205 & \ensuremath{5.17^{+0.07}_{-0.07}} \\
    \hline
\end{tabular}
\end{table}

\begin{figure*}
    \centering
    \begin{subfigure}{\linewidth}
        \centering
        \includegraphics[width=\linewidth]{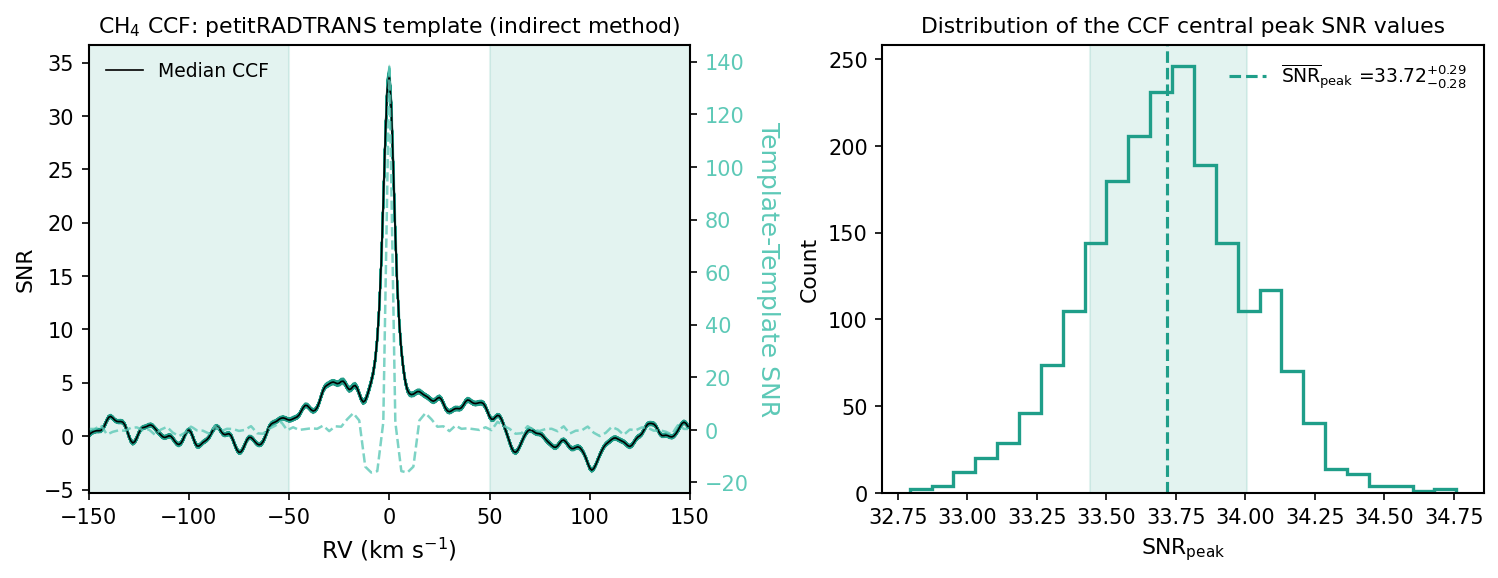}
    \end{subfigure}
    \begin{subfigure}{\linewidth}
        \centering
        \includegraphics[width=\linewidth]{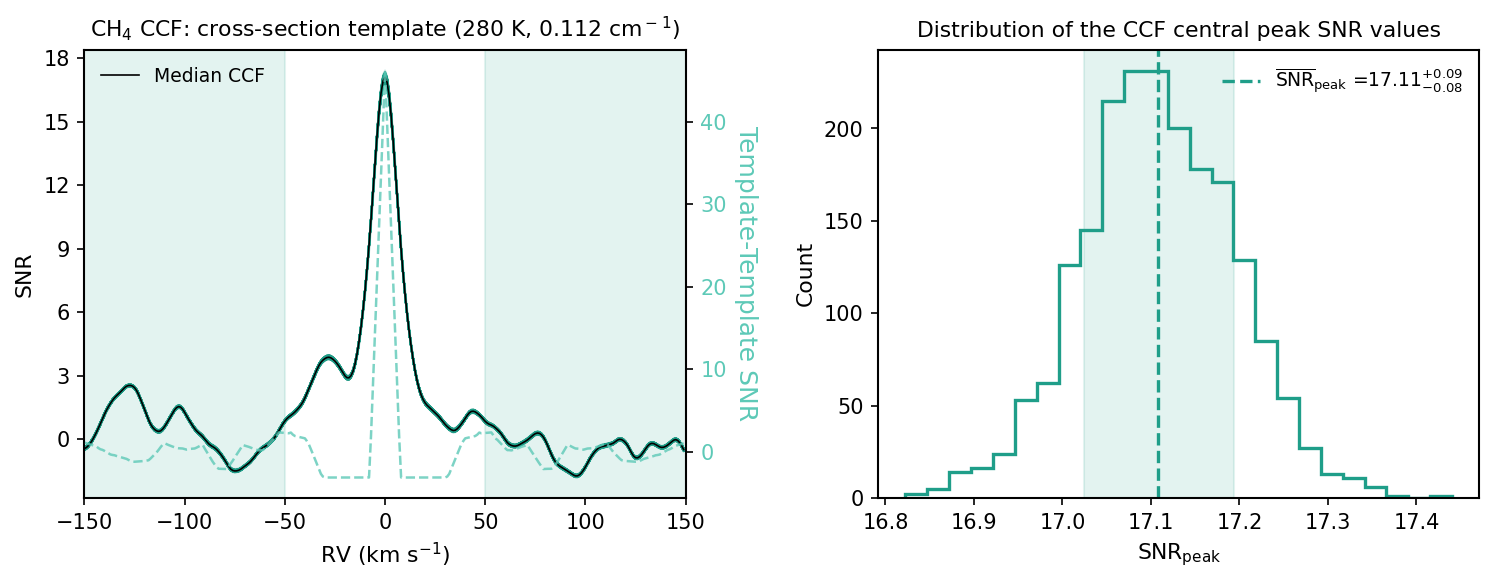}
    \end{subfigure}
    \begin{subfigure}{\linewidth}
        \centering
        \includegraphics[width=\linewidth]{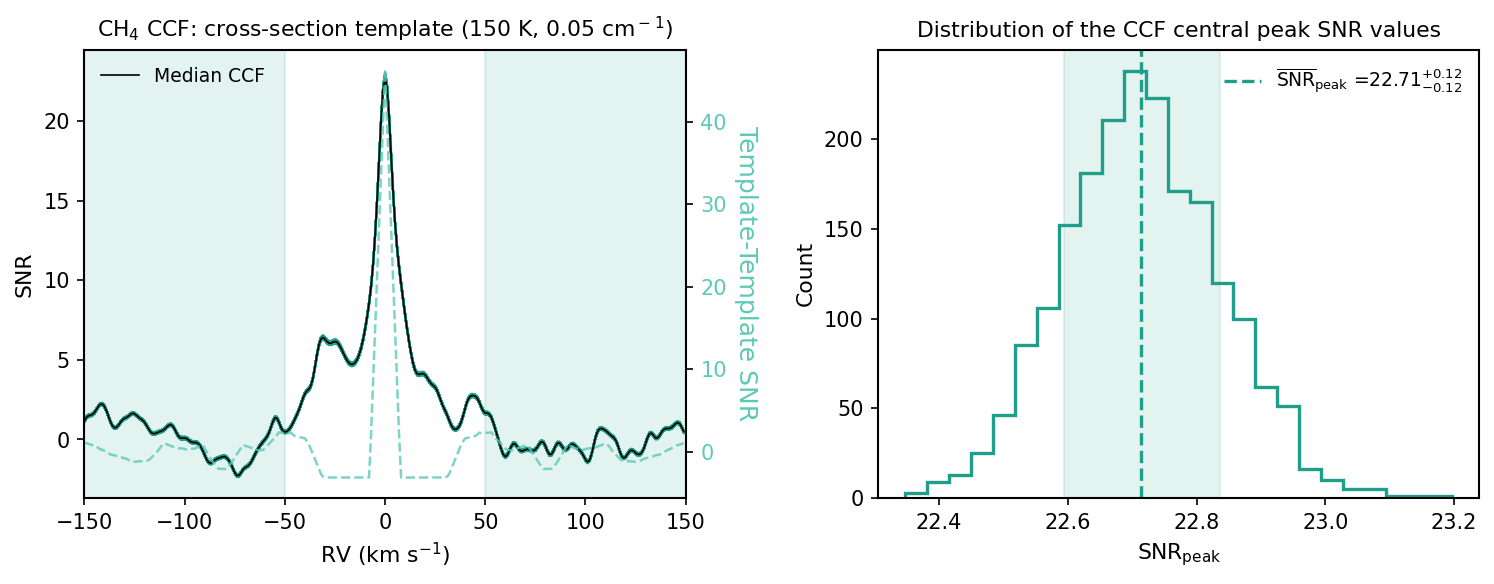}
    \end{subfigure}
    \caption{Cross-correlation results for \ce{CH4} using both \texttt{petitRADTRANS} line-by-line and cross-section-based templates. Left panels: CCFs obtained from 2000 Monte Carlo realisations using the \texttt{petitRADTRANS} template (top) and the ExoMol cross-section-based templates (middle and bottom figures). Each realisation is generated by perturbing the normalised spectrum within its flux uncertainties and recomputing the CCF. Black solid lines show the median CCF, while the dashed lines show the autocorrelation of the corresponding template. The shaded region marks the velocity interval used to estimate the noise statistics. Right panels: distributions of the central-peak SNR values (SNR$_{\text{peak}}$) extracted from all realisations. The vertical dashed lines mark the median SNR$_{\text{peak}}$ for each template.}
    \label{fig:CH4_results}
\end{figure*}

\begin{figure*}
    \centering
    \begin{subfigure}{\linewidth}
        \centering
        \includegraphics[width=\linewidth]{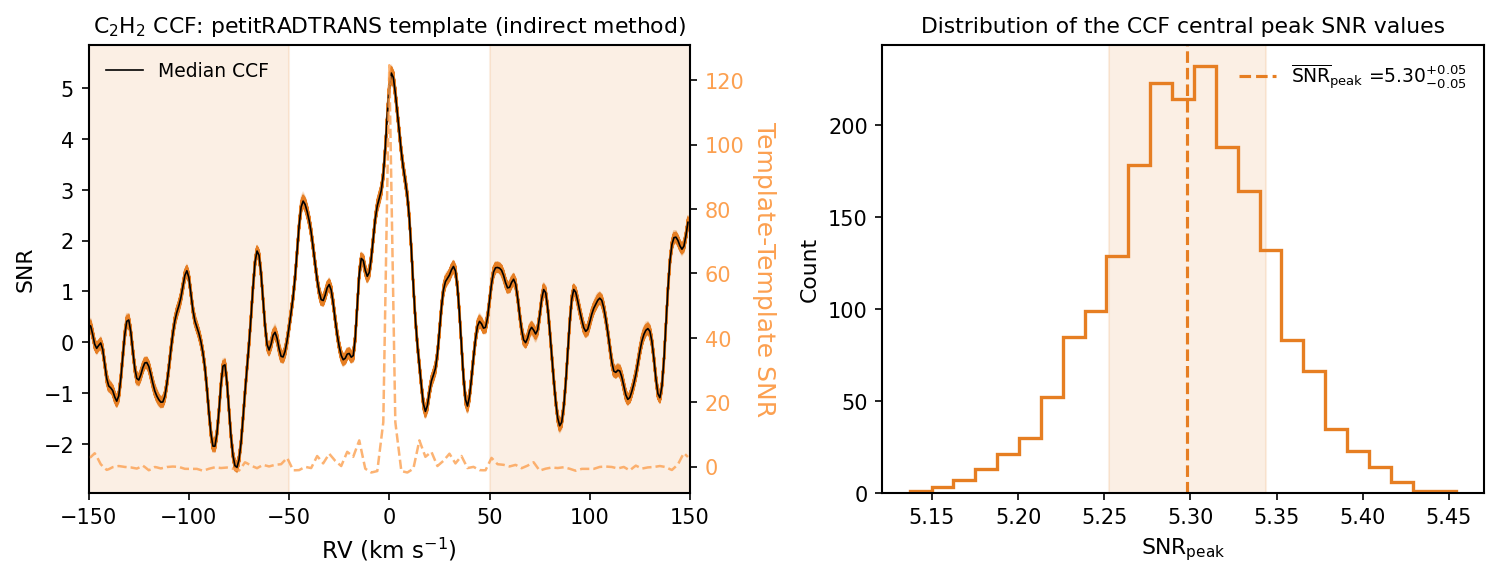}
    \end{subfigure}
    \begin{subfigure}{\linewidth}
        \centering
        \includegraphics[width=\linewidth]{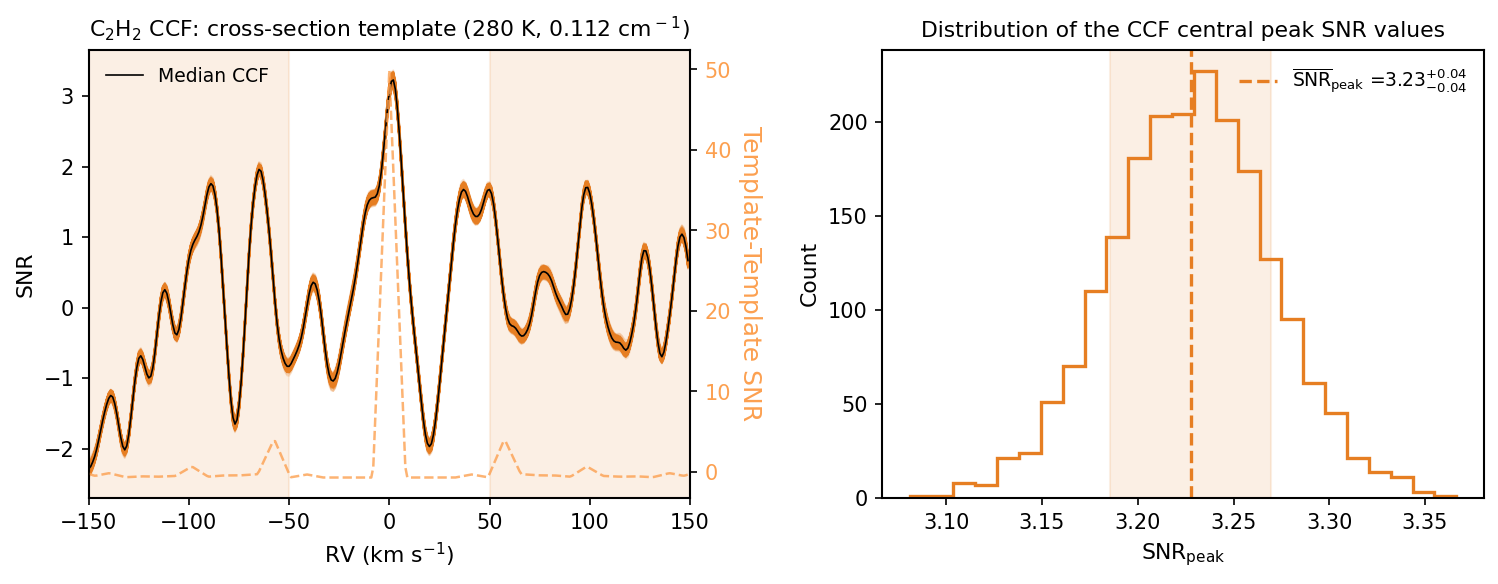}
    \end{subfigure}
    \begin{subfigure}{\linewidth}
        \centering
        \includegraphics[width=\linewidth]{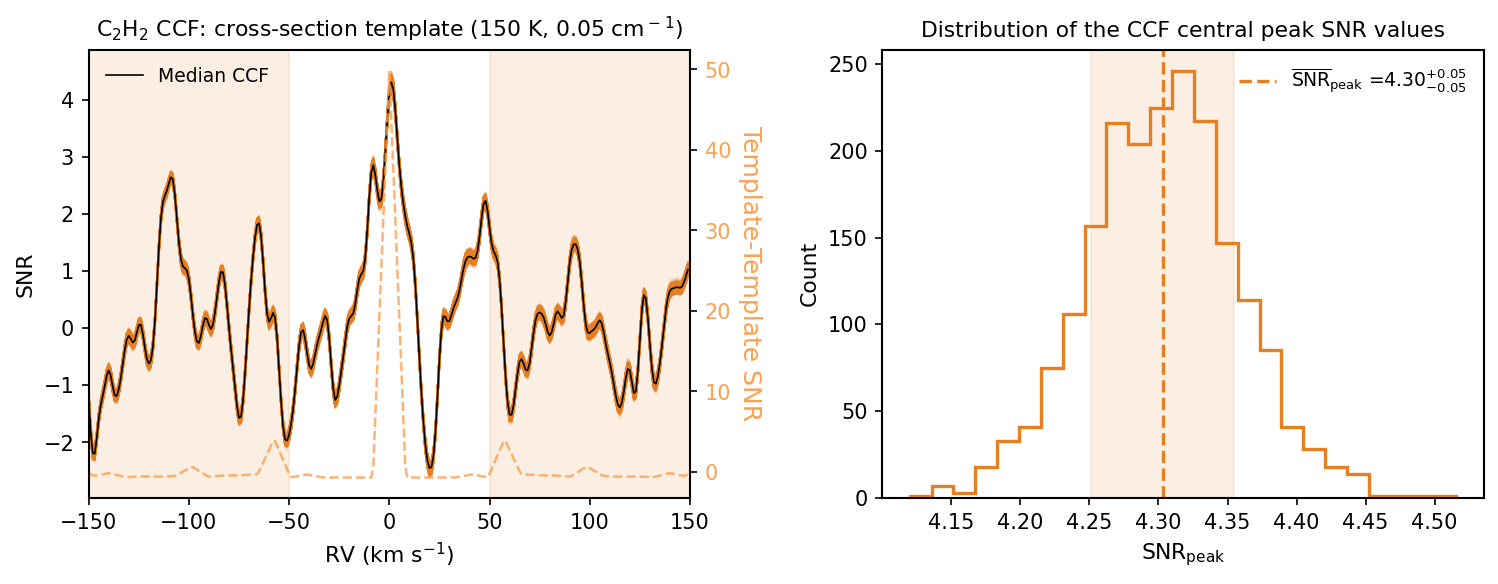}
    \end{subfigure}
    \caption{Cross-correlation results for \ce{C2H2} using both \texttt{petitRADTRANS} line-by-line and cross-section-based templates. Left panels: CCFs obtained from 2000 Monte Carlo realisations using the \texttt{petitRADTRANS} template (top) and the ExoMol cross-section-based templates (middle and bottom figures). Black solid lines show the median CCF, while dashed lines show the autocorrelation of the corresponding template. The shaded region marks the velocity interval used to estimate the noise statistics. Right panels: distributions of the central-peak SNR values (SNR$_{\text{peak}}$).}
    \label{fig:C2H2_results}
\end{figure*}

\section{Discussion}

\subsection{The role of cross-sections}

The major goal of this work was to evaluate whether absorption cross‑sections can serve as reliable substitutes for line lists in high-resolution cross-correlation spectroscopy (HRCCS). Our validation tests using ExoMol cross-sections for \ce{CH4} and \ce{C2H2}, computed under the same temperature and resolution as the HITRAN experimentally measured data, show that the indirect cross‑section method successfully recovers both molecules [Figures~\ref{fig:CH4_results} and \ref{fig:C2H2_results}]. However, laboratory-measured cross-sections are inherently limited by the conditions under which they are measured. HITRAN cross‑sections, for example, are available only at 278 K. These conditions are far warmer than Titan’s stratosphere and do not reflect so accurately the intrinsic line shapes expected at low temperature. The higher-temperature cross-sections also include transitions between energy levels which would not be present at lower temperatures, thereby increasing the noise.

Our tests using ExoMol cross‑sections computed at 150 K and 0.05 cm$^{-1}$ resolution demonstrate that template fidelity matters: when the cross‑sections more closely match Titan’s atmospheric conditions, the detection significance increases \(\sim 33 \%\) for both \ce{CH4} and \ce{C2H2}. This result underscores the need for laboratory cross‑sections measured across a wider range of temperatures, pressures, resolutions, and broadening environments, particularly for cold, \ce{N2}‑dominated atmospheres like Titan’s.

Beyond these validation tests, the most significant outcome of this work is the detection of \ce{C2H6}, which represents the first identification of this molecule using HRCCS. This detection was only possible through the use of experimentally measured cross‑sections, as no high‑resolution line list currently exists for ethane. The MOLLIST dataset of \citet{Hewett_2020}, measured at low temperature, high spectral resolution, and with \ce{N2} broadening, provides the most physically realistic representation of ethane’s opacity under Titan‑like conditions. After deconvolution to match the CRIRES+ resolving power, the resulting indirect template retains sufficient line‑contrast structure to produce a statistically significant CCF peak. This result demonstrates that cross‑sections can allow the detection of molecules that would otherwise be inaccessible to HRCCS and highlights the potential of expanding laboratory cross‑section databases to include a wider range of experimental conditions. The ethane detection, therefore, serves as a proof‑of‑concept for extending HRCCS to a much broader chemical space, including many hydrocarbons and prebiotic molecules for which no line lists currently exist.

\subsection{Direct vs. indirect templates: strengths and limitations}

\begin{figure*}
    \centering
    \begin{subfigure}{0.33\linewidth}
        \centering
        \includegraphics[width=\linewidth]{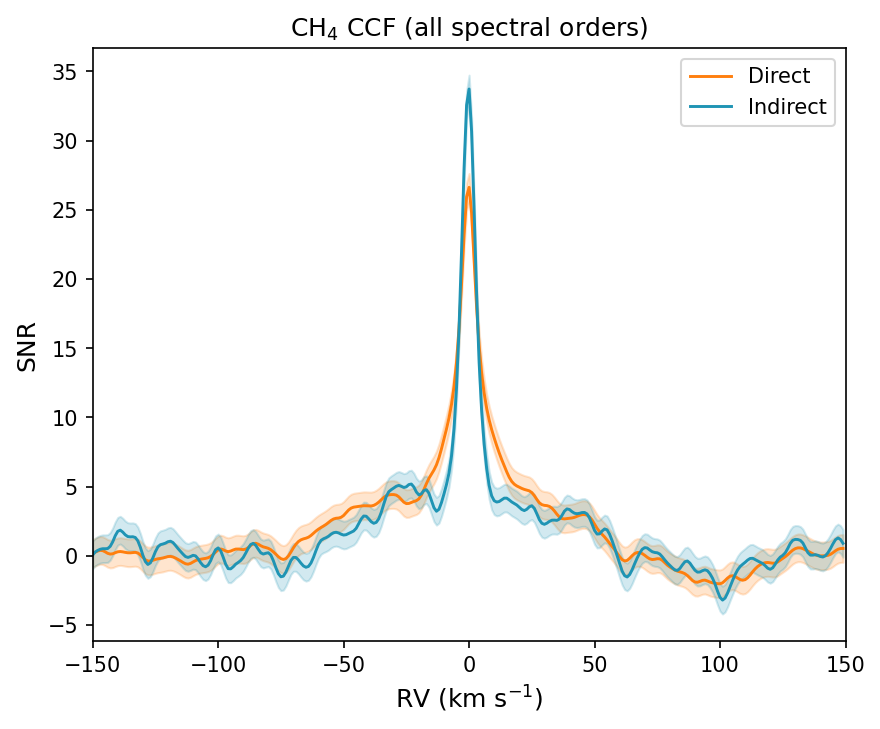}
        \caption{\ce{CH4} CCFs (all nights and spectral orders).}
        \label{fig:CCF_CH4}
    \end{subfigure}
    %\hspace{0.02\linewidth}
    \begin{subfigure}{0.33\linewidth}
        \centering
        \includegraphics[width=\linewidth]{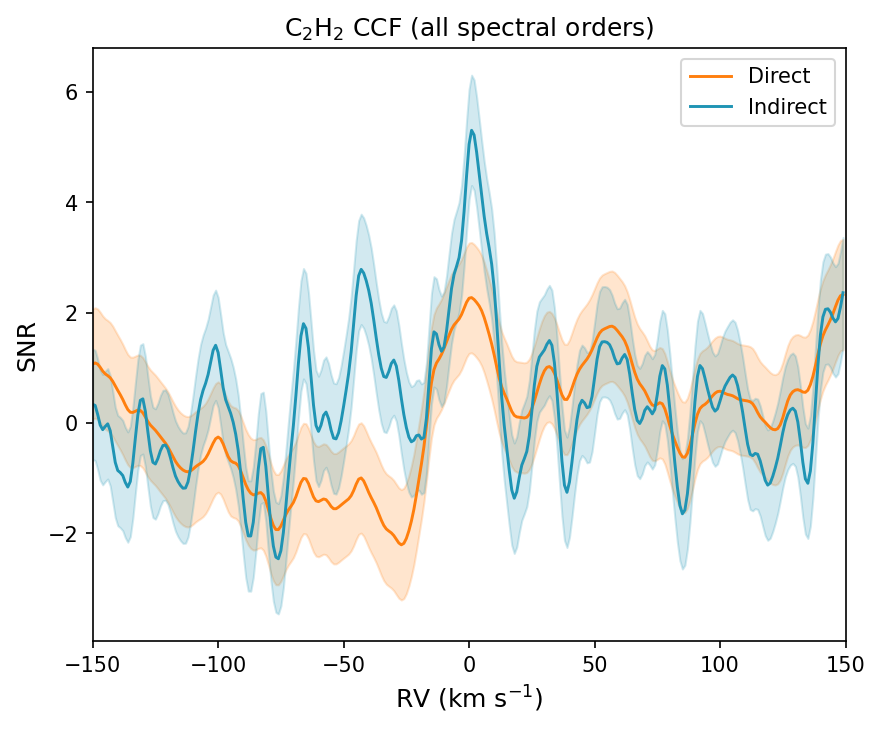}
        \caption{\ce{C2H2} CCFs (all nights and spectral orders).}
        \label{fig:CCF_C2H2}
    \end{subfigure}
    %\hspace{0.02\linewidth}
    \begin{subfigure}{0.33\linewidth}
        \centering
        \includegraphics[width=\linewidth]{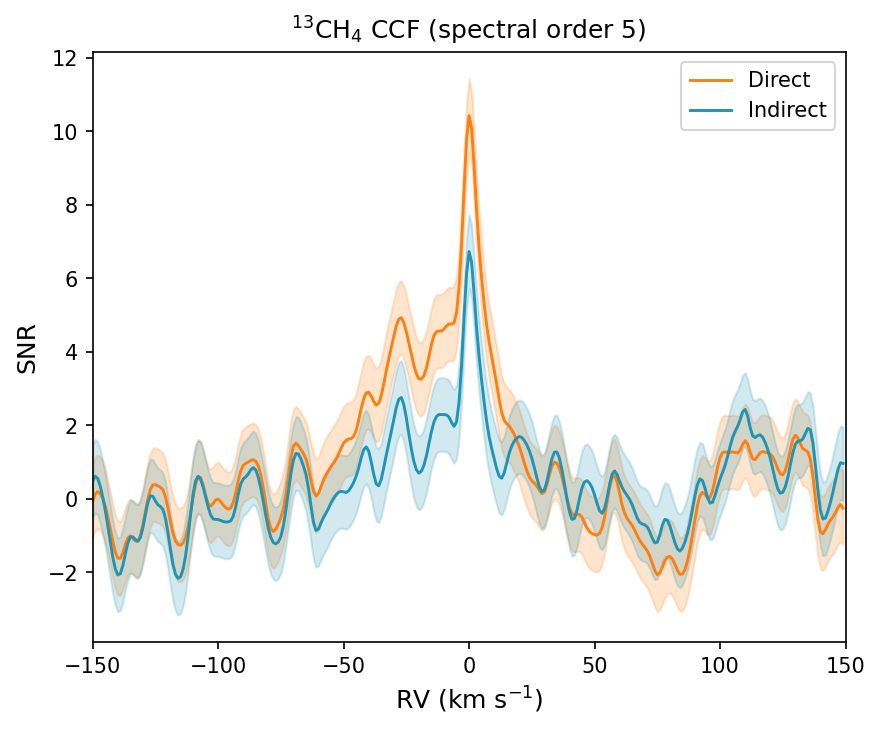}
        \caption{\(^{13}\)\ce{CH4} CCFs (all nights, spectral order~5).}
        \label{fig:CCF_13CH4}
    \end{subfigure}
    \caption{Cross-correlation results for \ce{CH4}, \ce{C2H2} and \(^{13}\)\ce{CH4} using both the direct (orange) and indirect (blue) \texttt{petitRADTRANS} line-by-line templates.}
    \label{fig:comparison}
\end{figure*}

The choice between the direct and indirect template methods has a measurable impact on detection significance, and the optimal strategy depends on the spectral complexity of each molecule. Figure~\ref{fig:comparison} compares the direct and indirect method results for \ce{CH4}, \ce{C2H2} and \(^{13}\)\ce{CH4}. The direct method preserves the full radiative‑transfer structure of the \texttt{petitRADTRANS} line-by-line model, making it the most physically complete representation of the molecular fingerprint. However, for molecules with extremely dense line lists, such as \ce{CH4}, the direct template contains hundreds of weak transitions that contribute little to the cross‑correlation signal while adding substantial noise. As a result, the direct \ce{CH4} template yields a lower SNR than the indirect counterpart, as shown by the CCFs in Figure~\ref{fig:CCF_CH4}.

The indirect method mitigates this problem by ranking absorption features by depth and retaining only the strongest lines. For \ce{CH4}, this approach substantially improves the detection significance, increasing the SNR by \(\sim 27\%\), confirming that the cross‑correlation is dominated by a subset of high‑contrast lines. \ce{C2H2} provides a similar example: although its line list is less crowded than methane’s, the indirect template still outperforms the direct one because the strongest \ce{C2H2} lines dominate this molecule's spectral fingerprint in the K-band [Figure~\ref{fig:CCF_C2H2}].

In contrast, the isotopologue $^{13}$\ce{CH4} illustrates a limitation of the indirect method. In spectral order 5, its absorption profile consists of many lines of nearly uniform intensity, which contribute comparably to the cross-correlation. Removing a subset of these lines therefore reduces the template fidelity and weakens the resulting signal. For this species, the direct template performs better, as shown in Figure~\ref{fig:CCF_13CH4}, demonstrating that the indirect method is not universally optimal and must be tuned to the line‑intensity distribution of each molecule.

This trend becomes clear when analysing how the cross-correlation SNR responds to different fractions of included lines in the indirect template, as illustrated in Figure~\ref{fig:SNRvsFraction} for \ce{CH4} and \ce{C2H6}. For each molecule, we find an optimal fraction of retained lines that maximises the SNR, and this value depends on the underlying line‑intensity distribution. Table~\ref{tab:line_fraction} presents the optimal fraction of lines included for each molecule. Overall, when using the indirect template, removing the weakest lines systematically improves the detection significance; however, the optimal fraction of retained lines is molecule‑specific rather than universal, reflecting differences in each species’ absorption profile.

A further advantage of the indirect method emerges when templates are constructed from high-resolution cross‑sections. High‑resolution cross‑sections often contain extremely dense forests of weak transitions. Including all of these features in a direct template dilutes the cross‑correlation signal: the strongest lines drive the CCF peak, while the vast majority of weak lines contribute only noise. Therefore, the indirect template naturally mitigates this problem by ranking features by intensity and retaining only those that meaningfully contribute to the detection.

\subsection{Impacts of wavelength coverage}

The CRIRES+ settings used for these observations impose an additional constraint on molecular detectability. Several molecules of interest, most notably CO and HCN, do not produce significant cross‑correlation peaks (as shown in Figure~\ref{fig:non_detections_petitRADTRANS}), and their non‑detection can be directly attributed to the limited spectral range sampled in the K-band configuration.

For CO, the strongest rovibrational band at 2.3--2.4 \textmu m is only partially sampled (as illustrated in Figure~\ref{fig:CO_temp}), and the number of lines is insufficient to produce any detectable signal. HCN presents an even more restrictive case. Its strongest near‑infrared features lie beyond 3~\textmu m, and the lines that do fall within the 2~\textmu m region lie largely outside the CRIRES+ orders selected for these observations, as shown in Figure~\ref{fig:HCN_temp}. In contrast, \ce{C2H2} has several strong K‑band lines that fall squarely within the CRIRES+ spectral orders used here, as shown in Figure~\ref{fig:C2H2_temp}, which is why \ce{C2H2} can be detected with the indirect template while HCN cannot. For HCN, the resulting template contains only isolated peaks with insufficient density to produce a detectable cross‑correlation signature. Although more recent line lists provide improved coverage at longer wavelengths, the distribution of HCN lines in the K‑band remains the limiting factor for these observations.

These examples highlight a key limitation of HRCCS when applied to instruments with discontinuous wavelength coverage: even molecules with strong intrinsic absorption may remain undetectable if their dominant features fall outside the observed spectral range.

\subsection{Telluric contamination}
\label{sec:telluric_contamination}

To verify that telluric absorption does not contribute to the molecular detections reported in this work, we computed the CCFs between the calibration‑star spectra and the optimal template for each species, which can be seen in Figure~\ref{fig:ccf_star}. The calibration star HD~198802 is a G‑type telluric standard whose spectrum contains both intrinsic stellar absorption lines and telluric features. Because the stellar atomic lines do not correlate with the line patterns of the molecular templates, they do not produce spurious CCF peaks. The calibration‑star CCFs therefore isolate the contribution of telluric absorption. For all molecules except methane, the CCFs show no significant peaks, confirming that telluric contamination does not drive the detections. Methane is the only exception: the calibration‑star spectra yield a clear CCF peak at 0~km~s\(^{-1}\), corresponding to absorption by terrestrial \ce{CH4}.

To demonstrate that Titan’s methane detection is not caused by this telluric contribution, we recomputed the CCFs using Titan spectra in the geocentric frame and the optimal template for methane, as shown in the first row of Figure~\ref{fig:ccf_geo}. In this frame, telluric absorption remains fixed at 0~km~s\(^{-1}\), whereas Titan’s methane lines appear Doppler‑shifted by the changing Titan--Earth radial velocity. The resulting CCFs show a stronger peak that shifts consistently from night to night, tracking Titan’s radial velocity, while a comparatively smaller peak is observed in Earth's rest frame, clearly separated from the Titan-originated peak. This behaviour confirms that the detected molecular signals originate in Titan’s atmosphere rather than in Earth’s.

The ability to distinguish Titan‑originated features from telluric absorption is enabled by the resolving power of CRIRES+. At \(R \approx 100{,}000\), the instrumental velocity resolution is \(\Delta v \approx 3\)~km~s\(^{-1}\). Thus, any two line populations separated by more than this threshold produce distinct, resolvable peaks in the cross‑correlation function. Across our observing nights, Titan’s barycentric velocity differs from the telluric rest frame by tens of~km~s\(^{-1}\), well above the CRIRES+ resolution limit. A telluric contribution would therefore manifest as a stationary peak at 0~km~s\(^{-1}\), clearly separable from Titan’s Doppler‑shifted signal. As a result, the observed night‑to‑night motion of the CCF peak is incompatible with a telluric origin and uniquely identifies Titan’s atmosphere as the source of the detected molecular absorption \citep{RiancoSilva_2024}.

\subsection{Molecular distinguishability}

The CRIRES+ observations analysed in this work probe the 1.99--2.48 \textmu m spectral region, which corresponds primarily to overtone and combination bands of hydrocarbons rather than their fundamental vibrational modes. This distinction is important for interpreting molecular detections. In the fundamental C--H stretching region (3.0--3.5 \textmu m), many hydrocarbons exhibit similar band shapes and line‑density patterns due to shared vibrational symmetries and closely related molecular structures \citep{Niraula_2025}. Such degeneracy can, in principle, lead to false-positive cross-correlation signals if two molecules share sufficiently similar line patterns.

To assess potential degeneracy, we cross‑correlated the optimal templates of \(^{13}\)\ce{CH4}, \ce{C2H2}, and \ce{C2H6} against the \ce{CH4} template. The resulting CCFs show no significant peaks at 0~km~s\(^{-1}\) for any of the three molecules, as shown in Figure~\ref{fig:ccf_with_CH4}. Instead, all cross‑correlations fluctuate around zero with amplitudes consistent with noise. This behaviour confirms that the spectral structures of \(^{13}\)\ce{CH4}, \ce{C2H2}, and \ce{C2H6} are not correlated with that of \ce{CH4} in the CRIRES+ wavelength range. The absence of spurious peaks demonstrates that the detections reported in this work are not artefacts of template degeneracy.

These results indicate that, within the overtone and combination band region probed by CRIRES+, the molecular fingerprints of hydrocarbons are sufficiently distinct to enable molecule‑specific detections through cross‑correlation, even for species with similar chemical structures. This contrasts with the fundamental C--H stretching region at 3.0--3.5~\textmu m, where degeneracies are known to be more common, although a full comparison would require dedicated analysis in that wavelength range.

\section{Conclusions}

In this work, we applied high-resolution cross-correlation spectroscopy (HRCCS) to K-band CRIRES+ observations of Titan (1.99--2.48 \textmu m), using Saturn's largest moon as a controlled testbed to develop and validate a new cross-section-based methodology for HRCCS template construction. Our analysis successfully recovers \ce{CH4}, $^{13}$\ce{CH4} and \ce{C2H2} and yields the first HRCCS detection of \ce{C2H6}, at a significance of SNR$_\text{peak}$ = $5.17\pm0.07$.

Five main conclusions emerge from this study. First, laboratory-measured absorption cross-sections are a viable and powerful alternative to line‑by‑line opacities used in radiative transfer models to generate HRCCS templates, enabling HRCCS analyses for molecules that lack high‑resolution line lists. Our validation tests using ExoMol computed cross-sections for \ce{CH4} and \ce{C2H2} confirm that the indirect cross-section method successfully recovers both species. More importantly, the detection of \ce{C2H6}, a molecule for which no high-resolution line list currently exists, was made possible exclusively through cross-section-based templates, using the experimentally measured MOLLIST dataset of \cite{Hewett_2020}. This result demonstrates that cross-section-based template construction can extend HRCCS to molecules that would otherwise remain inaccessible, opening a path to systematic searches across a much broader chemical space, including many hydrocarbons and prebiotic molecules.

Second, the way templates are constructed strongly influences detection sensitivity. For molecules with dense line forests such as \ce{CH4}, the indirect method, which ranks absorption features by depth and retains only the strongest, substantially outperforms the direct line-by-line template, as the inclusion of many weak lines adds substantial noise to the cross-correlation signal. For $^{13}$\ce{CH4}, the absorption profile in order~5 consists of many lines of nearly uniform intensity that contribute comparably to the cross-correlation signal. Removing a fraction of these lines reduces the fidelity of the template and weakens the cross‑correlation signal. The optimal strategy is therefore molecule-specific and must account for the underlying line-intensity distribution.

Third, template fidelity matters. Our comparison of ExoMol cross-sections computed under 280~K, 0.112~cm$^{-1}$ versus 150~K, 0.05~cm$^{-1}$ consistently yields higher SNR$_\text{peak}$ values in the latter case, for both \ce{CH4} and \ce{C2H2}. This underscores the importance of having laboratory cross-sections available across a wide range of temperatures, pressures, and spectral resolutions, particularly for cold, \ce{N2}-dominated atmospheres. Expanding such databases should be a priority for future studies.

Fourth, instrumental wavelength coverage can fundamentally limit detectability, independently of template quality. The non-detections of CO and HCN in our K-band dataset are not indicative of their absence from Titan's atmosphere, but rather reflect the fact that the spectroscopic features of these species are insufficient or fall outside the CRIRES+ orders used here. This highlights the need for multi-band or broader-coverage observations when attempting comprehensive molecular inventories.

Fifth, within the overtone and combination band region probed by CRIRES+, hydrocarbons exhibit sufficiently distinct line patterns to avoid template degeneracy. Although degeneracies are more commonly discussed in the fundamental C--H stretching region at 3.0--3.5~\textmu m,  where many hydrocarbons exhibit similar band shapes and line‑density distributions due to shared vibrational symmetries, establishing a direct comparison would require equivalent high-resolution cross-correlation analysis in that wavelength range. In our dataset, cross-correlations of $^{13}$CH$_4$, C$_2$H$_2$, and C$_2$H$_6$ against the CH$_4$ template show no significant peaks at 0~km~s$^{-1}$, demonstrating that the detections are not artefacts of template degeneracy.

Taken together, these results establish Titan as a benchmark for calibrating molecular detection techniques applicable to both solar system and exoplanet atmospheres. The cross-section-based methodology provides a practical framework for extending HRCCS beyond the limited set of species with complete line lists, as demonstrated by the ethane detection. Future applications of this approach to other ground-based high-resolution spectrographs, as well as to JWST's highest-resolution modes and next-generation facilities such as the ELT, could significantly expand the inventory of molecules detectable in planetary atmospheres across the solar system and beyond.

\section*{Acknowledgements}

The observations used in this work were obtained under ESO programme 110.23UK.001 (PI: Bruno B\'ezard). We thank the observing team for carrying out the observations. This work was supported by Fundação para a Ciência e Tecnologia (FCT) of reference PTDC/FIS-AST/29942/2017, through national funds and by FEDER through COMPETE 2020 of reference POCI-01-0145-FEDER-007672, and through the research grants UIDB/04434/2020, UIDP/04434/2020 and UID/04434/2025. It was also financially supported by IMS LA/P/0056/2020 (\url{https://doi.org/10.54499/LA/P/0056/2020}), CQE UID/00100/2025 (\url{https://doi.org/10.54499/UID/00100/2025}), UID/PRR/100/2025 (\url{https://doi.org/10.54499/UID/PRR/00100/2025}), and UID/PRR2/00100/2025 (\url{https://doi.org/10.54499/UID/PRR2/00100/2025}), funded by national funds through FCT/MECI (PIDDAC). This study is framed within the College on Polar and Extreme Environments (Polar2E) of the University of Lisbon. RRS acknowledges that this work was supported by FCT, I.P. by project reference 2024.02527.BD and DOI identifier \url{https://doi.org/10.54499/2024.02527.BD}. DG is funded by FCT through project ORIGINS (2022.05284.PTDC).

%%%%%%%%%%%%%%%%%%%%%%%%%%%%%%%%%%%%%%%%%%%%%%%%%%
\section*{Data availability}

The observational data used in this study are publicly available at the ESO Science Archive and correspond to program 110.23UK.001 (PI: Bruno Bézard). Titan was observed with CRIRES+ at the VLT on 9, 20 and 24 October 2022, using the K2166 setting.

\section*{Conflicts of interest}

Authors declare no conflicts of interest.

%%%%%%%%%%%%%%%%%%%% REFERENCES %%%%%%%%%%%%%%%%%%

\bibliographystyle{rasti}
\bibliography{example} % if your bibtex file is called example.bib

%%%%%%%%%%%%%%%%%%%%%%%%%%%%%%%%%%%%%%%%%%%%%%%%%%

%%%%%%%%%%%%%%%%% APPENDICES %%%%%%%%%%%%%%%%%%%%%

\appendix

\section{Non-detections using absorption cross-sections}

\begin{figure*}
    \centering
    \begin{subfigure}{0.32\linewidth}
        \centering
        \includegraphics[width=\linewidth]{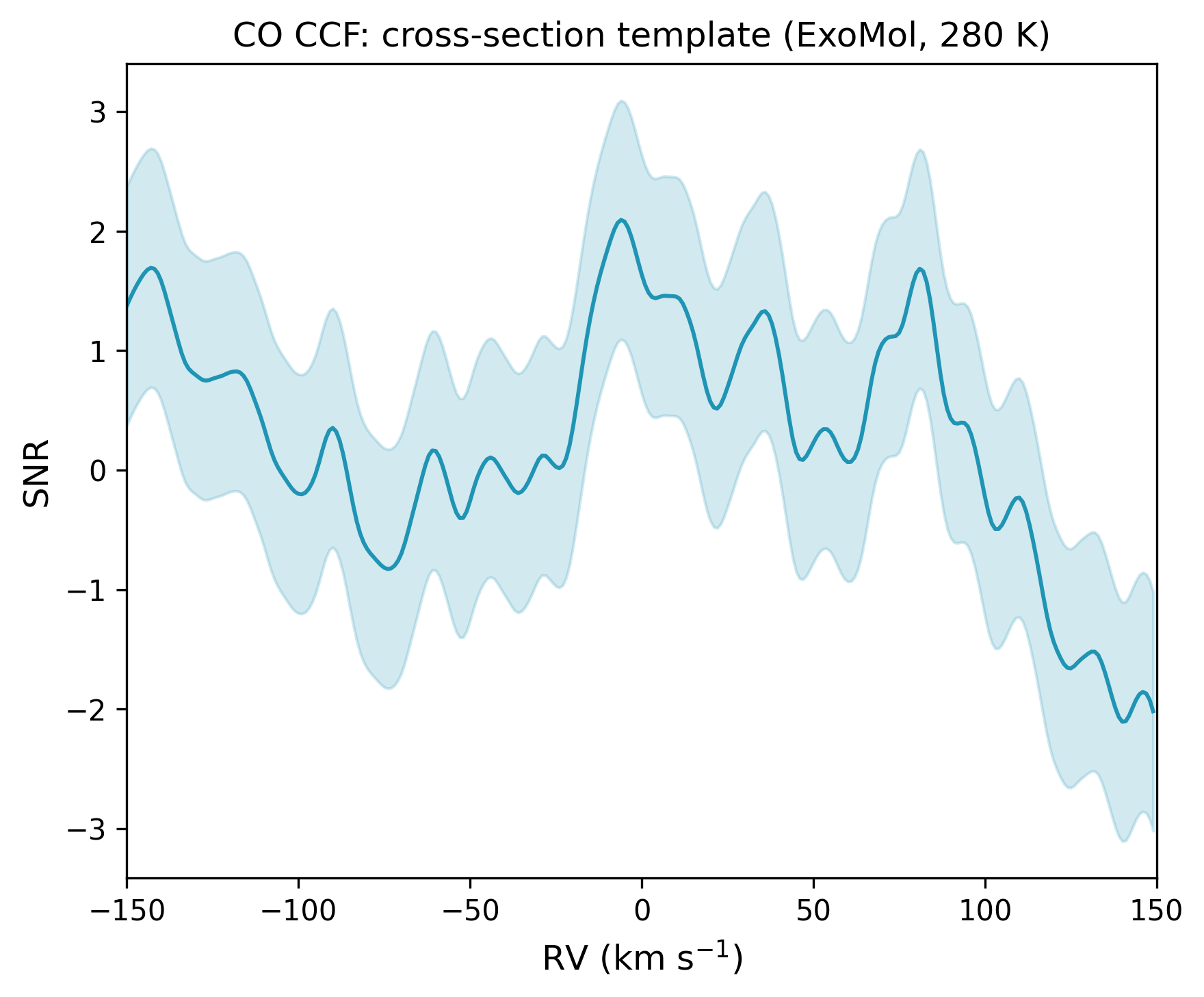}
        \caption{Carbon monoxide (280 K)}
    \end{subfigure} \hspace{0.01\linewidth}
    \vspace{0.01\linewidth}
    \begin{subfigure}{0.32\linewidth}
        \centering
        \includegraphics[width=\linewidth]{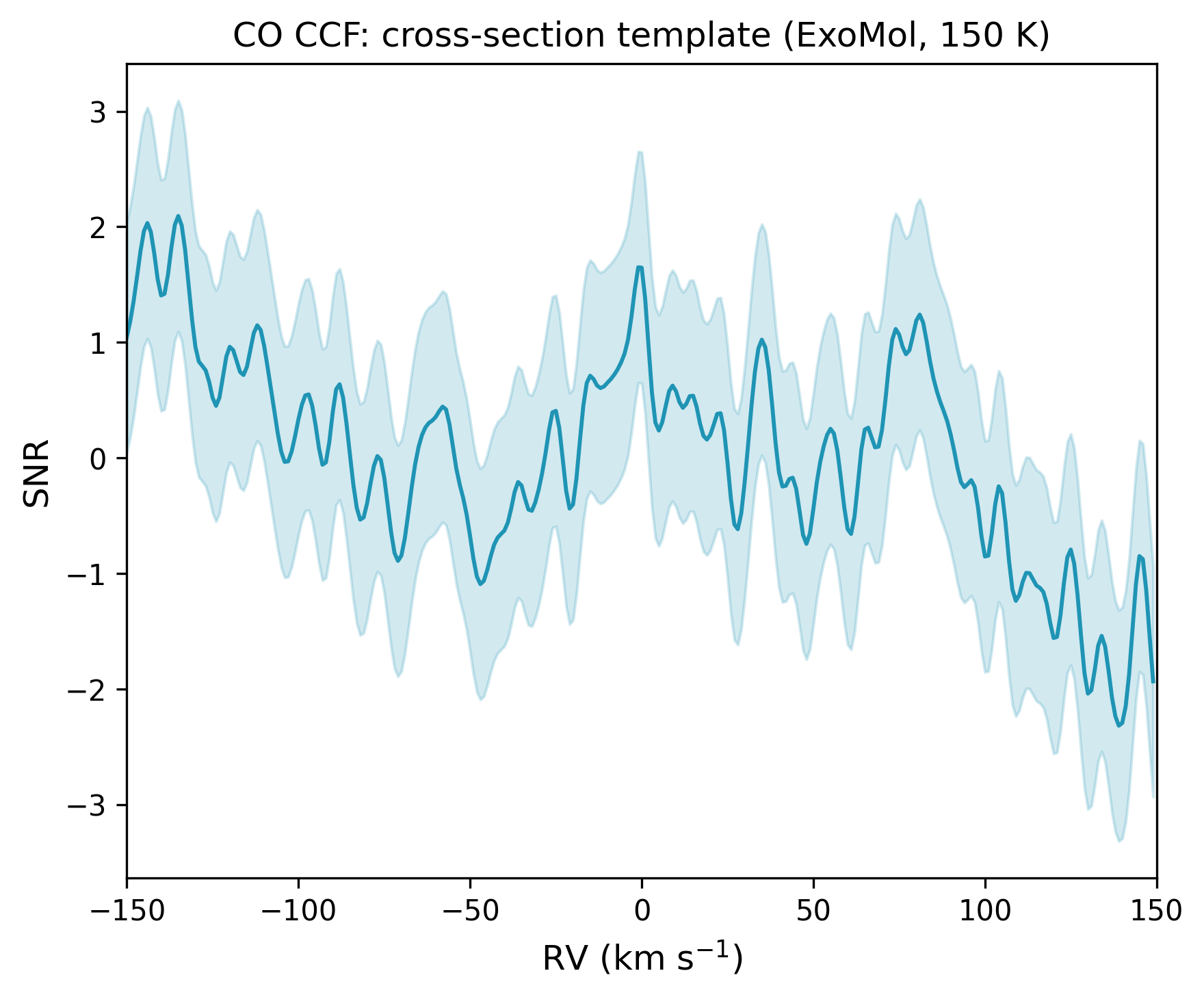}
        \caption{Carbon monoxide (150 K)}
    \end{subfigure} \hspace{0.01\linewidth}
    \begin{subfigure}{0.32\linewidth}
        \centering
        \includegraphics[width=\linewidth]{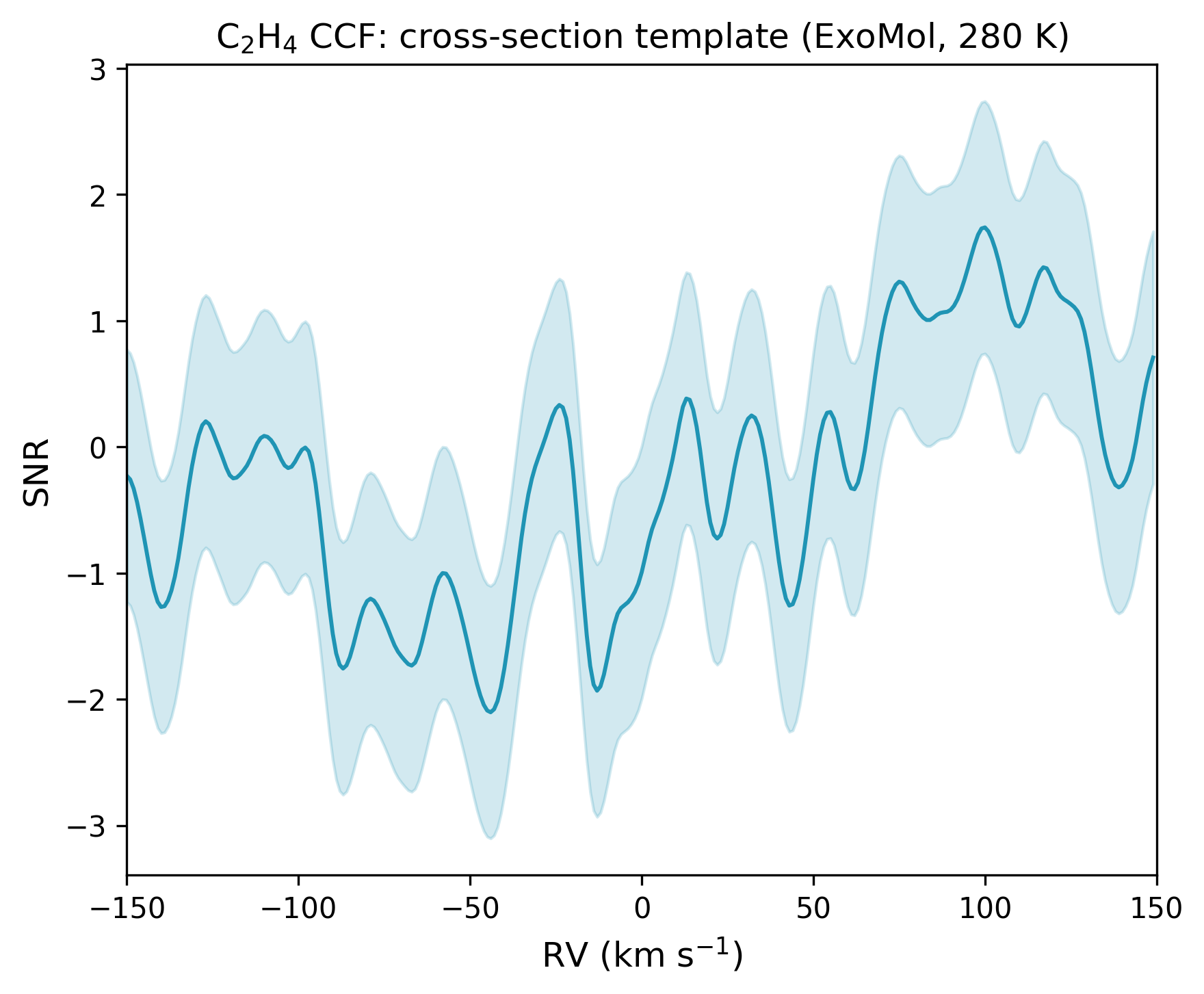}
        \caption{Ethylene (280 K)}
    \end{subfigure} \vspace{0.01\linewidth}
    \begin{subfigure}{0.32\linewidth}
        \centering
        \includegraphics[width=\linewidth]{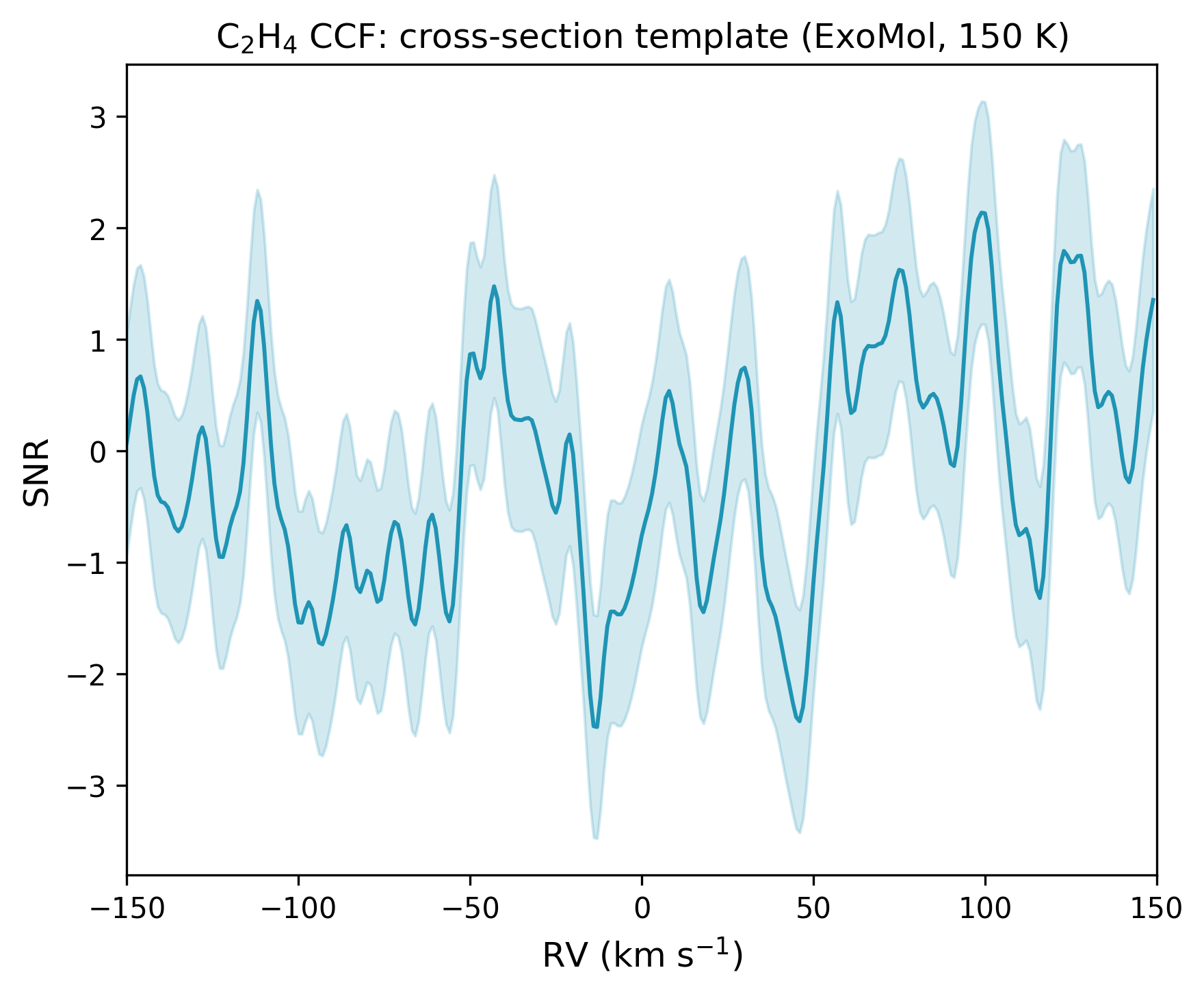}
        \caption{Ethylene (150 K)}
    \end{subfigure} \hspace{0.01\linewidth}
    \begin{subfigure}{0.32\linewidth}
        \centering
        \includegraphics[width=\linewidth]{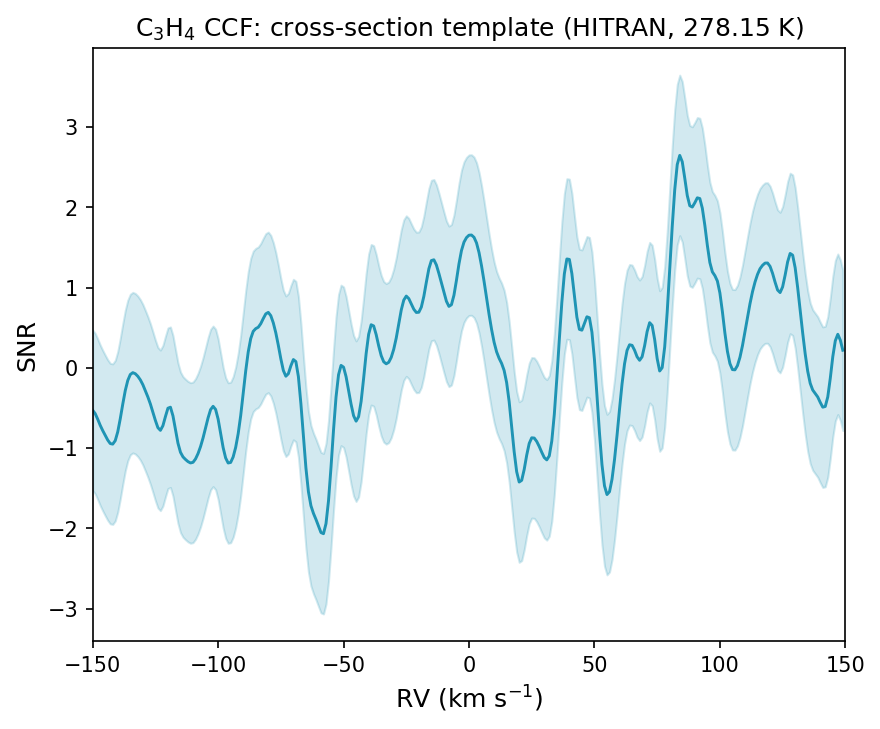}
        \caption{Propadiene}
    \end{subfigure} \hspace{0.01\linewidth}
    \begin{subfigure}{0.32\linewidth}
        \centering
        \includegraphics[width=\linewidth]{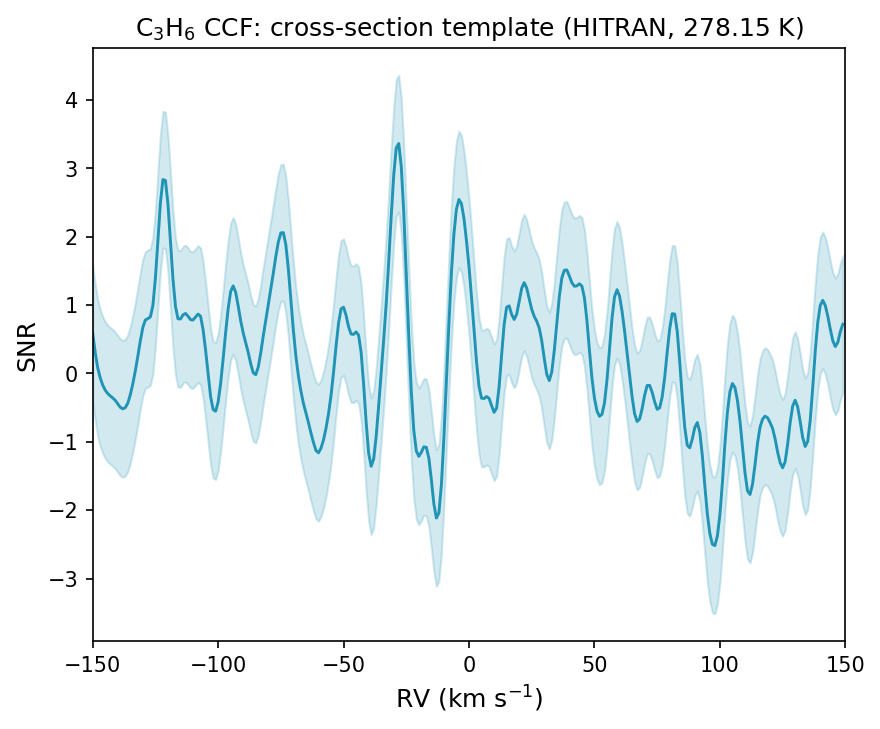}
        \caption{Propene}
    \end{subfigure} \vspace{0.01\linewidth}
    \begin{subfigure}{0.32\linewidth}
        \centering
        \includegraphics[width=\linewidth]{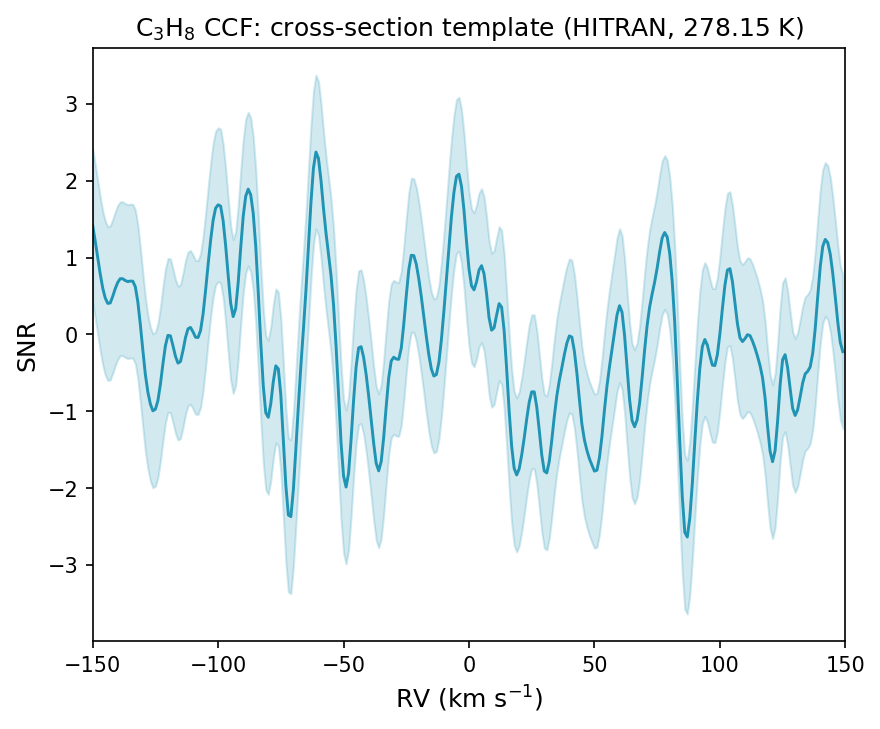}
        \caption{Propane}
    \end{subfigure} \hspace{0.01\linewidth}
    \begin{subfigure}{0.32\linewidth}
        \centering
        \includegraphics[width=\linewidth]{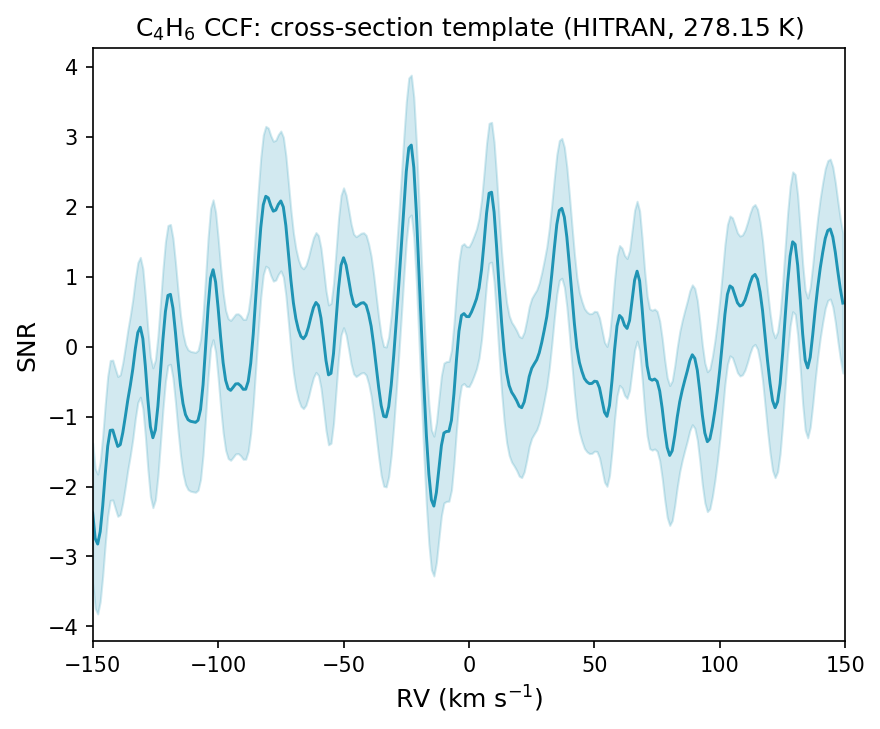}
        \caption{1,3-Butadiene}
    \end{subfigure} \hspace{0.01\linewidth}
    \begin{subfigure}{0.32\linewidth}
        \centering
        \includegraphics[width=\linewidth]{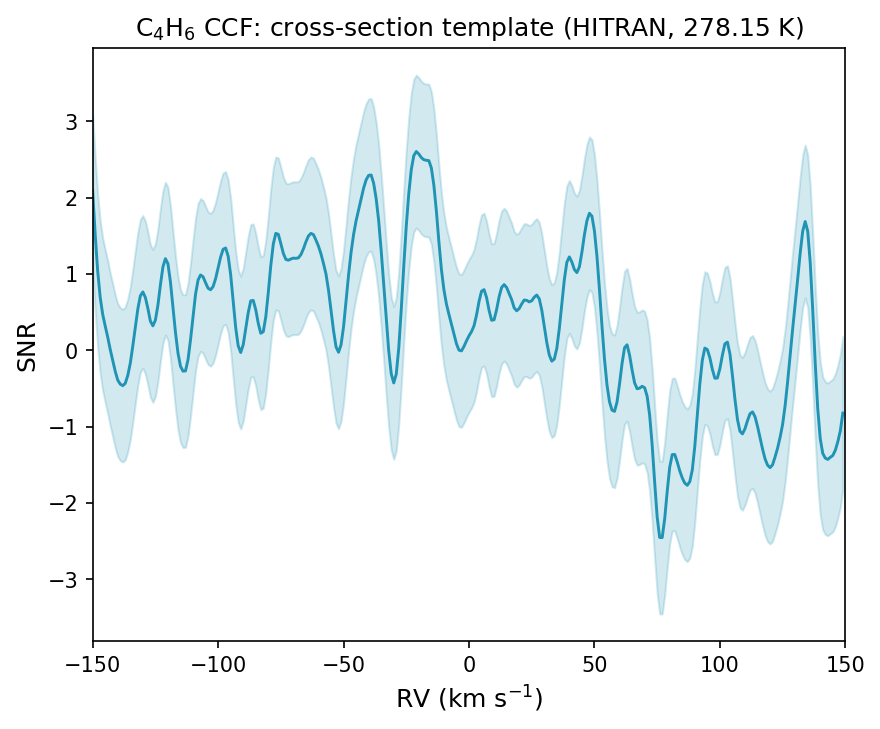}
        \caption{1-Butyne}
    \end{subfigure}
    \begin{subfigure}{0.32\linewidth}
        \centering
        \includegraphics[width=\linewidth]{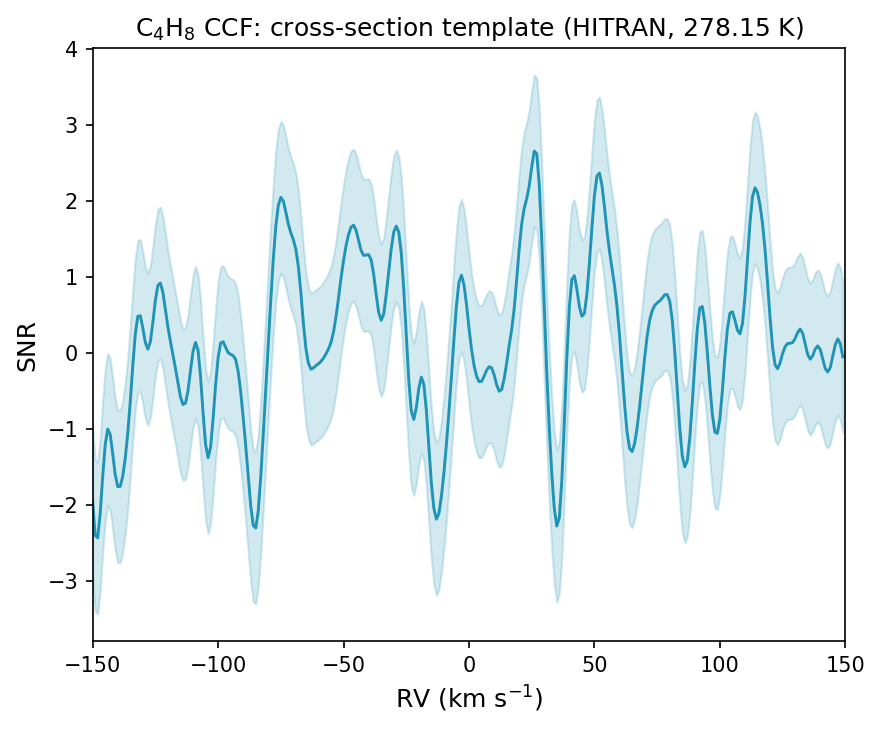}
        \caption{1-Butene}
    \end{subfigure} \hspace{0.01\linewidth}
    \begin{subfigure}{0.32\linewidth}
        \centering
        \includegraphics[width=\linewidth]{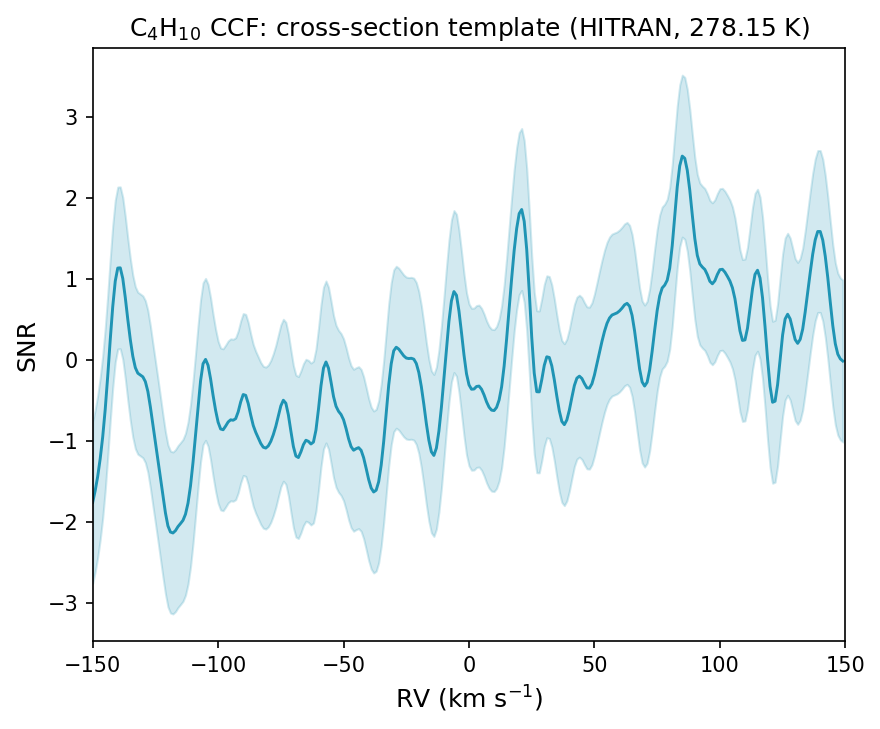}
        \caption{Butane}
    \end{subfigure} \hspace{0.01\linewidth}
    \begin{subfigure}{0.32\linewidth}
        \centering
        \includegraphics[width=\linewidth]{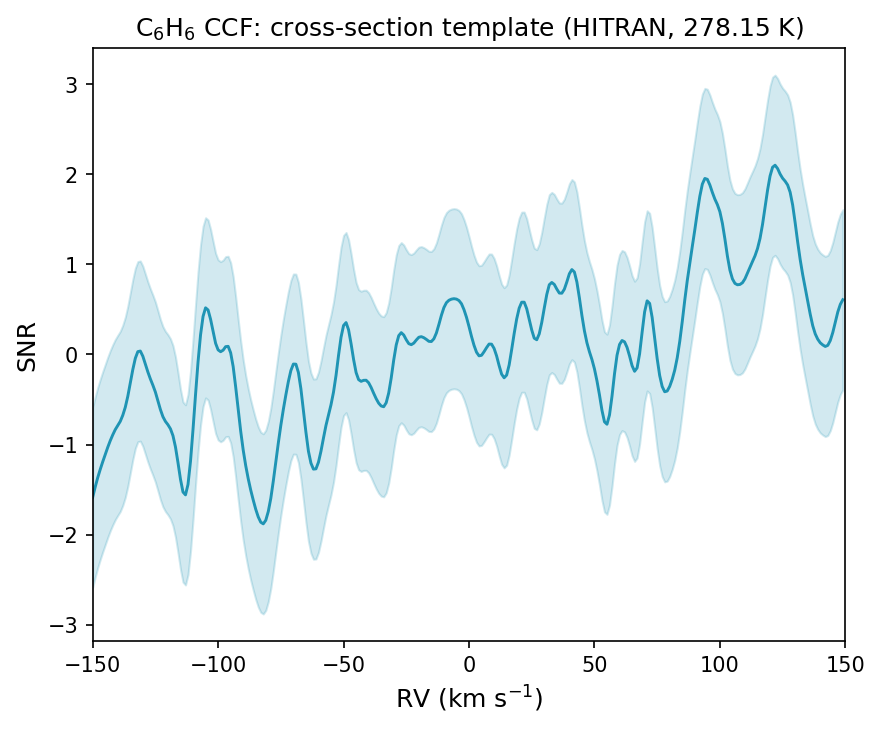}
        \caption{Benzene}
    \end{subfigure}
    \caption{Examples of non-detections for several hydrocarbons using HITRAN (278.15~K, 0.112~cm\(^{-1}\)) and ExoMol (280~K, 0.112~cm\(^{-1}\) and 150~K, 0.05~cm\(^{-1}\)) absorption cross-sections.}
    \label{fig:non_detections_hydrocarbons}
\end{figure*}

\begin{figure*}
    \centering
    \begin{subfigure}{0.32\linewidth} 
        \centering
        \includegraphics[width=\linewidth]{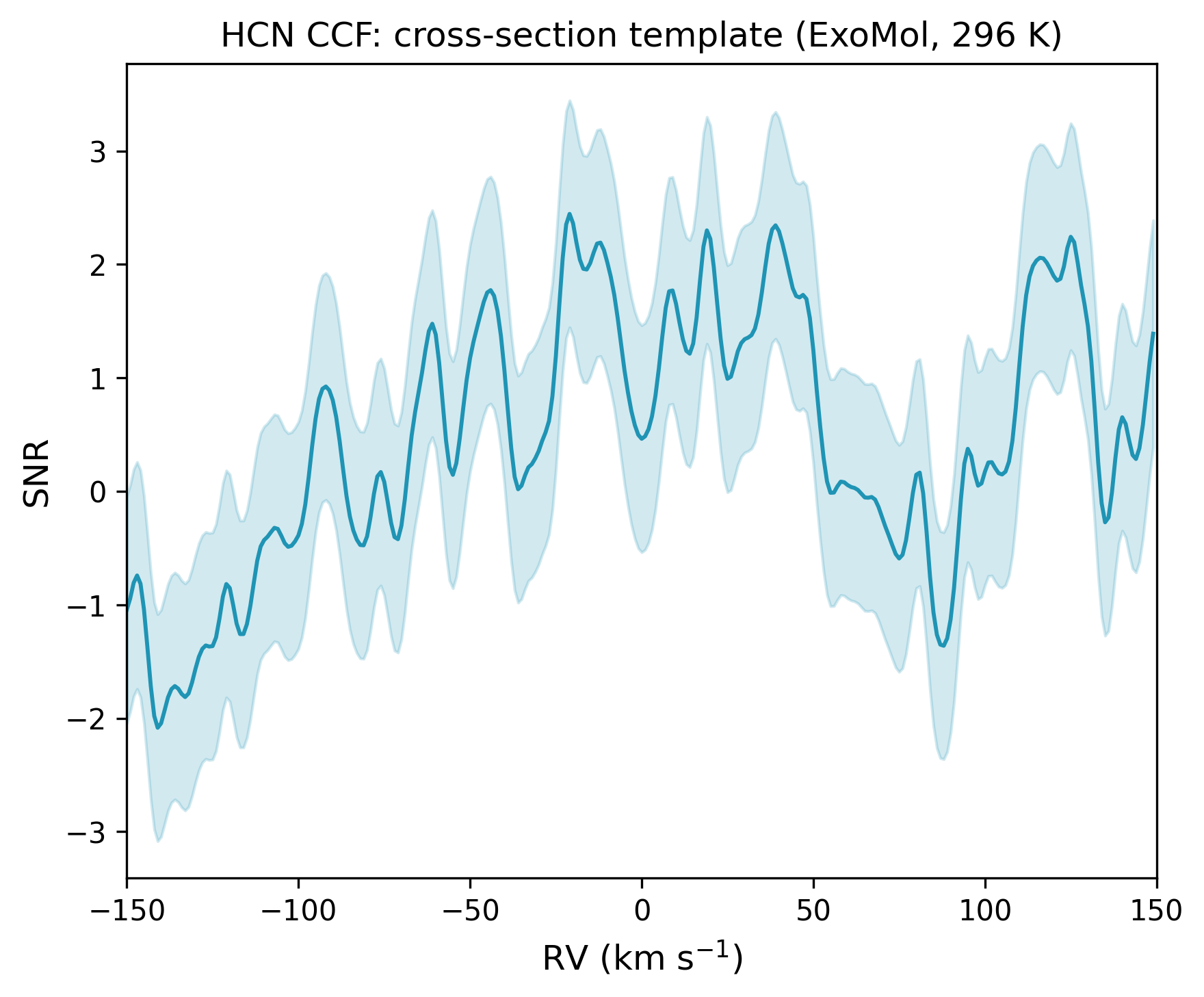}
        \caption{Hydrogen cyanide}
    \end{subfigure} \hspace{0.01\linewidth}
    \begin{subfigure}{0.32\linewidth} 
        \centering
        \includegraphics[width=\linewidth]{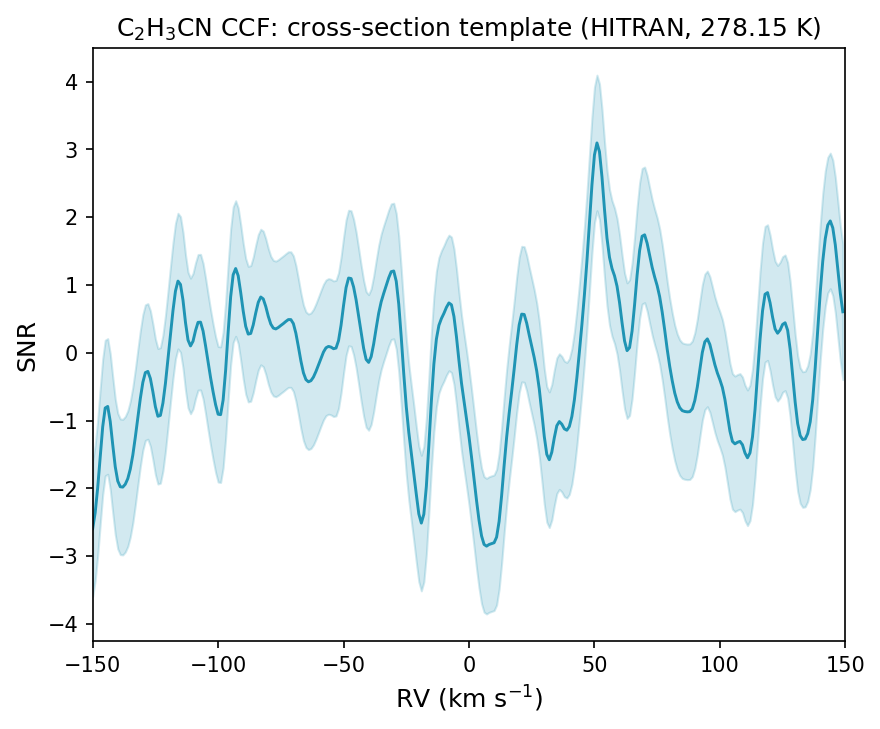}
        \caption{Acrylonitrile}
    \end{subfigure} \hspace{0.01\linewidth}
    \begin{subfigure}{0.32\linewidth} 
        \centering
        \includegraphics[width=\linewidth]{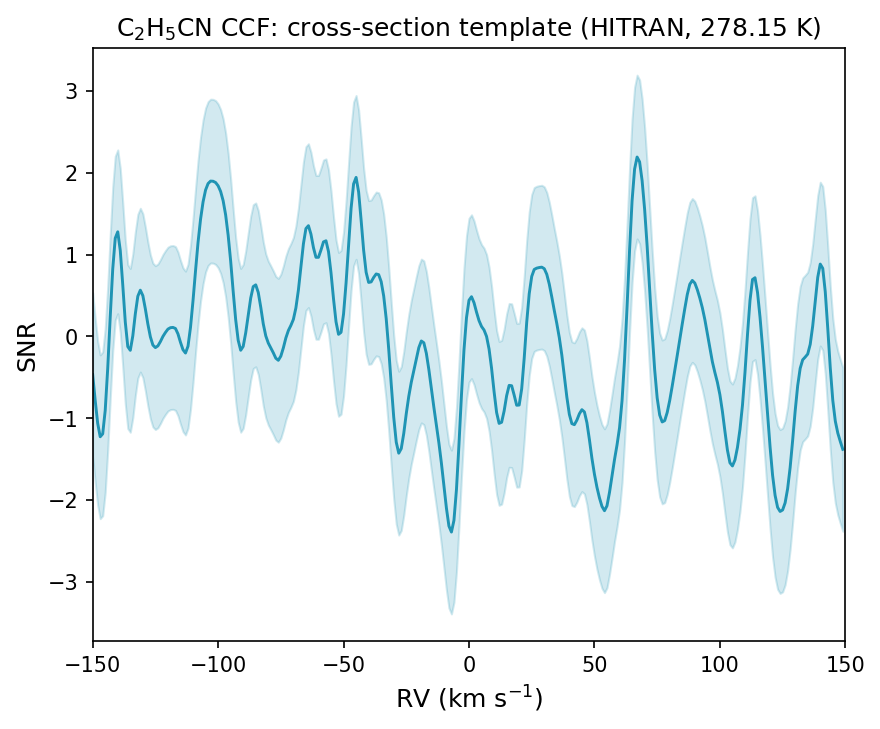}
        \caption{Ethyl cyanide}
    \end{subfigure}
    \caption{Examples of non-detections for several nitriles using HITRAN (0.112 cm\(^{-1}\)) and ExoMol (0.05 cm\(^{-1}\)) absorption cross-sections.}
    \label{fig:non_detections_nitriles}
\end{figure*}

\section{Monte-Carlo resampling results}

\begin{figure*}
    \centering
    \begin{subfigure}{.75\linewidth} 
        \centering
        \includegraphics[width=\linewidth]{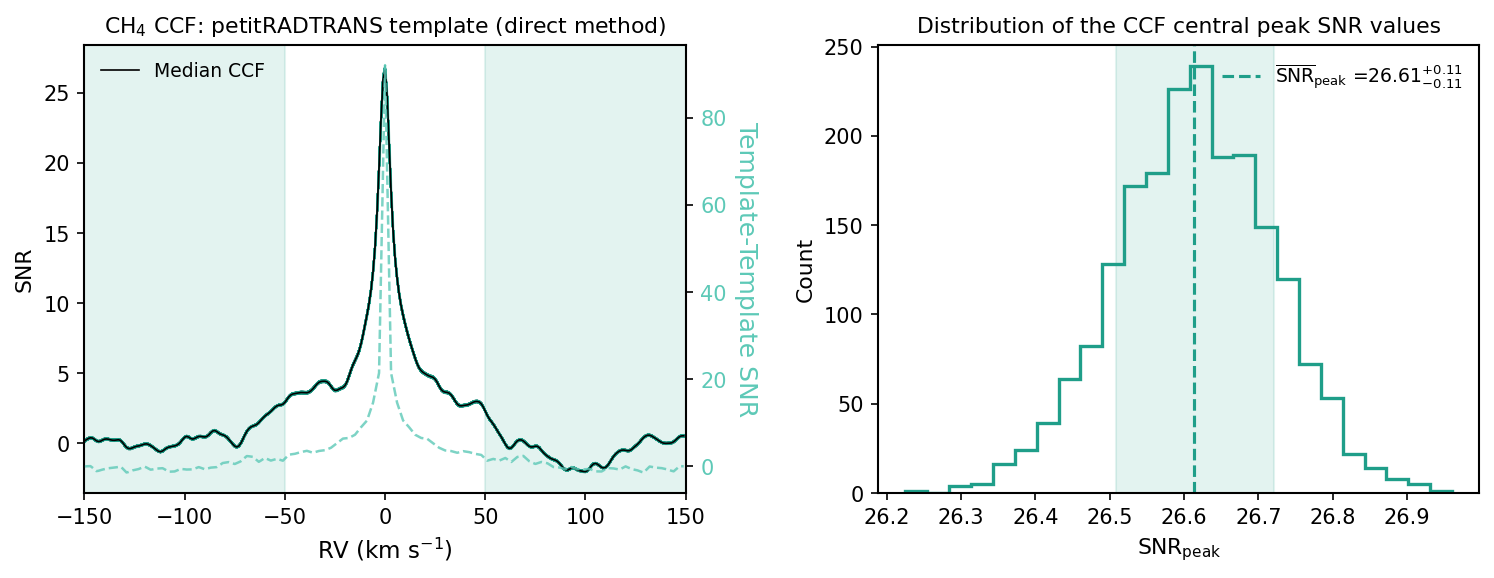}
    \end{subfigure}
    \begin{subfigure}{.75\linewidth} 
        \centering
        \includegraphics[width=\linewidth]{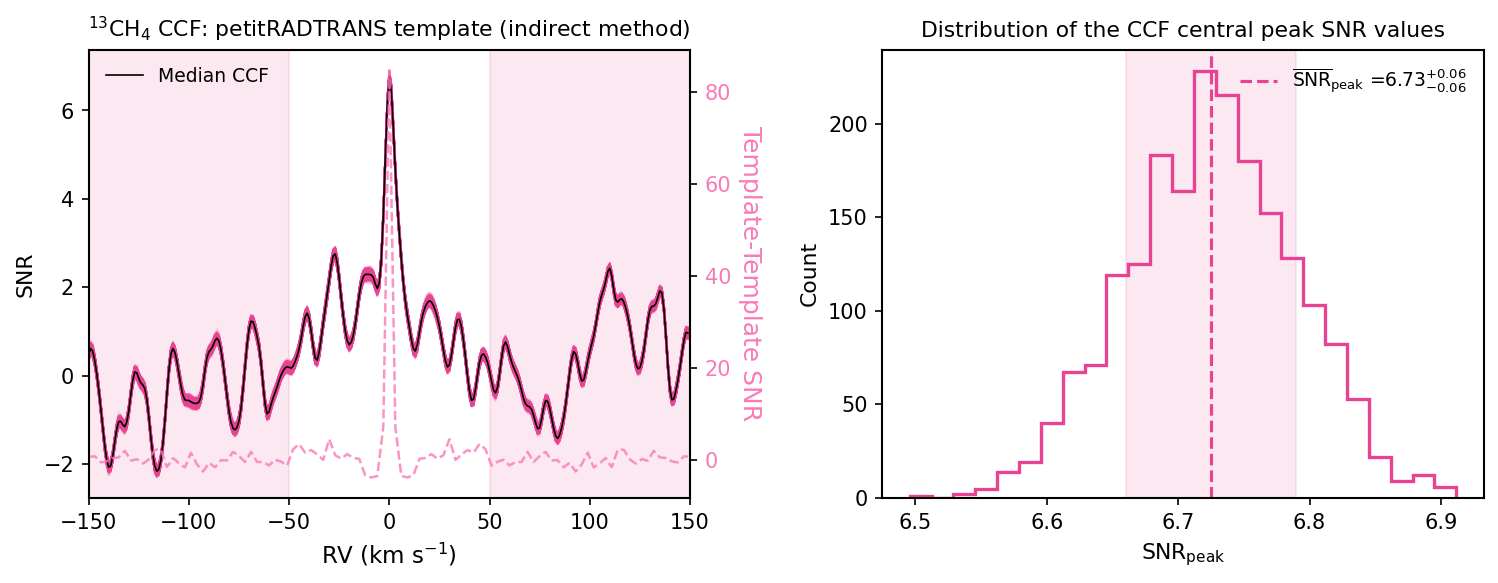}
    \end{subfigure}
    \begin{subfigure}{0.35\linewidth} 
        \centering
        \includegraphics[width=\linewidth]{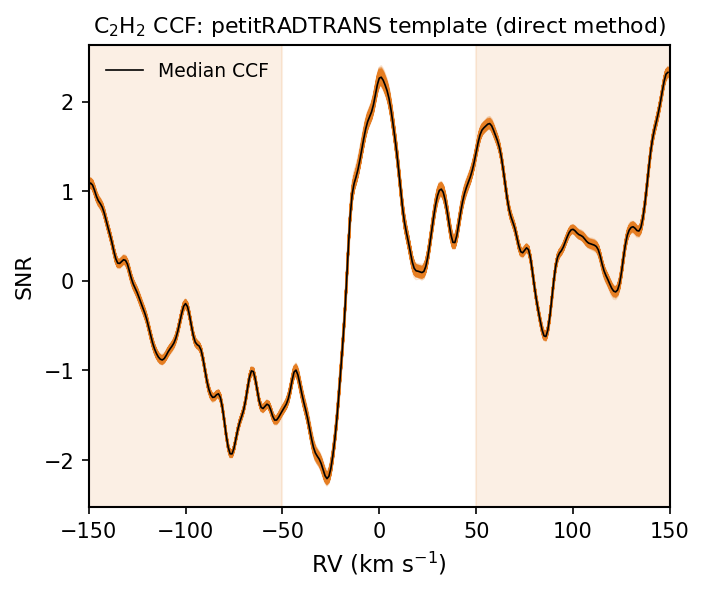}
    \end{subfigure} \hspace{0.03\linewidth}
    \begin{subfigure}{0.35\linewidth}
        \centering
        \includegraphics[width=\linewidth]{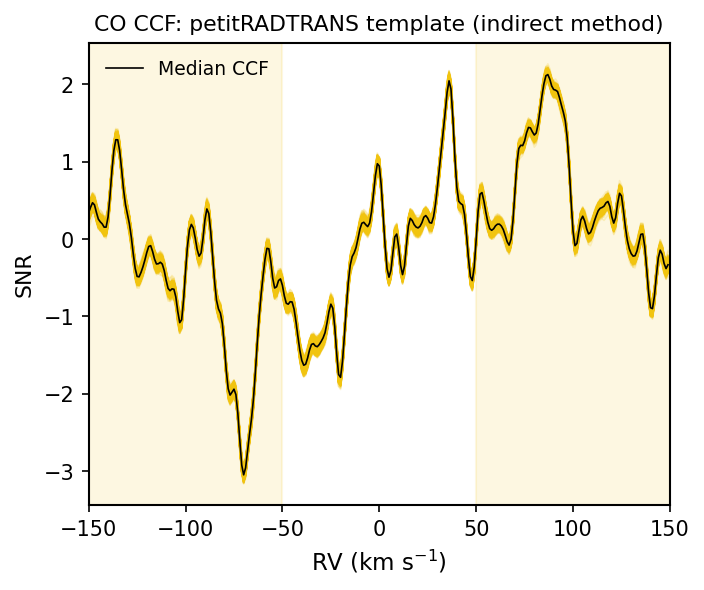}
    \end{subfigure} \hspace{0.03\linewidth}
    \begin{subfigure}{0.35\linewidth} 
        \centering
        \includegraphics[width=\linewidth]{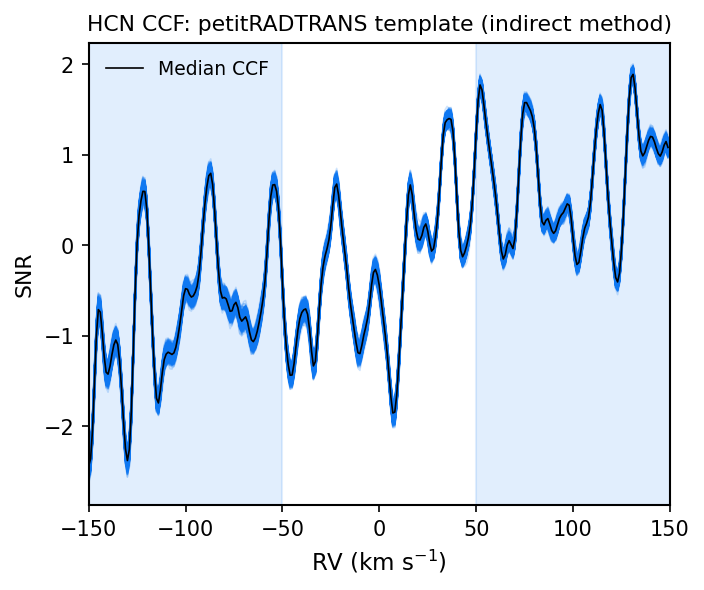}
    \end{subfigure}
    \caption{Cross‑correlation results obtained from the Monte Carlo realisations of the \texttt{petitRADTRANS}‑based templates. Panels show: \ce{CH4} with the direct template, $^{13}$\ce{CH4} with the indirect template, \ce{C2H2} with the direct template, CO with the indirect template, and HCN with the indirect template. \ce{CH4} and $^{13}$\ce{CH4} yield significant CCF peaks at Titan’s rest‑frame velocity, while \ce{C2H2}, CO, and HCN do not produce detectable signals. For C$_2$H$_2$, the non-detection presented here reflects the performance of the direct line-by-line template only; the molecule is detected in the main analysis using the indirect template, which suppresses weak features and retains only the strongest lines.}
    \label{fig:non_detections_petitRADTRANS}
\end{figure*}

\section{Detection of $\mathbf{^{13}}$\ce{CH4}}

\begin{figure*}
    \centering
    \includegraphics[width=.4\linewidth]{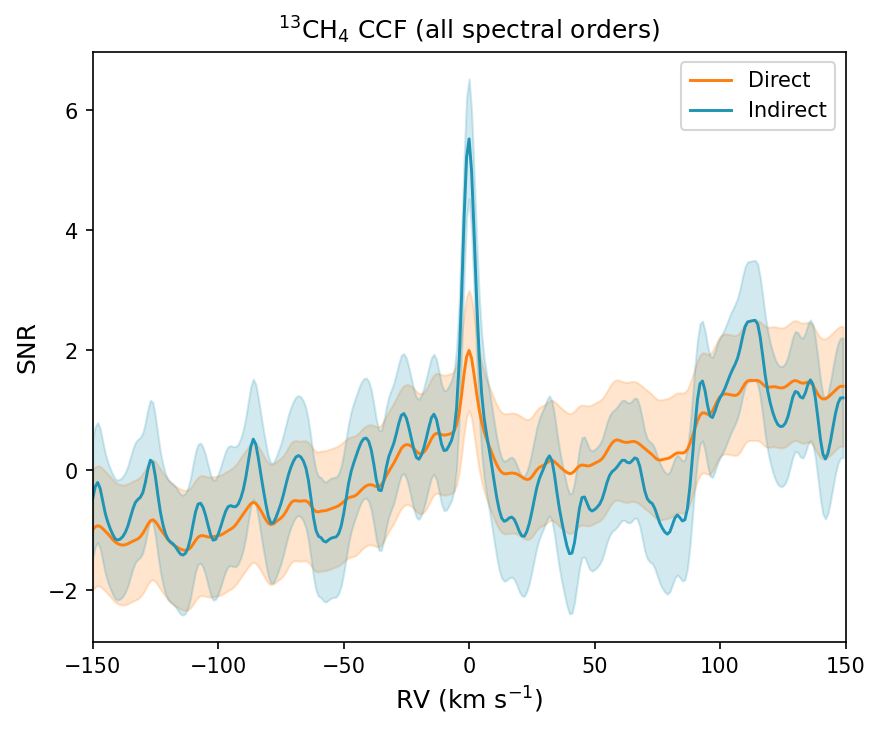}
    \caption{$^{13}$\ce{CH4} CCFs obtained with the direct (orange) and indirect (blue) \texttt{petitRADTRANS} templates after summing all nights and all spectral orders. The indirect template yields a clear and statistically significant peak at Titan’s rest‑frame velocity. In contrast, the direct template shows only a weak peak: including all the weaker features introduces noise that dilutes the overall signal.}
    \label{fig:CCF_13CH4_all_orders}
\end{figure*}

\section{SNR dependence on the fraction of lines included}

\begin{figure*}
    \centering
    \begin{subfigure}{0.49\linewidth}
        \centering
        \includegraphics[width=\linewidth]{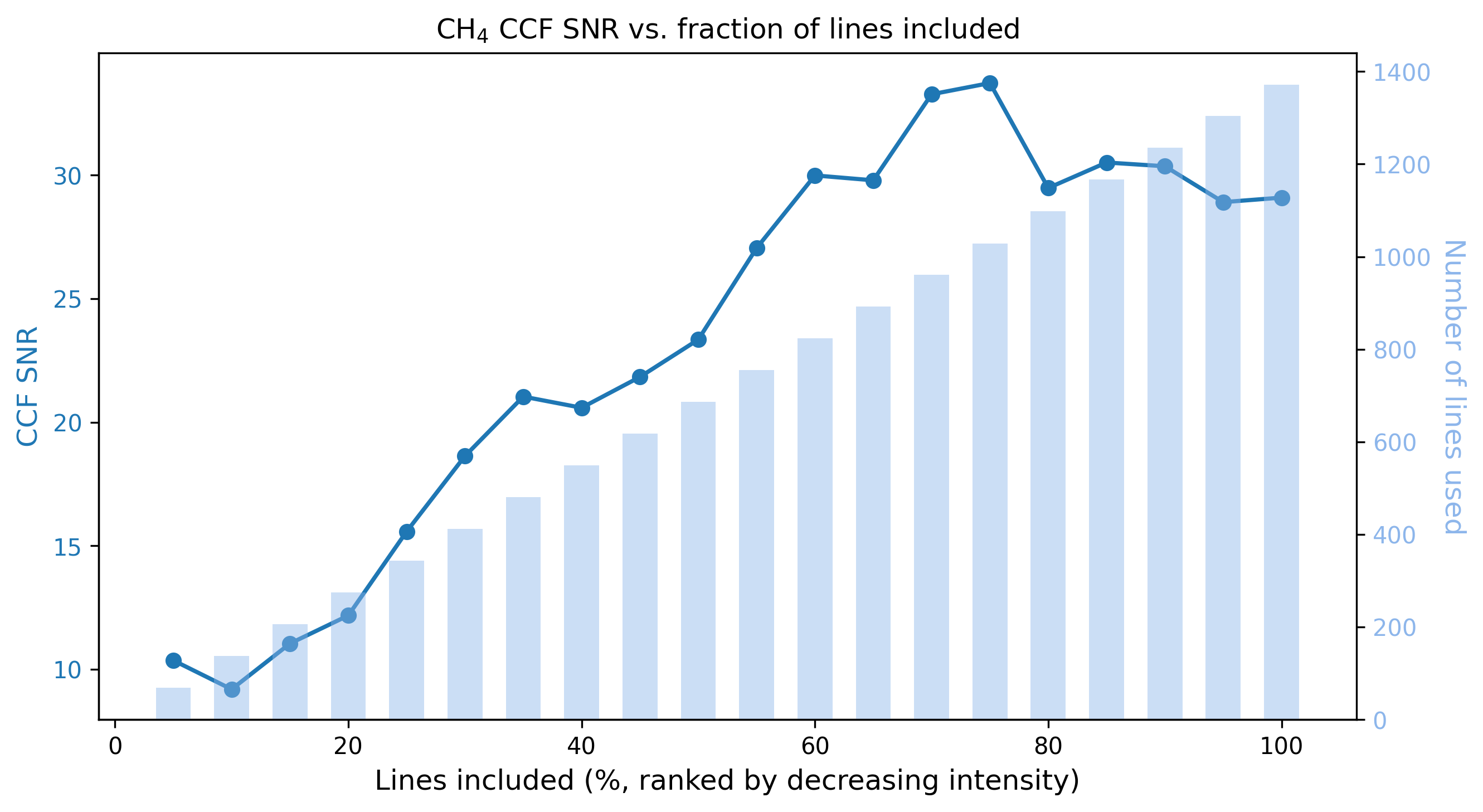}
    \end{subfigure} \hspace{0.01\linewidth}
    \begin{subfigure}{0.49\linewidth}
        \centering
        \includegraphics[width=\linewidth]{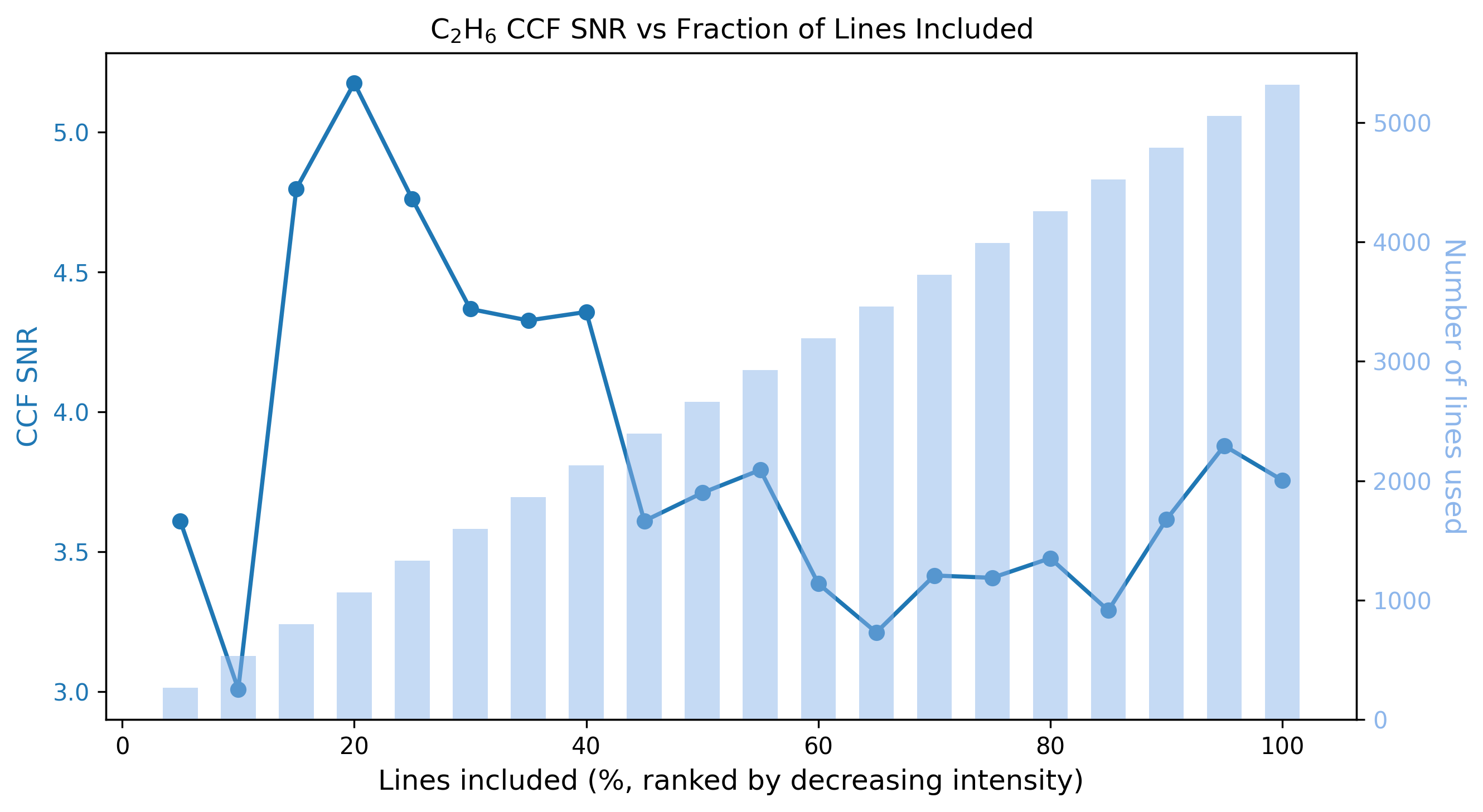}
    \end{subfigure}
    \caption{Cross‑correlation signal‑to‑noise ratio (SNR) as a function of the fraction of spectral lines included in the indirect template for \ce{CH4} and \ce{C2H6}. The blue curve shows the CCF SNR obtained when progressively larger fractions of the strongest absorption lines are incorporated into the template, while the bars indicate the corresponding number of lines used. For \ce{CH4}, the SNR reaches a clear maximum when 75\% of the ranked lines are included, whereas for \ce{C2H6} the optimal value is 20\% of lines. This demonstrates that there exists an optimal fraction of lines that maximises the detection significance, reflecting a balance between adding genuine line information and introducing noise from weaker features. This optimum is molecule-dependent, varying with the absorption profile of each species.}
    \label{fig:SNRvsFraction}
\end{figure*}

\begin{table*}
    \centering
    \caption{Overview of the indirect template configurations used in the cross‑correlation analysis. For each molecule, we list the opacity source, the temperature of the corresponding cross‑sections or line‑by‑line model, and the fraction of spectral lines included in the final indirect template. The optimal fraction of lines included is molecule-dependent, varying with the absorption profile of each species.}
    \begin{tabular}{l l l l c}
        \hline
        Molecule & Template & Opacity data & T (K) & Fraction of Lines Included \\
        \hline
        CH$_4$ & Indirect & \texttt{petitRADTRANS} line-by-line model & 150 & 0.75 \\
        & Indirect & ExoMol cross-section & 280 & 0.60 \\
        & Indirect & ExoMol cross-section & 150 & 0.60 \\
        C$_2$H$_2$ & Indirect & \texttt{petitRADTRANS} line-by-line model & 150 & 0.60 \\
        & Indirect & ExoMol cross-section & 280 & 0.15 \\
        & Indirect & ExoMol cross-section & 150 & 0.20 \\
        C$_2$H$_6$ & Indirect & MOLLIST laboratory cross-section & 205 & 0.20 \\
        \hline
    \end{tabular}
    \label{tab:line_fraction}
\end{table*}

\section{Line-by-line templates for \ce{C2H2}, CO and HCN}

\begin{figure*}
    \centering
    \includegraphics[width=0.9\linewidth]{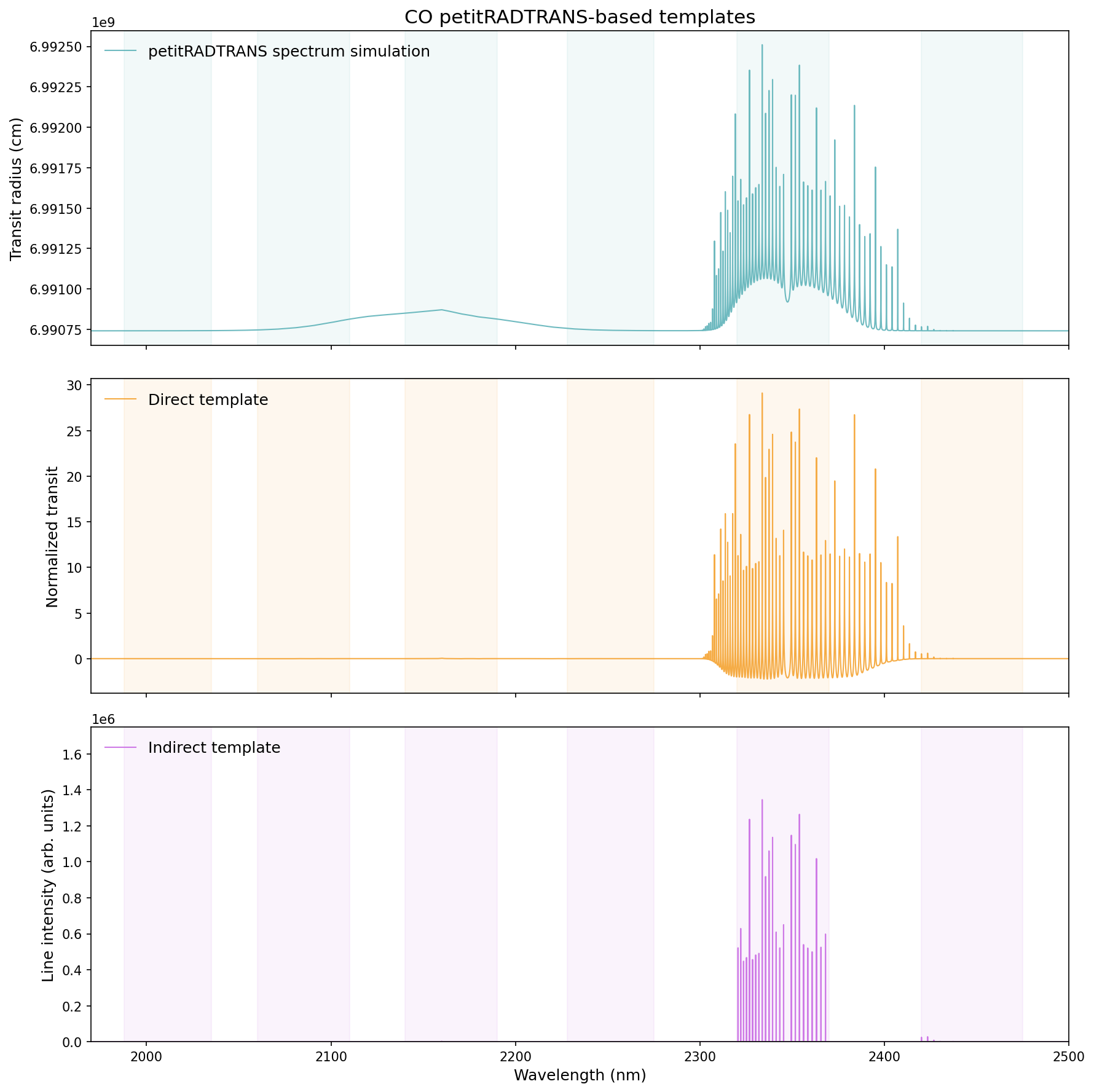}
    \caption{\texttt{petitRADTRANS} line-by-line transit spectrum, and corresponding direct and indirect templates for \ce{CO} across the CRIRES+ wavelength range. The top panel shows the transit spectrum computed with the radiative‑transfer model. The middle panel displays the corresponding direct template, obtained by normalising the transit radius with a Gaussian filter. The bottom panel shows the indirect template constructed from the strongest absorption features. Only lines that fall within the CRIRES+ spectral orders are included in this step, ensuring that the intensity‑based ranking reflects the subset of lines actually used in the cross‑correlation analysis.}
    \label{fig:CO_temp}
\end{figure*}

\begin{figure*}
    \centering
    \includegraphics[width=0.9\linewidth]{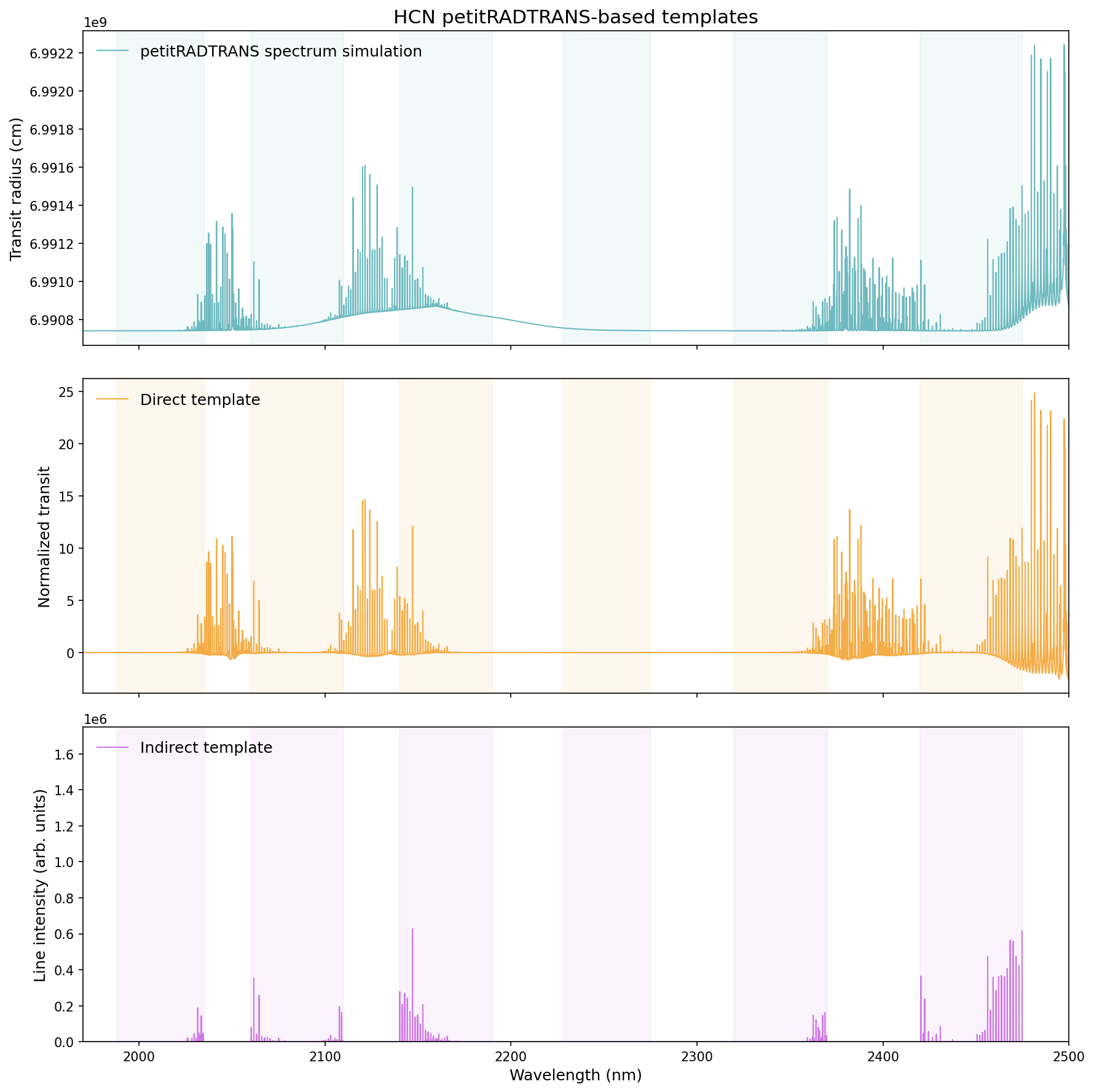}
    \caption{\texttt{petitRADTRANS} line-by-line transit spectrum, and corresponding direct and indirect templates for \ce{HCN} across the CRIRES+ wavelength range. The shaded regions indicate the wavelength coverage of each CRIRES+ spectral order.}
    \label{fig:HCN_temp}
\end{figure*}

\begin{figure*}
    \centering
    \includegraphics[width=0.9\linewidth]{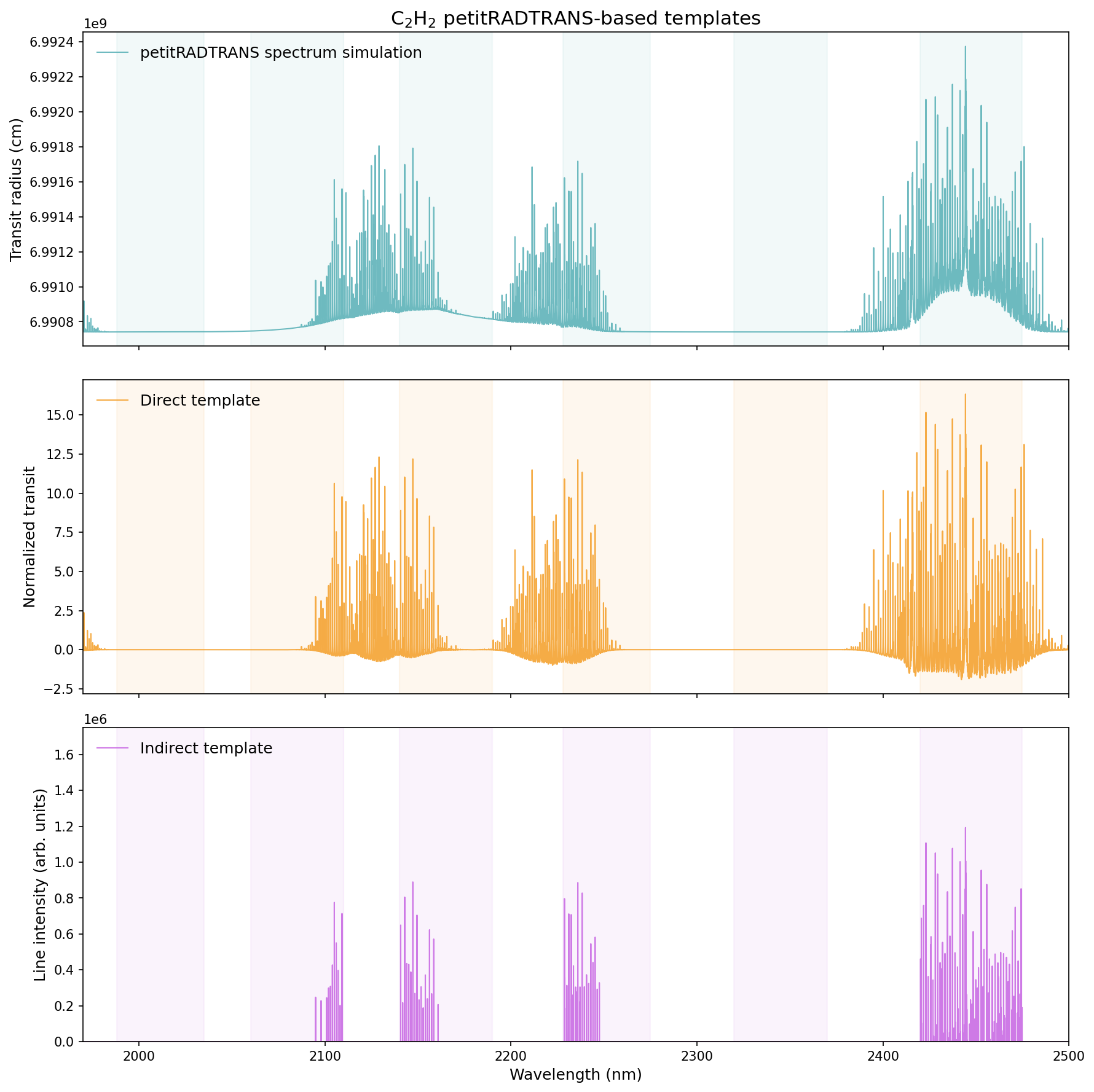}
    \caption{\texttt{petitRADTRANS} line-by-line transit spectrum, and corresponding direct and indirect templates for \ce{C2H2} across the CRIRES+ wavelength range. The shaded regions indicate the wavelength coverage of each CRIRES+ spectral order.}
    \label{fig:C2H2_temp}
\end{figure*}

\section{Cross-correlations with the calibration star spectra}

\begin{figure*}
    \centering
    \begin{subfigure}{0.4\linewidth} 
        \centering
        \includegraphics[width=\linewidth]{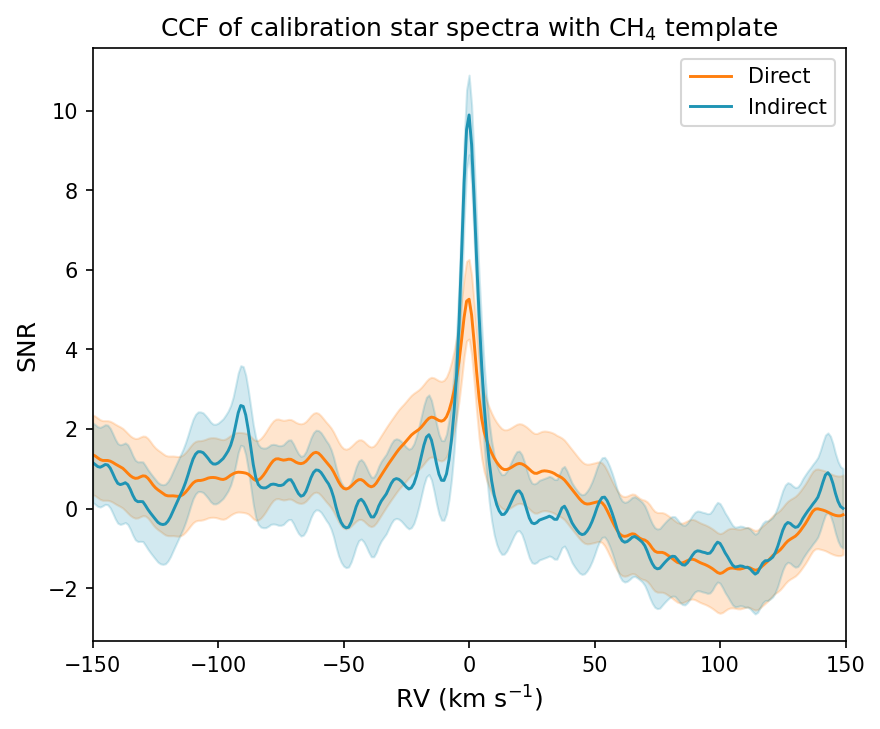}
    \end{subfigure} \hspace{0.03\linewidth}
    \begin{subfigure}{0.4\linewidth} 
        \centering
        \includegraphics[width=\linewidth]{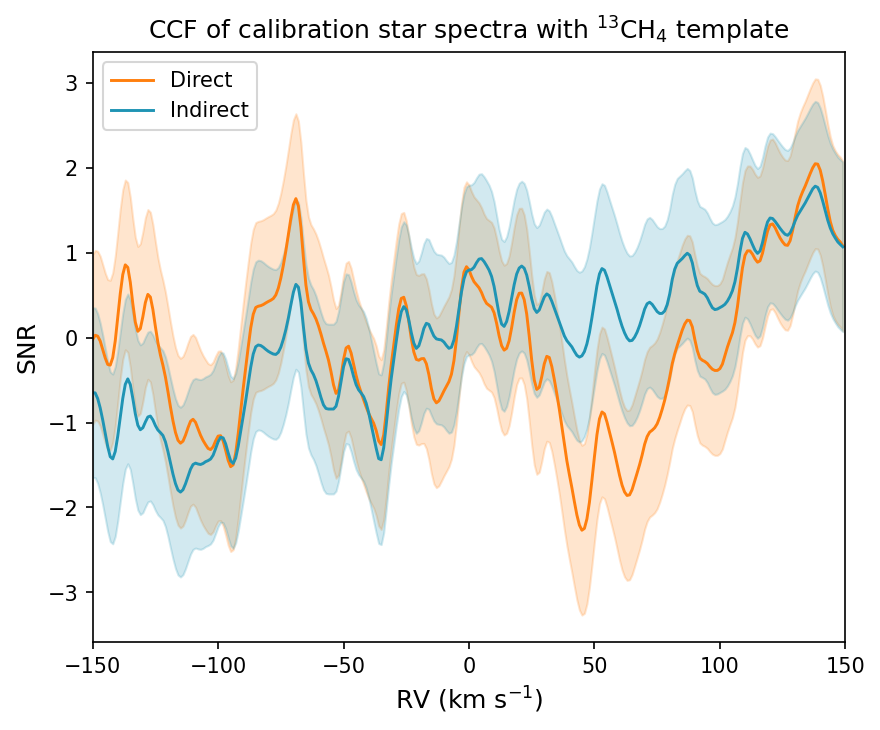}
    \end{subfigure}    
    \begin{subfigure}{0.4\linewidth} 
        \centering
        \includegraphics[width=\linewidth]{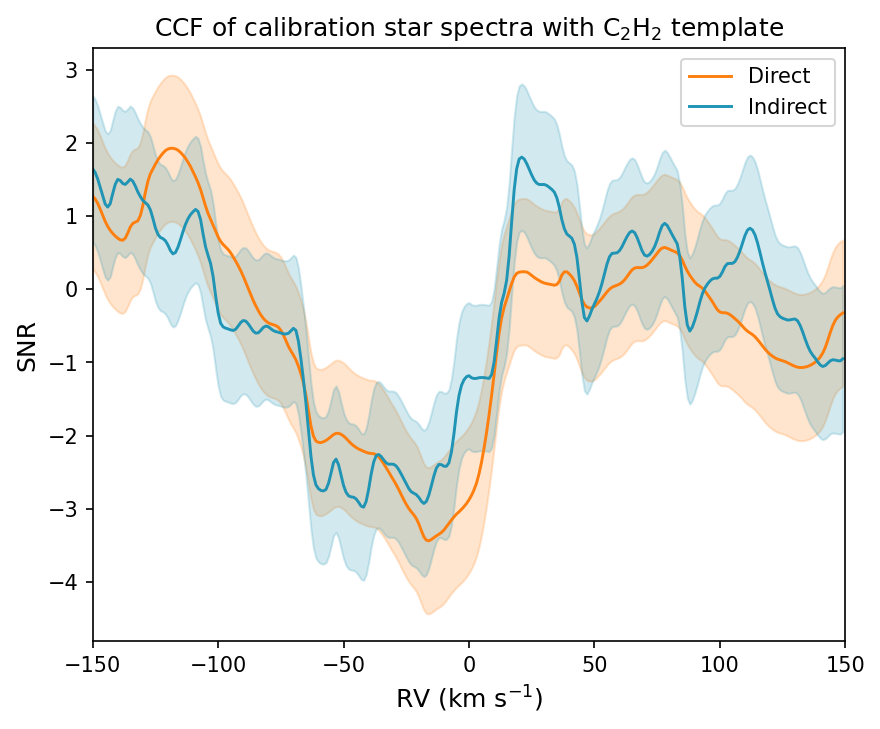}
    \end{subfigure} \hspace{0.03\linewidth}
    \begin{subfigure}{0.4\linewidth} 
        \centering
        \includegraphics[width=\linewidth]{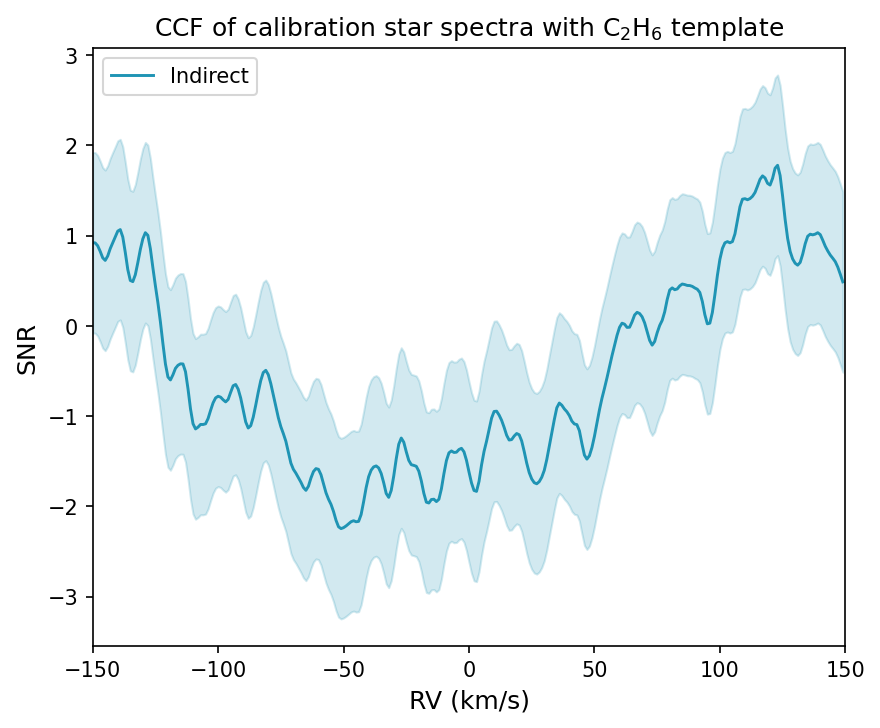}
    \end{subfigure}
    \caption{Cross‑correlation functions computed using the calibration‑star spectra and the optimal template for each molecular species. For all molecules except methane, the calibration‑star CCFs show no significant peak, confirming that telluric contamination does not contribute to the detections reported in this work. Methane is the only exception: the calibration‑star spectra yield a clear CCF peak at 0~km~s$^{-1}$, corresponding to absorption by terrestrial \ce{CH4}. To verify that the methane detection in Titan’s spectra is not driven by this telluric contribution, we recomputed the CCFs using Titan spectra in the geocentric frame. In this case, the CCF peak shifts consistently with Titan’s changing radial velocity across the three observing nights, demonstrating that the detected signal originates from Titan’s atmosphere rather than from Earth’s [see Figure~\ref{fig:ccf_geo}].}
    \label{fig:ccf_star}
\end{figure*}

\section{Cross-correlations in the geocentric frame}

\begin{figure*}
    \centering
    \begin{subfigure}{0.32\linewidth}
        \centering
        \includegraphics[width=\linewidth]{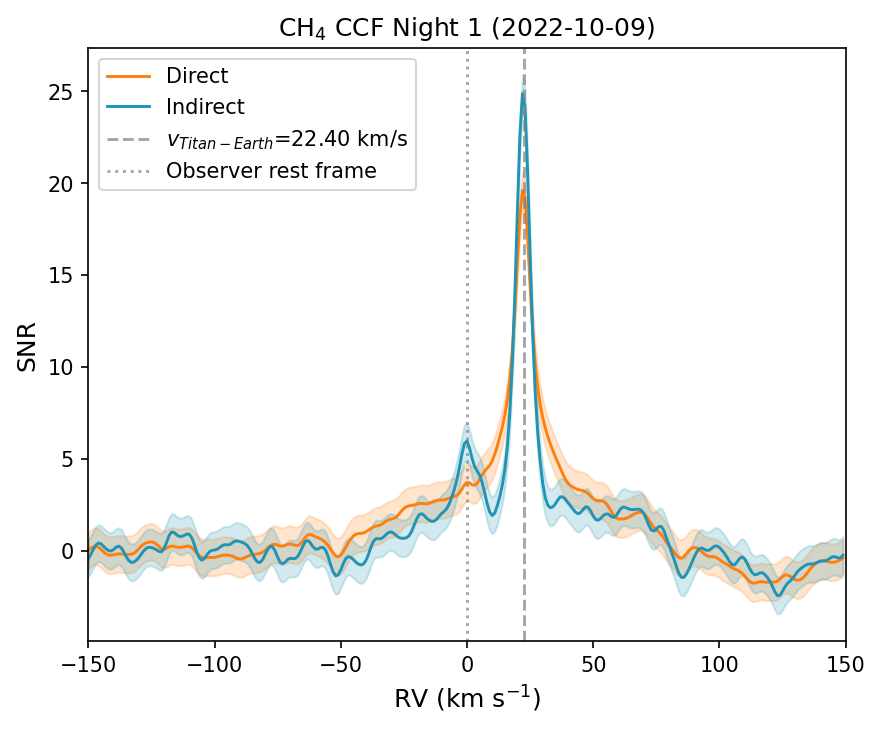}
    \end{subfigure} \hspace{0.01\linewidth}
    \begin{subfigure}{0.32\linewidth}
        \centering
        \includegraphics[width=\linewidth]{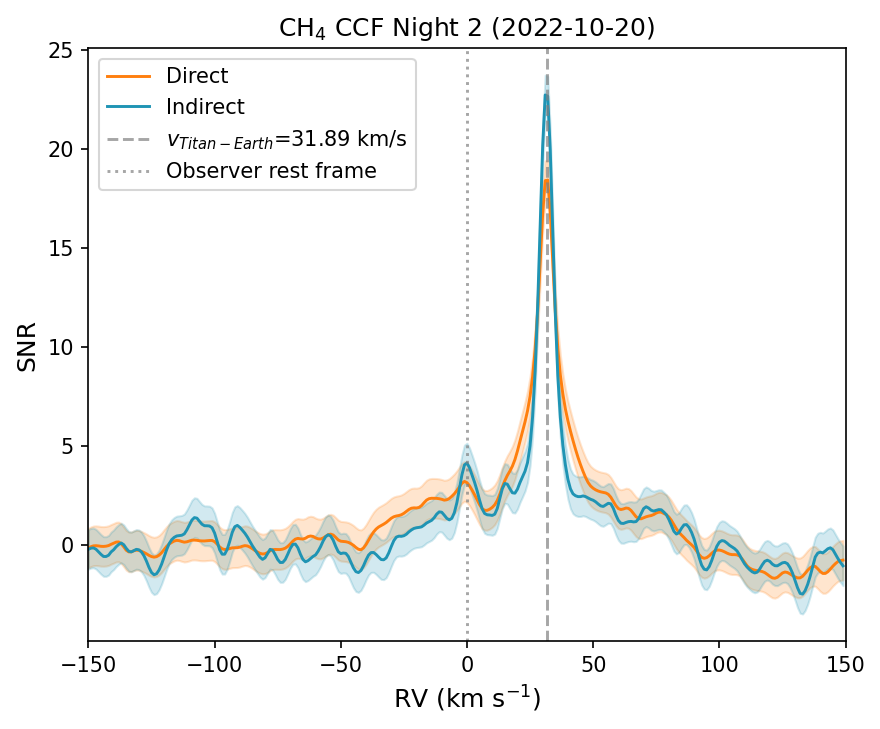}
    \end{subfigure} \hspace{0.01\linewidth}
    \begin{subfigure}{0.32\linewidth}
        \centering
        \includegraphics[width=\linewidth]{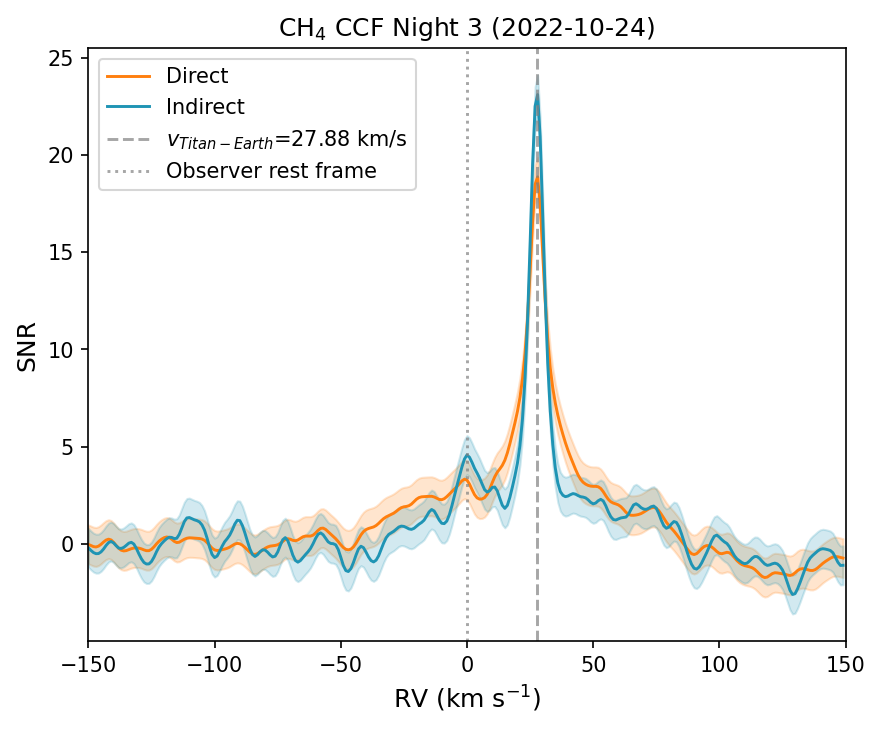}
    \end{subfigure}
    \begin{subfigure}{0.32\linewidth}
        \centering
        \includegraphics[width=\linewidth]{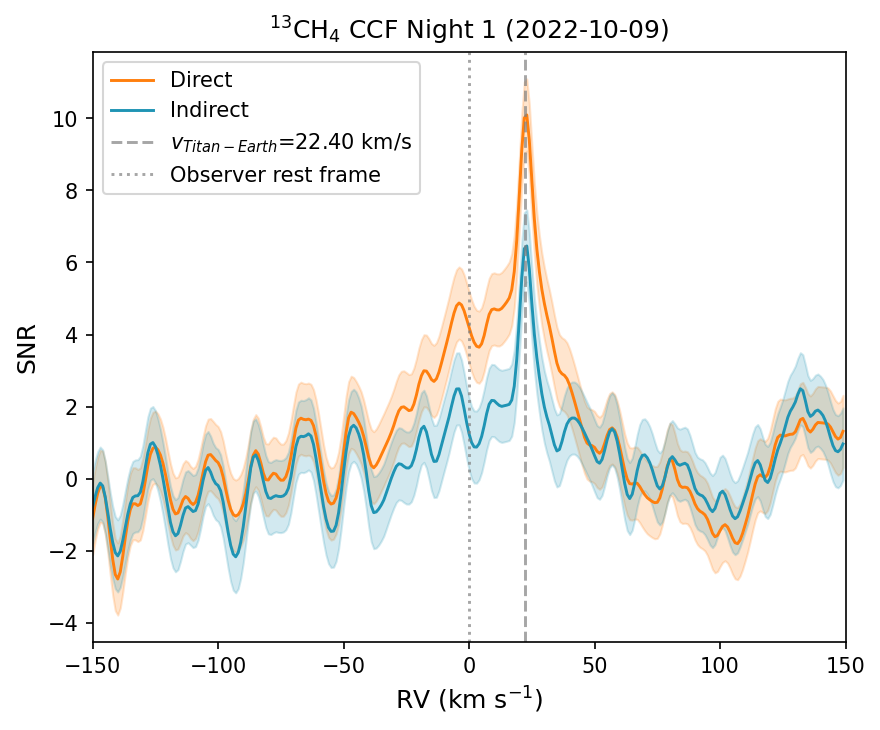}
    \end{subfigure} \hspace{0.01\linewidth}
    \begin{subfigure}{0.32\linewidth}
        \centering
        \includegraphics[width=\linewidth]{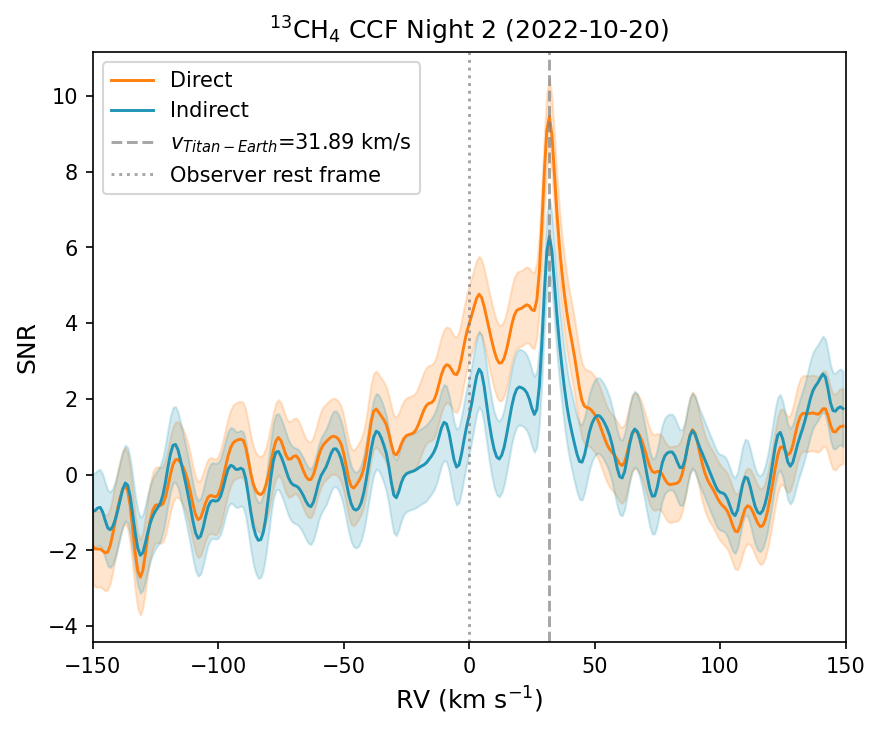}
    \end{subfigure} \hspace{0.01\linewidth}
    \begin{subfigure}{0.32\linewidth}
        \centering
        \includegraphics[width=\linewidth]{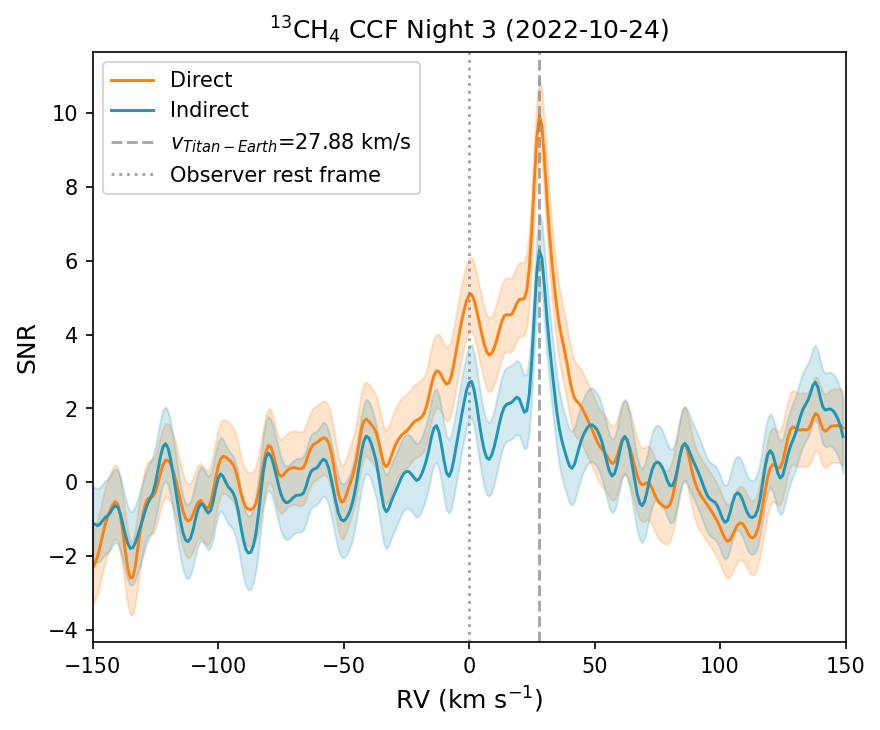}
    \end{subfigure}
    \begin{subfigure}{0.32\linewidth}
        \centering
        \includegraphics[width=\linewidth]{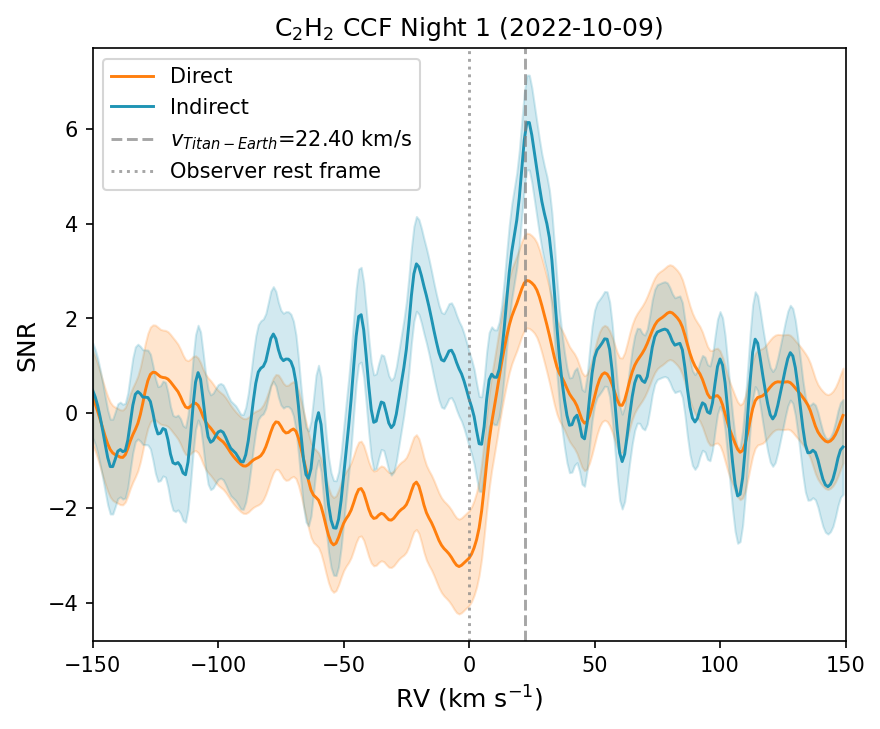}
    \end{subfigure} \hspace{0.01\linewidth}
    \begin{subfigure}{0.32\linewidth}
        \centering
        \includegraphics[width=\linewidth]{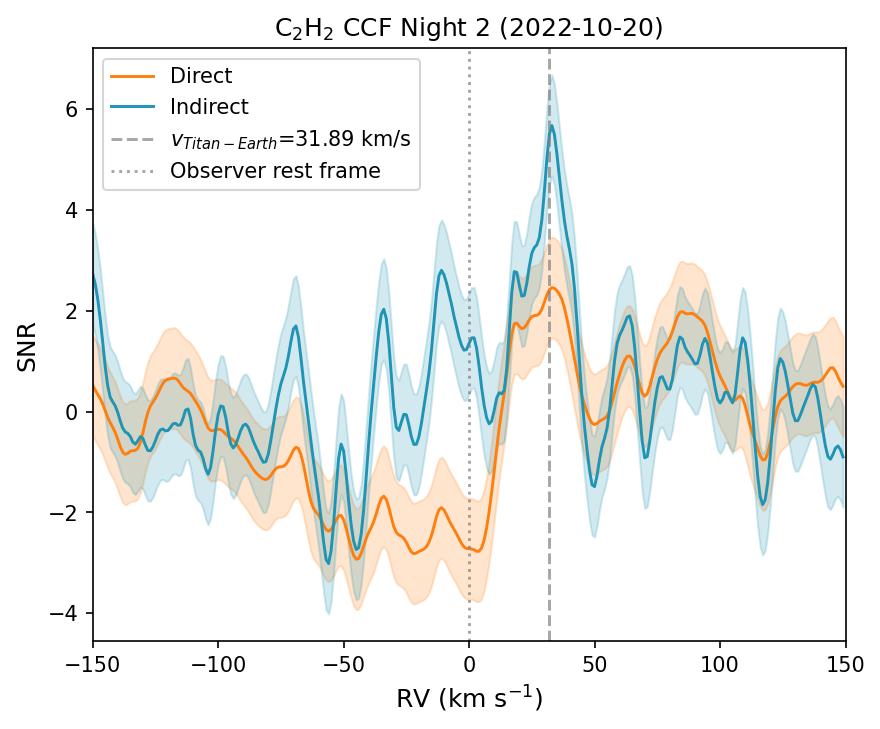}
    \end{subfigure} \hspace{0.01\linewidth}
    \begin{subfigure}{0.32\linewidth}
        \centering
        \includegraphics[width=\linewidth]{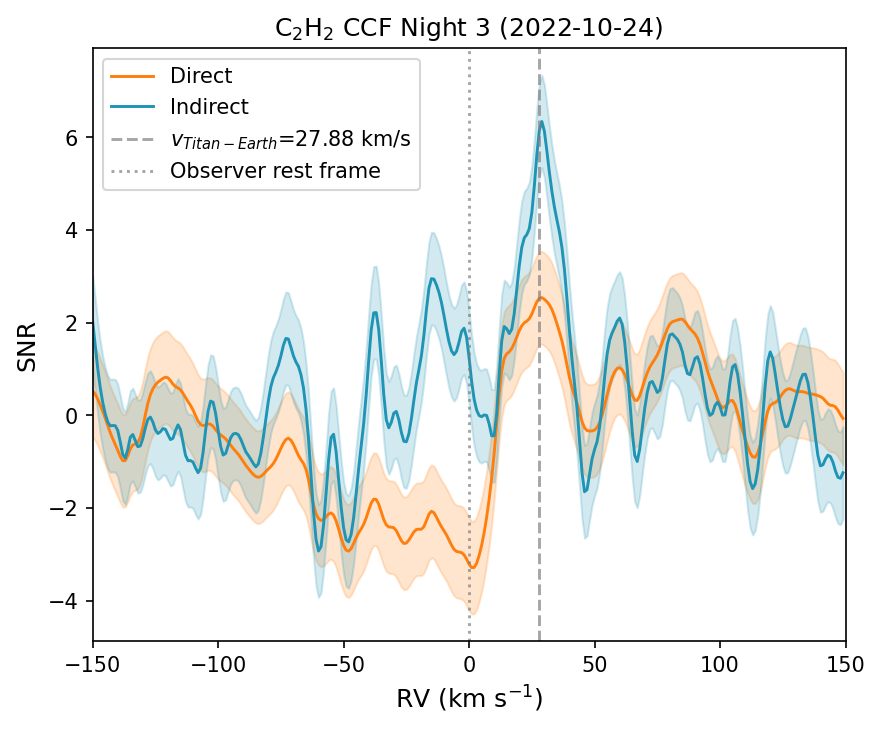}
    \end{subfigure}
    \begin{subfigure}{0.32\linewidth}
        \centering
        \includegraphics[width=\linewidth]{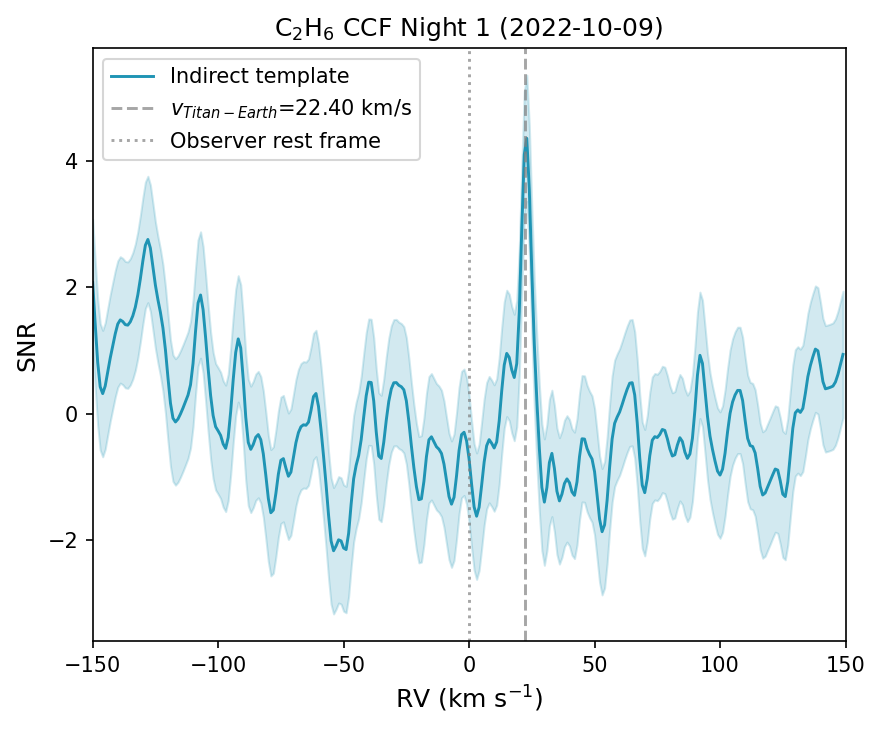}
    \end{subfigure} \hspace{0.01\linewidth}
    \begin{subfigure}{0.32\linewidth}
        \centering
        \includegraphics[width=\linewidth]{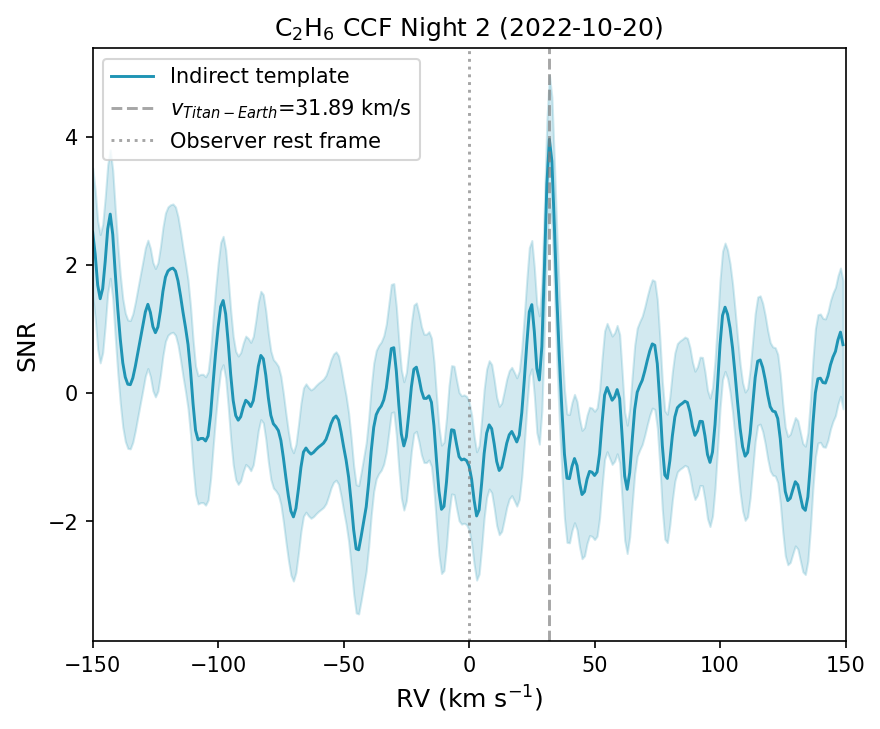}
    \end{subfigure} \hspace{0.01\linewidth}
    \begin{subfigure}{0.32\linewidth}
        \centering
        \includegraphics[width=\linewidth]{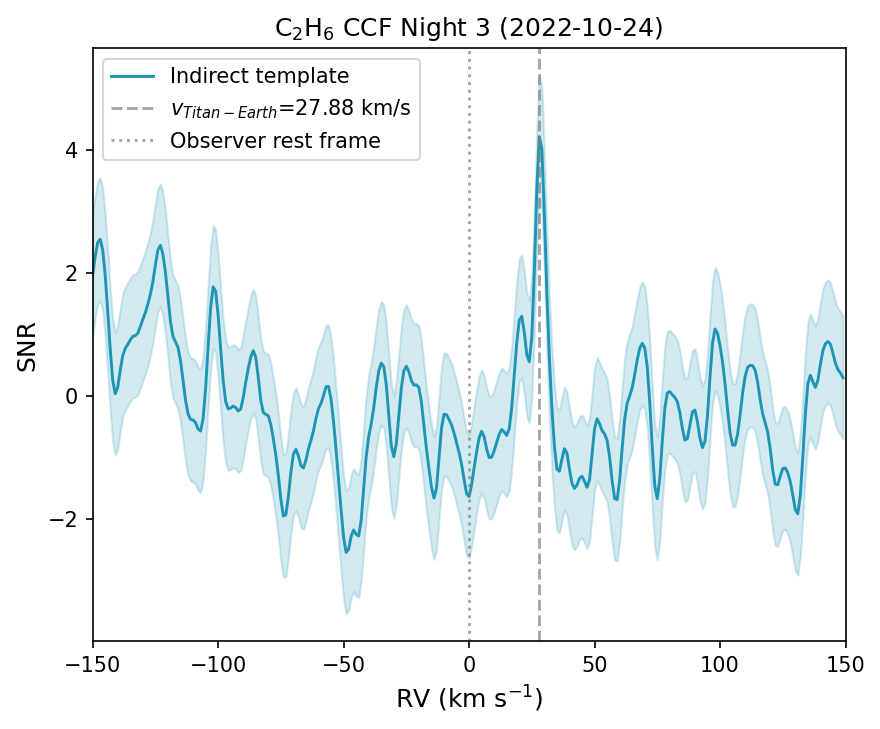}
    \end{subfigure}
    \caption{Cross‑correlation functions computed using Titan spectra in the geocentric frame and the optimal template for each molecular species. Each panel shows the CCF for one observing night, with the vertical dashed line marking the predicted Titan-Earth radial velocity. In the geocentric frame, telluric lines remain fixed at 0~km~s$^{-1}$, whereas Titan's lines appear Doppler‑shifted by the changing Titan-Earth velocity. The CCF peaks move consistently from night to night, tracking Titan’s varying radial velocity rather than remaining stationary at 0~km~s$^{-1}$. This behaviour demonstrates that the detected molecular signals, including methane, originate in Titan’s atmosphere: a telluric contribution would produce a peak fixed at 0~km~s$^{-1}$ and would not shift between nights.}
    \label{fig:ccf_geo}
\end{figure*}

\section{Cross-correlation between the molecular templates and the \ce{CH4} template}

\begin{figure*}
    \centering
    \begin{subfigure}{0.32\linewidth} 
        \centering
        \includegraphics[width=\linewidth]{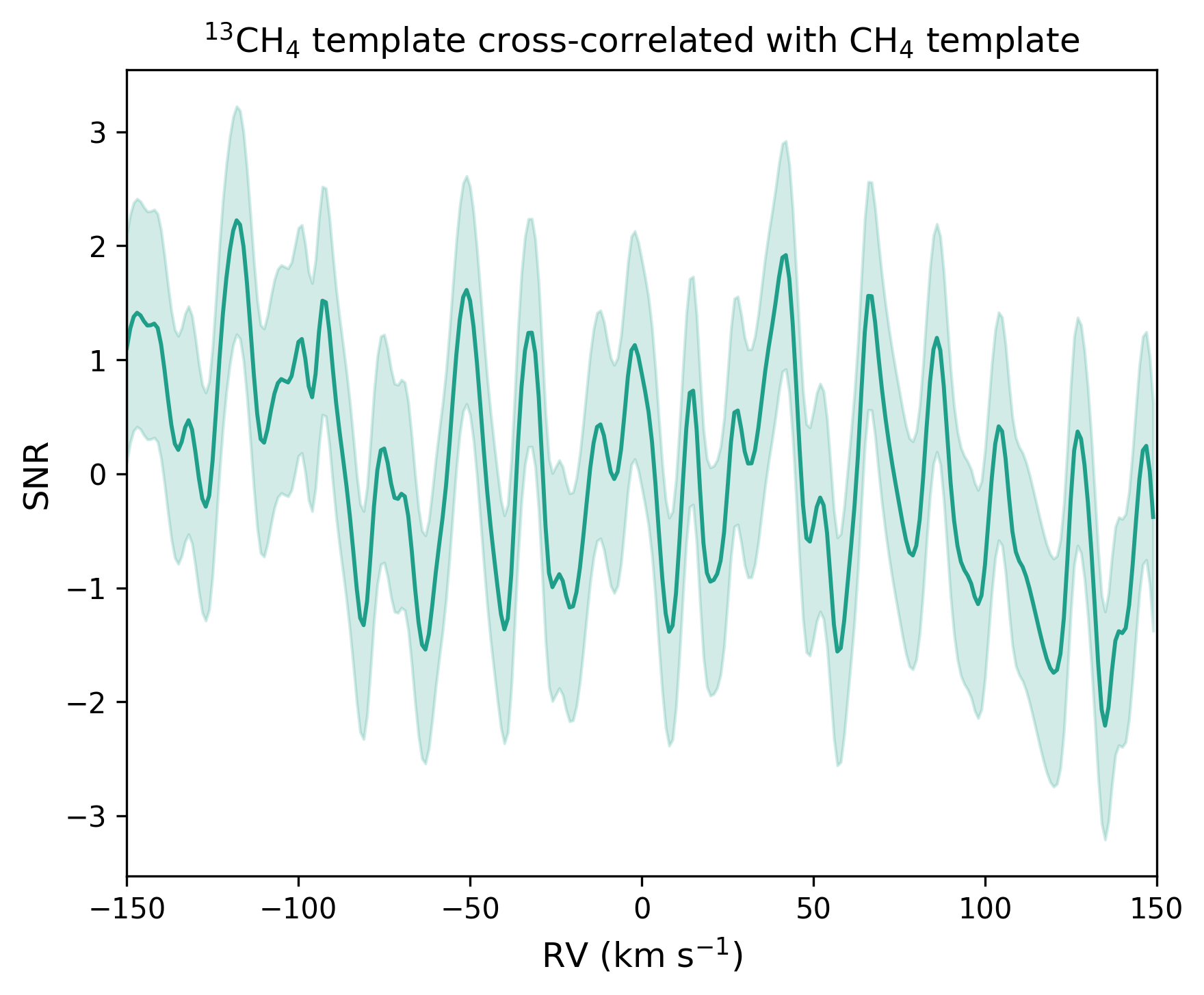}
    \end{subfigure} \hspace{0.01\linewidth}   
    \begin{subfigure}{0.32\linewidth} 
        \centering
        \includegraphics[width=\linewidth]{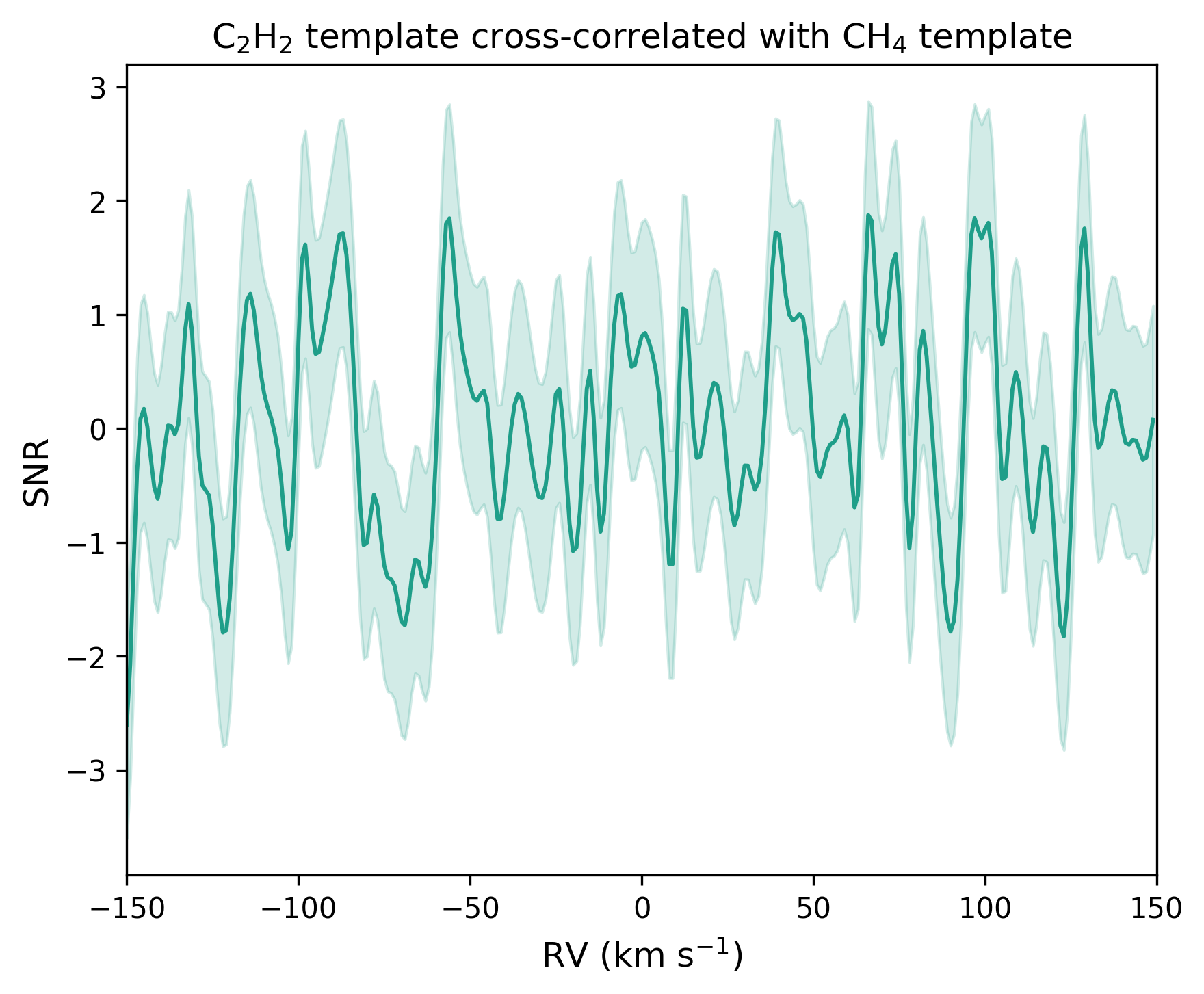}
    \end{subfigure} \hspace{0.01\linewidth}
    \begin{subfigure}{0.32\linewidth} 
        \centering
        \includegraphics[width=\linewidth]{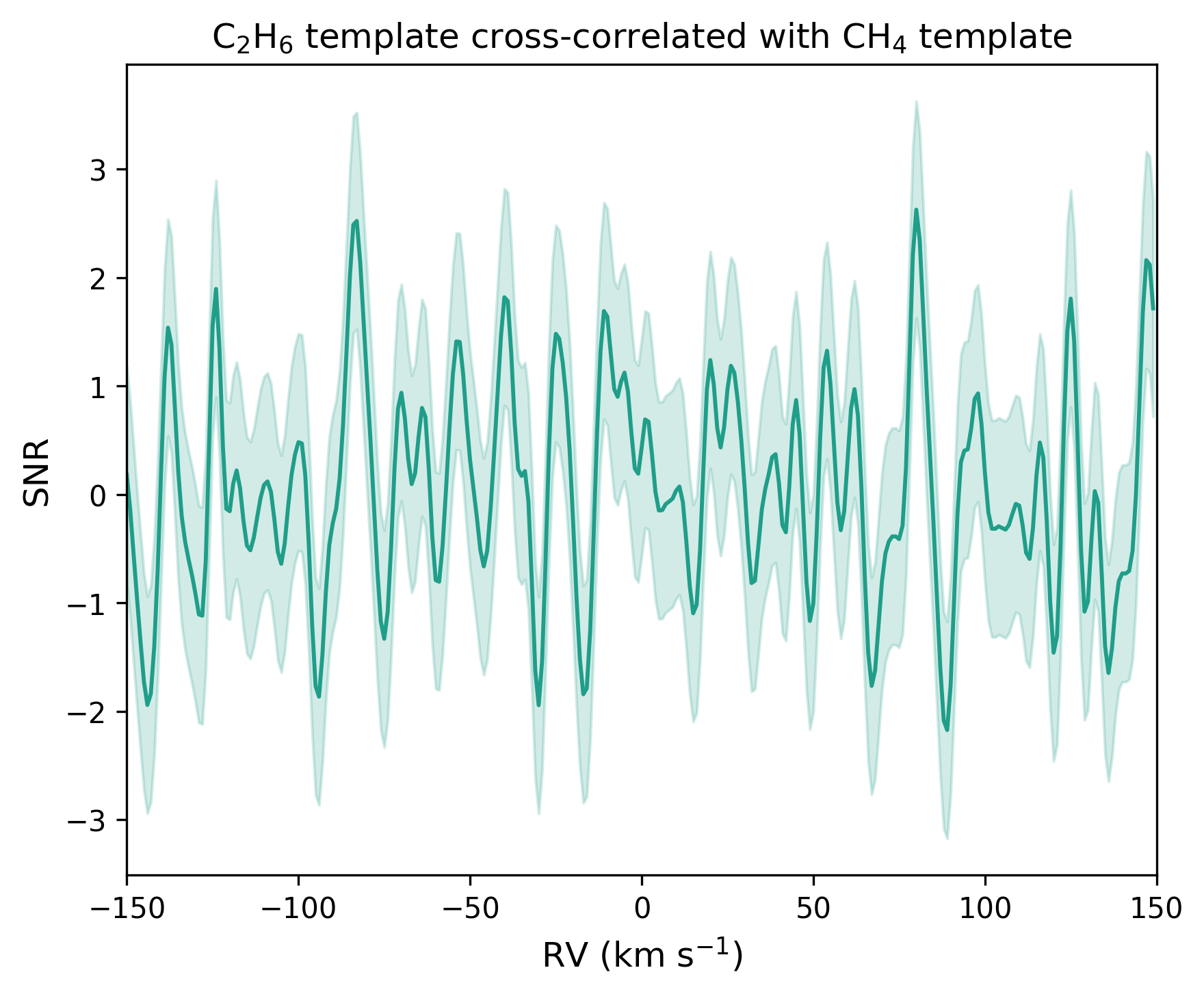}
    \end{subfigure}
    \caption{Cross‑correlation between each optimal molecular template and the \ce{CH4} template. These plots show the CCFs obtained by correlating the templates of $^{13}$\ce{CH4}, \ce{C2H2} and \ce{C2H6} with the \ce{CH4} template across a radial‑velocity range of \(\pm150\)~km~s\(^{-1}\). The absence of significant peaks at 0~km~s\(^{-1}\) demonstrates that the spectral structure of these molecules is not degenerate with that of methane in the CRIRES+ wavelength range. This behaviour is expected because the 4000--5000 cm\(^{-1}\) region corresponds primarily to overtone and combination bands rather than fundamental vibrational modes; these higher-order bands produce more complex and molecule‑specific line patterns, reducing the similarity between hydrocarbons even when their fundamental C--H stretching bands overlap. The lack of correlation with the \ce{CH4} template therefore confirms that the detections of $^{13}$\ce{CH4}, \ce{C2H2} and \ce{C2H6} in Titan’s atmosphere are genuine and not artefacts of methane contamination or template degeneracy.}
    \label{fig:ccf_with_CH4}
\end{figure*}

%%%%%%%%%%%%%%%%%%%%%%%%%%%%%%%%%%%%%%%%%%%%%%%%%%

% Don't change these lines
\bsp	% typesetting comment
\label{lastpage}
\end{document}